\documentclass[aps,twocolumn,reprint,tightenlines,floatfix]{revtex4-1}
 \usepackage{graphicx}
\usepackage{amsmath,amssymb,times} 
\usepackage{amsthm}
\usepackage{color}
\usepackage{mathtools}
\usepackage[T1]{fontenc} 
\usepackage[english]{babel} 
\usepackage{xcolor}
\usepackage[T1]{fontenc}

\newcommand{\ket}[1]{\left| #1 \right\rangle}
\newcommand{\bra}[1]{\left\langle #1 \right|}

\newcommand{\Tr}{\mathrm{Tr}}
\newcommand{\tr}{\mathrm{Tr}}

\newcommand{\mX}{\mathbb{X}}
\newcommand{\mY}{\mathbb{Y}}
\newcommand{\mZ}{\mathbb{Z}}
\newcommand{\mI}{\mathbb{I}}
\newcommand{\mG}{\mathbb{G}}
\newcommand{\mS}{\mathbb{S}}
\newcommand{\mP}{\mathcal{P}}
\newcommand{\beq}{\begin{equation}}
\newcommand{\eeq}{\end{equation}}

\newcommand{\expect}[1]{\left\langle#1\right\rangle}

\newcommand{\Seq}{\mathcal{S}}

\makeatletter
\def\@ssect@ltx#1#2#3#4#5#6[#7]#8{%
  \def\H@svsec{\phantomsection}%
  \@tempskipa #5\relax
  \@ifdim{\@tempskipa>\z@}{%
    \begingroup
      \interlinepenalty \@M
      #6{%
       \@ifundefined{@hangfroms@#1}{\@hang@froms}{\csname @hangfroms@#1\endcsname}%
       {\hskip#3\relax\H@svsec}{#8}%
      }%
      \@@par
    \endgroup
    \@ifundefined{#1smark}{\@gobble}{\csname #1smark\endcsname}{#7}%
  }{%
    \def\@svsechd{%
      #6{%
       \@ifundefined{@runin@tos@#1}{\@runin@tos}{\csname @runin@tos@#1\endcsname}%
       {\hskip#3\relax\H@svsec}{#8}%
      }%
      \@ifundefined{#1smark}{\@gobble}{\csname #1smark\endcsname}{#7}%
      \addcontentsline{toc}{#1}{\protect\numberline{}#8}%
    }%
  }%
  \@xsect{#5}%
}%
\makeatother
\begin{document}
\title{Testing the context-independence of quantum gates}
\author{Andrzej Veitia}
\email[email: ]{aveitia@gmail.com}
\author{S.J. van Enk}
\affiliation{Department of Physics and Center for Optical, Molecular and Quantum Sciences,
University of Oregon, Eugene, OR 97403}
\begin{abstract}
The actual gate performed on, say, a qubit in a quantum computer may depend, not just on the actual laser pulses and voltages we programmed to implement the gate, but on its {\em context} as well. For example, it may depend on what gate has just been applied to the same qubit, or on how much a long series of previous laser pulses has been heating up the qubit's environment. This paper analyzes several tests to detect such context-dependent errors (which include various types of non-Markovian errors). A key feature of these tests is that they are robust against both state preparation and measurement (SPAM) errors and gate-dependent errors. Since context-dependent errors are expected to be small in practice, it becomes important to carefully analyze the effects of statistical fluctuations and so we investigate the power and precision of our tests as functions of the number of repetitions and the length of the sequences of gates. From our tests an important quantity emerges: the logarithm of the determinant (log-det) of a probability (relative frequency) matrix $\mP.$ For this reason, we derive the probability distribution of the log-det estimates which we then use to examine the performance of our tests for various single- and two-qubit sets of measurements and initial states. Finally, we emphasize the connection between the log-det and the degree of reversibility (the unitarity) of a context-independent operation.
\end{abstract}
\maketitle
\section{Introduction} Precise control of operations on single microscopic systems lies at the heart of the progress in experimental quantum computing and quantum simulation~\cite{zhang2017observation,bernien2017probing,king2018observation,colless2018computation,hempel2018quantum,gambetta2017building}.
A given operation is implemented by programming a particular set of instructions for laser pulses (polarization, intensity and phase as explicit functions of time) and/or voltages or currents (as explicit functions of time). A given set of instructions, however, may not lead every time to the same microscopic implementation. Apart from fluctuations in laser and electronic properties, there is another, more subtle error possible. The actual gate performed may depend on the previous operation (e.g., because the tail end of a laser pulse may still be lingering around) or on how long ago the system was reset (because, for example, the temperature increases slowly but steadily after each reset operation), or because the system may have been interacting with a (quantum) memory that kept partial track of previous operations.  In Ref.~\cite{veitia2017macroscopic}  we and our collaborators summarized all these errors as ``context-dependence'' and preferred that term to ``non-Markovianity'' mostly because various different inequivalent definitions of the latter exist  (see Ref.~\cite{rivas2014quantum} for a review). Later on in this paper we will point out these definitions when our tests happen to detect one of the inequivalent instances of non-Markovian errors.

{\indent}In Ref.~\cite{veitia2017macroscopic}  three tests were introduced for detecting context-dependent errors. These tests involve preparing the system in a state $\rho_{i},$ running a specific sequence of instructions $\Seq$, and then recording  the outcome of a two-outcome measurement $k$. The measurement $k$ can be regarded as yielding either a ``click'' or not, and we use the positive operator $\Pi_k$ to describe the ``click'' outcome. (And so the no-click outcome corresponds to $I-\Pi_k$.) Such measurement, repeated $N_{s}$ times, gives the probability (relative frequency) $\mP_{k|i}(\Seq)$, which is defined as the probability with which click $k$ occurs given the input state $\rho_{i}.$ The approach of Ref.~\cite{veitia2017macroscopic} is to focus on the probability matrix $\mP(\Seq)$ (with entries $\mP_{k|i}(\Seq))$ obtained by preparing $d^{2}$ input states $\rho_{i}$ and measuring $d^{2}$ observables $\Pi_{k}$, where $d$ is the system's dimension. The tests in question consist in comparing sets of probability matrices $\mP(\Seq_{1}), \mP(\Seq_{2}),\ldots$ corresponding to specific (and structured) sets of sequences $\Seq_{1}, \Seq_{2},\ldots$ These sequences have the property that, in the absence of context-dependence, the corresponding probability matrices should exhibit certain well-defined symmetries. So, the tests  work by checking whether these symmetries are broken. An essential point here is that testing these symmetries does not require precise knowledge of either the actual input states or the observables measured. That is, these tests for context-dependence are robust against  state preparation and measurement (SPAM) errors \cite{Merkel2013}.

Unlike self-consistent quantum tomography (see \cite{Merkel2013,Stark2014,blume2013,sugiyamaself}), the tests discussed here do not attempt to (and cannot) reconstruct a set of gates. Nonetheless, in the absence of context-dependent errors, one of the tests can be used as a protocol for characterizing the degree of reversibility (i.e., the unitarity) of an operation. This protocol allows us to estimate the unitarity of a gate by examining the decay rate of the logarithm of the determinant (the log-det) of a sequence of probability matrices. This gives the unitarity thus defined a clear operational meaning.\\
{\indent}Due to the high quality (fidelity) of quantum gates in current state-of-the-art experiments (see e.g.~\cite{barends2014superconducting,blume2017demonstration,allcock2014}), most errors -- especially the context-dependent errors we are considering here -- are expected to be small. It is crucial, therefore, to understand how from a finite set of experimental data one can distinguish a context-dependent error from what is merely a statistical fluctuation (due to the finite number of experimental runs $N_{s}$). We present here a careful analysis of all three tests, with the aim of answering questions like (a) how many repetitions $N_{s}$ of gate sequences are needed to detect a certain size error? (b) how long should the gate sequences be? (c) what are the best (most efficient) ways of implementing the tests? (d) what is the precision of the unitarity estimates?

In this paper we examine these questions by including statistical fluctuations into the model for context-dependence considered previously in Ref.~\cite{veitia2017macroscopic}. This model contains standard gate errors like energy relaxation and dephasing  (characterized by $T_1$ and $T_2$ times) and SPAM errors. Context-dependent errors are modeled by having our computational qubit interact with a memory qubit. We assume we cannot do measurements on the memory qubit and so we must (and will) infer the context-dependence purely from measurements performed on the computational qubit, i.e., from $\mP(\Seq)$. In addition, independently of this model, we present a set of practical tools aimed at assessing the effect of statistical fluctuations for a given choice of gate sequences and SPAM scheme. We then show via simulations how these very same tools can be used to improve the power and precision of our tests.

While there are no crisp precise answers to the aforementioned questions in all generality, here are the four main takeaways from our investigations: The log-det of the measured probability matrix (see Eq.~(\ref{pd-test2}) in Section \ref{Sec:context}) is a remarkably useful quantity to characterize quantum gates. Under mild assumptions, testing longer sequences is more effective in reducing statistical errors in our estimates than increasing the number of experimental runs $N_{s}$ (see Section \ref{sec:uni} for details). The $F$ statistic, suitable for {\em nested} hypotheses, is a handy tool for deciding on the null hypothesis of {\em no} context-dependent errors (see Section \ref{Sec:fluct} for details). Symmetrically informationally complete measurements are more efficient than the standard Pauli measurements for both detecting context-dependent errors and improving the unitarity estimates (see Section \ref{Sec:SIC}).

This paper is organized as follows. We begin by reviewing, in Sec.~\ref{Sec:context}, the Liouville representation of a quantum map. We then show that application of this representation to a minimal tomographic scheme (requiring precisely $d^{2}\times d^{2}$ measurement configurations) leads to a family of SPAM-insensitive tests for context-dependence and CP-indivisibility, based on spectral properties of probability matrices. The determinant-based unitarity of a gate $u'(G)$ is discussed in detail in Sec.~\ref{subsec:iter-uni}, including its connection with the Lindblad master equation. Finally, with the purpose of elucidating our tests, we consider in Sec.~\ref{subsec:toy-model} an exactly solvable toy-model of context-dependence. In Sec.~\ref{Sec:zz} we describe a natural generalization of this toy-model, which includes dissipative effects and SPAM errors. This model is then used in subsequent sections to explore the impact of statistical fluctuations on the performance of our tests. In Sec.~\ref{Sec:fluct}, we discuss and apply various statistical tools to test the context-independence hypothesis, and we also check whether the premises underlying these (known) statistical methods are indeed fulfilled. 

In Sec.~\ref{sec:uni}, we focus on estimating the unitarity of a context-independent gate. More specifically, we provide bounds for the precision of the determinant-based unitarity estimates in terms of the lengths of the sequences employed in the tests, while taking into account the heteroskedasticity of the observations (mainly caused by decoherence). In addition, we compare our unitarity measure with that introduced in Ref.~\cite{wallman2015} and show that these will typically yield nearly the same values for high-fidelity gates.

Section~\ref{subsec:log-det-dist} deals with the problem of determining the distribution of the log-det estimates. We derive a handy expression for the variance of the log-det estimates as a function of the true probability matrix $\mP$ and $N_{s}.$ Based on this result, we show in Sec.~\ref{Sec:SIC} how using single-qubit and qutrit Symmetric Informationally Complete (SIC) \cite{appleby2014symmetric,renes2004symmetric,wootters2006quantum} sets (as input states $\rho_i=\ket{\psi_{i}}\bra{\psi_{i}}$ and measurement directions $\Pi_{k}=\ket{\psi_{k}}\bra{\psi_{k}}$) typically leads to smaller error bars (and hence is more efficient) than using more standard sets (eigenstates of Pauli operators). Furthermore, we use these findings in sections~\ref{Q} and \ref{QQ} to improve the precision of the log-det-based unitarity estimates of single- and two-qubit (context-independent) gates. Finally, in the Appendix we provide some useful facts concerning the statistics employed in this paper for hypothesis testing.
\section{Context-independence tests}
\label{Sec:context}
We say an operation $G,$ resulting from the physical implementation of an instruction $\mathcal{G}$, is context-independent if its action on the quantum system can be described by a map $\rho\rightarrow \rho'=G(\rho)$ that does not depend on $\mathcal{G}$'s position in any sequence of instructions $\mathcal{S}$. In the next subsection we discuss a useful representation of quantum maps that will later allow us to design SPAM-independent tests for context-dependence and/or non-Markovianity.
\subsection{The Liouville representation of quantum maps}
\label{subsec:Liouv}
{\indent}Let us consider a $d$-dimensional quantum system and a  linear map $ S: {L}(\mathcal{H}_{d})\rightarrow {L}(\mathcal{H}_{d}),$ where $L(\mathcal{H}_{d})$ denotes the vector space of linear operators acting on the system's Hilbert space $\mathcal{H}_{d}.$ To describe compositions of maps, it will prove convenient to work in the Liouville representation, wherein all the information about the linear map $S$ is contained in a $d^{2}\times d^{2}$ matrix with entries given by 
\beq 
\label{matrep}
S_{nm}=\frac{1}{d}\tr[P_{n}S(P_{m})], 
\eeq
Here,~$\{P_{n}\}_{n=1}^{d^{2}}$ is a Hermitian operator basis such that its elements are orthogonal with respect to the Hilbert-Schmidt inner product, i.e., $(P_{n}|P_{m}):=\tr(P_{n}P_{m})=d\delta_{nm}.$ (We shall abuse notation by using the same symbol to denote both a map and its matrix representation.) From the above definition, it follows that any matrix $S$ representing a  hermiticity-preserving map, i.e., $S(A)^{\dagger}=S(A^{\dagger}),$ has real entries. It is also worth noting that the matrix representation $S', $ of the map $S(\cdot)$, in a different orthogonal basis $\{P'_{n}\}_{n=1}^{d^{2}}$, such that $P'_{n'}=\sum_{n}O_{n'n}P_{n}$ (with $O\in \mathbb{R}^{d^{2}\times d^{2}}$ satisfying $OO^{T}=I_{d^{2}}$), is related to the matrix Eq.~(\ref{matrep}) by the transformation 
\beq
\label{orto}
S'=OSO^{T}. 
\eeq
Since a context-independence test should not depend on the (orthogonal) operator basis we choose to describe a map $S$, any such test should be insensitive to a transformation of the form Eq.~(\ref{orto}). In addition, in the language of Gate Set Tomography (GST) \cite{blume2013}, a gate-set admits a more general transformation; specifically, GST allows a gauge transformation $G_{i}\rightarrow T_{\text{gauge}}G_{i}T^{-1}_{\text{gauge}}$, $T_{\text{gauge}} \in \text{GL}_{d^{2}}(\mathbb{R})$, which is compatible with the observation (i.e., the data) and does not alter the predictions generated by that gate-set (see \cite{blume2013} for details). The tests for context-independence introduced in Ref.~\cite{veitia2017macroscopic} are based on spectral properties of the matrices representing a series of sequences and are, therefore, gauge invariant  (in the sense of GST) and, in particular, independent of the choice of the operator basis $\{P_{n}\}_{n=1}^{d}.$\\
{\indent}Now, expressing an operator $A:\mathcal{H}_{d}\rightarrow \mathcal{H}_{d}$ as $A=1/d \sum_{k}\tr[P_{k}A]P_{k}$ and making use of the definition Eq.~(\ref{matrep}), we readily find that $S(A)$ is given by
\beq
\label{mapaction}
S(A)=\frac{1}{d}\sum_{n,m}S_{nm}\tr[P_{m}A]P_{n}, 
\eeq
where we made use of the fact that $S$ is linear.~Furthermore, the above equation implies that the action of the composition $S_{2}\circ S_{1},$ of two linear maps $S_{2}$ and $S_{1},$ on an operator $A$ is
\beq 
\label{matmul}
S_{2}(S_{1}(A))=\frac{1}{d} \sum_{n,m}(S_{2}S_{1})_{nm}\tr[P_{m}A]P_{n}
\eeq
A comparison between equations (\ref{mapaction}) and (\ref{matmul}) reveals that the matrix representing the composition $S_{2}\circ S_{1}$ is simply given by the matrix product $S_{2}S_{1}$. Unsurprisingly, this property of the Liouville representation will turn out to be especially convenient when testing for context-dependence. \\
 {\indent}The following are some additional aspects of the Liouville representation that are relevant to this work. If a linear map $S$ is trace-preserving (TP), that is $\Tr[S(A)]=\Tr[A]$, then it is clear that it must preserve the traces of all the basis elements $\{P_{n}\}_{n=1}^{d^{2}}.$~Using Eq.~(\ref{mapaction}) we find that the trace-preservation condition can be succinctly stated -- in a basis-independent fashion -- as follows 
\beq 
\label{TP}
S^{T}{\tau}_{r}={\tau}_{r},
\eeq
where $\tau_{r}=1/d[\tr[P_{1}],\ldots, \tr[P_{d^{2}}]]^{T}.$~Hence, a TP map must have an eigenvalue equal to 1. An alternative way of expressing the trace preservation conditions is $S^{\dagger}(I_{d})=I_{d}$, where $S^{\dagger}$ denotes the Hermitian conjugate of the map $S$ with respect to the Hilbert-Schmidt inner product, that is, $(A|S(B))=(S^{\dagger}(A)|B).$ On the other hand, a unital map, i.e., a map that preserves the identity $I_{d}$, must satisfy the condition $S\tau_{r}=\tau_{r}$, which can be easily proved using Eq.~(\ref{mapaction}), together with the orthogonality relation $\tr[P_{n}P_{m}]=d\delta_{nm}.$ Note that thus far, we have not made use of a particular choice of basis $\{P_{n}\}_{n=1}^{d^{2}}$ (the context-independence tests that will be described here do not require choosing a specific basis).~A common choice is $P_{1}=I_{d},$ for which the  remaining basis element must be traceless (e.g., the generalized Pauli matrices). The condition Eq.~(\ref{TP}) implies that in such bases, the matrix representation of a TP map must assume the form~\cite{king2001minimal}
\beq
\label{matTP}
S=
\left[
\begin{array}{c|c}
1 & 0_{1\times(d^{2}-1)} \\
\hline
\vec{\kappa} & W_{S}
\end{array}
\right],
\eeq
where the $(d^{2}-1)\times (d^{2}-1)$ matrix $W_{s}$ and the $(d^{2}-1)\times 1$ vector $\vec{\kappa}$ are called the unital and non-unital parts of $S$, respectively. Finally, from the definition Eq.~(\ref{matrep}) it follows that the Hermitian conjugate $S^{\dagger}$, of a map $A\rightarrow S(A),$ is represented by the matrix $S^{T}.$ This implies that the matrix representing a unitary operation $S_{U}(A):=UAU^{\dagger},$ where $UU^{\dagger}=I_{d}$, satisfies $S_{U}S_{U}^{T}=I_{d^{2}}$ and thus, $|\det(S_{U})|=1.$ Furthermore, the fact that a unitary matrix can be written as $U=\sqrt{U}\sqrt{U}$, where $\sqrt{U}$ is also unitary, implies that $\det(S_{U})=1.$
\subsection{Quantum process tomography revisited}
\label{subsec:QPT}
\begin{figure}[t!]
\includegraphics[width=8.5cm]{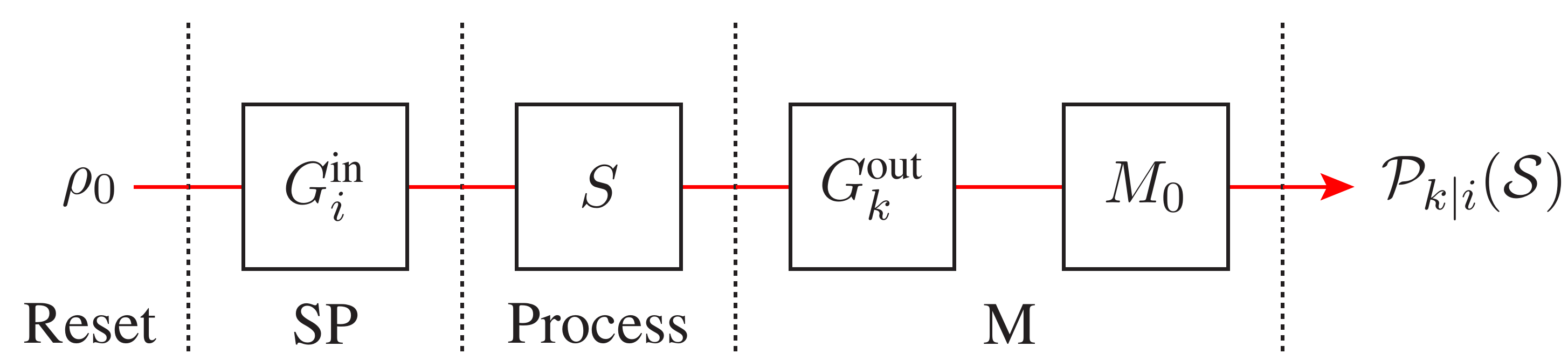}
\caption{Schematic description of the state preparation (SP), process  and measurement (M) stages of a generic quantum process tomography (QPT) scheme. The initial state $\rho_{0}$ of a $d$-dimensional system and the POVM effect $M_{0}$ are assumed to be fixed. In this scheme, the $d^4$ sequences $\mathcal{G}_{k}^{\text{out}}\circ \Seq\circ \mathcal{G}_{i}^{\text{in}}$, where $i,k=1,\ldots,d^{2},$ lead to a $d^{2}\times d^{2}$ probability matrix $\mP(\Seq)$ having entries $\mP_{k|i}(\Seq)$, which in the absence of context-dependence are given by $\mP_{k|i}(\Seq)=\tr[M_{k}^{\text{exp}}S(\rho_{i}^{\text{exp}})].$ 
\label{fig:spam-circuit}}
\end{figure}

The property Eq.~(\ref{matmul}) implies that if one could reliably determine the matrix representations $\{G_{i}\}$  corresponding to a  set of instructions $\{\mathcal{G}_{i}\}$, then testing for context-dependence would simply amount to verifying whether each product $G_{i}G_{j}$ matches -- within some tolerance --  the operation resulting from the implementation of the sequence of instructions $\mathcal{G}_{i}\circ\mathcal{G}_{j}$.~To explain how these tests work, let us consider the \emph{minimal} tomographic scheme depicted in Fig.~\ref{fig:spam-circuit}. The set  of gates $\{G_{i}^{\text{in}}\}_{i=1}^{d^{2}}$ is used to  prepare $d^{2}$ input states $\{\rho_{i}^{\text{exp}}\}_{i=1}^{d^{2}}$ while the gates  $\{G_{k}^{\text{out}}\}_{k=1}^{d^{2}}$ rotate the measurement axes, before measuring some POVM effect $M_{0}.$ As in Randomized Benchmarking (RB) \cite{Emerson2005,Knill2008,Magesan2011} and GST protocols, we take the initial state of the system $\rho_{0}$ and the POVM effect $M_{0}$ to be fixed \footnote{Note this is just a convenient mathematical assumption, made to simplify our discussion. Due to possible fluctuations in both $\rho_{0}$ and $M_{0},$ a more realistic assumption is that for any sequence $\Seq'_{ki}=\mathcal{G}_{k}\circ\Seq \circ \mathcal{G}_{i}$ the probability estimates $\hat{\mathcal{P}}_{k|i}$ are given by $\Tr[\bar{M}_{0}\bar{S'}_{ki}(\bar{\rho}_{0})]$ (in the infinite sample-size limit), where $\bar{\rho}_{0}$ and ${\bar{M}_{0}}$ represent the (fixed) averages of the operators
 $\rho_{0}$ and $M_{0}$ over experimental runs}. Furthermore, if we assume that all the operations involved in our tomographic scheme are context-independent, then this scheme yields a $d^{2}\times d^{2}$ probability matrix whose entries are (in the infinite sample-size limit $N_{s}\rightarrow \infty$), given by $\mathcal{P}_{k|i}(\Seq)=\tr[M_{k}^{\text{exp}}S(\rho_{i}^{\text{exp}})]$, where $\rho^{\text{exp}}_{i}:=G^{\text{in}}_{i}(\rho_{0})$ and 
$M^{\text{exp}}_{k}:={G^{\text{out}}_{k}}^{\dagger}(M_{0})$ (which satisfy $0\leq M_{k}^{\text{exp}}\leq I_{d}$~\cite{busch2003quantum}). Now, choosing an operator basis $\{P_{n}\}_{n=1}^{d^{2}}$ and making use of Eq.~(\ref{mapaction}) we can readily express the probability matrix $\mP(\Seq)$ as 
\beq 
\label{main}
\mP(\Seq)=\Phi_{\text{out}}^{T}S\Phi_{\text{in}},
\eeq
where the matrix $S$ is given by Eq.~(\ref{matrep}) and $\Phi_{\text{in (out)}}$ are real-valued $d^{2}\times d^{2}$ matrices with entries 
\begin{eqnarray}
\label{Phiin}
&(\Phi_{\text{in}})_{mi}&=\frac{1}{\sqrt{d}}\tr[P_{m}\rho^{\text{exp}}_{i}]=\frac{1}{\sqrt{d}}\tr[P_{m}G_{i}^{\text{in}}(\rho_{0})], \\  
\label{Phiout}
&(\Phi_{\text{out}})_{nk}&=\frac{1}{\sqrt{d}}\tr[P_{n}M_{k}^{\text{exp}}]=\frac{1}{\sqrt{d}}\tr[P_{n}{G^{\dagger}}_{k}^{\text{out}}(M_{0})].
\end{eqnarray}
Thus, if we knew the actual sets $\{\rho_{i}^{\text{exp}}\}$ and $\{M_{k}^{\text{exp}}\}$, then we could invert the matrix equation Eq.~(\ref{main}) to estimate the process $S$ from the experimental data $\hat{\mP}(\Seq)$ (i.e., the estimate of the true matrix $\mP(\Seq)$), provided both sets of operators $\{\rho_{i}^{\text{exp}}\}$ and $\{M_{k}^{\text{exp}}\}$ are linearly independent, i.e., 
\beq
\label{cond:inv}
\det(\Phi_{\text{in}}\Phi_{\text{out}})\neq 0. 
\eeq
This condition can be verified experimentally by computing  $\det(\mP(\mathcal{I}_{\emptyset}))=\det(\Phi_{\text{out}}\Phi_{\text{in}})$, where $\mathcal{I}_{\emptyset}$ represents the null instruction, i.e., the instruction to do nothing between the state preparation (SP) and measurement (M) stages (see Fig.~\ref{fig:spam-circuit}). In the limit $N_{s}\rightarrow \infty$, the context-independence tests presented in Ref.~\cite{veitia2017macroscopic} do not require any knowledge about the actual input states $\{\rho_{i}^{\text{exp}}\}$ and observables $\{M_{k}^{\text{exp}}\}$ and are, therefore, insensitive to the SPAM specifics, as long as the (verifiable) condition Eq.~(\ref{cond:inv}) is met. In practice, however, because of statistical fluctuations of the probability estimates $\hat{\mP}(\Seq)$ (due to finite sampling), some degree of control of the input states and measurements will be required to guarantee both the power and precision of the tests.~In fact, as discussed in Sec.~\ref{Sec:SIC}, certain target tomographic sets $\{\rho_{i}=\ket{\phi_{i}}\bra{\phi_{i}},\Pi_{k}=\ket{\varphi_{k}}\bra{\varphi_{k}}\}$ will, typically, lead to a better  performance of the tests than others. Thus, it will sometimes be convenient (though not essential) to consider the raw map \cite{veitia2017macroscopic}
\beq 
\label{sraw}
S^{\text{raw}}:=({\Phi^{\text{ideal}}_{\text{out}}}^{T})^{-1}\mP(\Seq)(\Phi^{\text{ideal}}_{\text{in}})^{-1},
\eeq
where $\Phi^{\text{ideal}}_{\text{out}}=\Phi_{\text{out}}\rvert_{M_{k}^{\text{exp}}=\Pi_{k}}$ and $\Phi^{\text{ideal}}_{\text{out}}=\Phi_{\text{in}}\rvert_{\rho_{i}^{\text{exp}}=\rho_{i}}.$ So, in the absence of SPAM errors, $S^{\text{raw}}=S$, up to statistical fluctuations.\\
{\indent}The key observation that allows us to test for context-dependence is that although the matrices $\Phi_{\text{in(out)}}$, in Eq.~(\ref{main}) are unknown, the relation $\mP(\Seq)=\Phi_{\text{out}}^{T}S\Phi_{\text{in}}$ -- which links the data $\mP(S)$ with the map $S$ -- should hold for any sequence $\Seq$, in the absence of context-dependence. Note that the matrices $\Phi_{\text{in}}$ and $\Phi_{\text{out}}$ are not unique.~For example, the transformation $\Phi_{\text{in}}\rightarrow \alpha \Phi_{\text{in}}$ and $\Phi_{\text{out}}\rightarrow \alpha^{-1} \Phi_{\text{out}}$ (with $\alpha\neq 0$) does not modify the relation Eq.~(\ref{main}). Moreover, $\alpha$ could be taken to depend on $S.$ The important point, however, is that the data $\mP(\Seq_{1}), \mP(\Seq_{2}),\ldots$, corresponding to the set of instructions $\Seq_{1}, \Seq_{2}, \ldots$, should be generated via the relation Eq.~(\ref{main}), using two fixed matrices $\Phi_{\text{in}}$ and $\Phi_{\text{out}}$ (e.g., those in Eqs.~(\ref{Phiin}) and (\ref{Phiout})). Finally, note that relation Eq.~(\ref{main}) may also be expressed in the following, somewhat different, form
\beq
\text{vec}[\mP(\Seq)]=\Phi\text{vec}[S], \quad \text{where}\quad \Phi=\Phi_{\text{in}}^{T}\otimes \Phi_{\text{out}}^{T},
\eeq
where $\text{vec}[A]$ denotes the vector obtained by stacking the columns of the matrix $A$.~However, this vectorized form of Eq.~(\ref{main}) does not serve our purposes because it obscures the relationship between compositions of maps and the data. On the other hand, working with expression Eq.~(\ref{main}) will allow us to take advantage of some well-known facts concerning squared matrices, as shown in the next subsections.   
\subsection{Permutational tests}
\label{subsec:perm}
{\indent}We will now establish a family of tests for context-dependence by checking whether a set of probability matrices $\{\mP(\Seq_{k})\}_{k=1}^{N}$ is compatible with the relation Eq.~(\ref{main}) and the assumptions that the gates in the sequences are context-independent.~For example, let us consider the sequences of instructions $\Seq=\mathcal{G}_{2}\circ\mathcal{G}_{1}$ and $\Seq'=\mathcal{G}_{1}\circ \mathcal{G}_{2}.$ Then, according to our definition of context-independence, these sequences of instructions should result in the operations $S=G_{2}G_{1}$ and $S'=G_{1}G_{2}.$ From Eq.~(\ref{main}) we readily see that context-independence imposes the following constraint on the probability matrices $\mP(\Seq)$ and $\mP(\Seq')$: $\det(\mP(\Seq))=\det(\mP(\Seq')).$ In general, for any sequence of instructions $\Seq_{1}=\mathcal{G}_{m}\circ \mathcal{G}_{m-1} \circ \ldots \circ \mathcal{G}_{1}$, we can consider the permuted sequence $\Seq_{\sigma}= \mathcal{G}_{\sigma({m})}\circ \mathcal{G}_{\sigma(m-1)} \circ \ldots \circ \mathcal{G}_{\sigma(1)}.$ Then for any permutation $\sigma$, the probability matrix $\mP_{\sigma}:=\mP(\Seq_{\sigma})$ will obey \cite{veitia2017macroscopic}
\beq 
\label{pd-test}
\det(\mP_{\sigma})=\det(\Phi_{\text{out}}\Phi_{\text{in}})\prod_{i=1}^{m}\det(G_{i})=\det(\mP_{1}),
\eeq
where $\mP_{1}:=\mP(\Seq_{1})$, provided the relevant gates are context-independent. Note that the constant $\det(\Phi_{\text{out}}\Phi_{\text{in}})$ is representation-independent and equal to the determinant of the matrix $\tr[M_{k}^{\text{exp}}\rho_{i}^{\text{exp}}].$ It is clear that this permutational determinant test (PD-test) is insensitive to SPAM errors as these cannot trigger a statistically significant difference between $\det(\mP_{\sigma})$ and $\det(\mP_{1}).$ Furthermore, note this test does not make use of a particular matrix representation (see Eq.~(\ref{matrep})) of the process $S.$ As mentioned in \cite{veitia2017macroscopic}, this test will not detect unitary context-dependent errors because $\det(G)=1$ when $G$ represents a unitary operation (as shown in Sec.~\ref{subsec:Liouv}). Finally, it will be prove convenient to express the PD-test as the invariance of the log-det $L_{\sigma}:=\log(|\det(S_{\sigma}^{\text{raw}})|).$ That is,
\begin{eqnarray}
\label{pd-test2} 
L_{\sigma}&=&\log(|\det(S^{\text{raw}}_{\sigma})|)\nonumber \\
                    &=&-\log(|\det(\mP_{0}^{\text{ideal}})|)+\log(|\det(\mP_{\sigma})|)=L_{1},
\end{eqnarray}
for any permutation $\sigma.$ Here we made use of Eq.~(\ref{sraw}) and introduced the constant matrix $\mP_{0}^{\text{ideal}},$ with entries given by $(\mP_{0}^{\text{ideal}})_{k|i}:=\Tr[\Pi_{k}\rho_{i}],$ where $\{\rho_{i}, \Pi_{k}\}$ is our target tomographic set.\\
{\indent}We can further exploit relation (\ref{main}) by considering a ``reference'' sequence $\mathcal{S}_{0},$ whose meaning will be become clear below. Let us apply the tomographic scheme Fig.~\ref{fig:spam-circuit} to two sequences $\Seq_{0}$ and $\Seq_{1}$ and let $\mP(\Seq_{0})$ and $\mP(\Seq_{1})$ be the resulting probability matrices. Then, using the relation~Eq.~(\ref{main}) we find that $\mP(\Seq_{1})\mP^{-1}(\Seq_{0})=\Phi_{\text{out}}^{T}S_{1}S_{0}^{-1} (\Phi_{\text{out}}^{T})^{-1}$, which implies that $\text{Spec}[\mP(\Seq_{1})\mP^{-1}(\Seq_{0})]=\text{Spec}(S_{1}S_{0}^{-1}).$ Choosing $\Seq_{1}= \Seq \circ \Seq_{0}$ and assuming context-independence (i.e., $S_{1}=SS_{0}$) we obtain the following expression for the spectrum of the map $S$, in terms of probability matrices
\beq
\label{spec}
\text{Spec}(S)=\text{Spec}[\mP(\Seq\circ \Seq_{0}) {\mP^{-1}(\Seq_{0})}]. 
\eeq
The reference sequence $\Seq_{0}$ can be chosen arbitrarily as long as the matrix $\mP(\Seq_{0})$ is invertible, which will the case for a sufficiently short sequence $\Seq_{0}$ \footnote{The determinant of the matrix representing a short sequence $\Seq_{0}$ of high fidelity gates will be close to 1. Therefore, $\det(\mP(\Seq_{0}))\approx \det(\Phi_{\text{out}})\det(\Phi_{\text{in}})\neq 0$.}. In particular, we can choose $\Seq_{0}=\mathcal{I}_{\emptyset}$, that is, the instruction to do nothing between SP and M -- which should not be confused with $\mathcal{I},$ the instruction to apply an idle gate of finite duration. Substituting $\mP(\Seq)=\mP(\Seq\circ \mathcal{I}_{\emptyset})$ and $\mP_{0}:=\mP(\mathcal{I}_{\emptyset})$ in Eq.~(\ref{spec}), we obtain the expression $\text{Spec}(S)=\text{Spec}[\mP(\Seq)\mP_{0}^{-1}]$ which allows us to establish the following permutational test for context-dependence. Consider a sequence of $m$ instructions $\Seq_{1}=\mathcal{G}_{m}\circ \mathcal{G}_{m-1}\circ \ldots \circ \mathcal{G}_{1}$ and cyclic permutations thereof, i.e., $\Seq_{\sigma'}=\mathcal{G}_{\sigma'(m)}\circ \mathcal{G}_{\sigma'(m-1)}\circ \ldots \mathcal{G}_{\sigma'(1)}$ ($\sigma'$ is a cyclic permutation of ${1,2,\ldots m}$).~Then, if the operations involved are context-independent, we have the permutational symmetry \cite{veitia2017macroscopic}
\begin{eqnarray}
\label{spec-cycle}
\text{Spec}(\mP_{\sigma'}\mP_{0}^{-1})&=&\text{Spec}[G_{\sigma'(m)}G_{\sigma'(m-1)}\ldots G_{\sigma'(1)}] \nonumber \\
&=&\text{Spec}(\mP_{1}\mP_{0}^{-1}),
\end{eqnarray}
where $\mP_{1}:=\mP(\Seq_{1})$ and $\mP_{\sigma'}:=\mP(\Seq_{\sigma'}).$ In deriving Eq.~(\ref{spec-cycle}) we made use 
of the fact that the spectrum of product of matrices is invariant under cyclic permutations of the matrices. Since the spectrum of a $d^{2}\times d^{2}$ matrix $A$ is completely specified by the set of traces $\{\tr(A^{r})\}_{r=1}^{d^{2}}$ (these specify the coefficients of the characteristic polynomial $\chi(z)=\det(zI_{d^{2}}-A)$), we can express the test Eq.~(\ref{spec-cycle}) in terms of the invariance of the fidelities 
\beq 
\label{cycle-test-trace}
\mathcal{F}^{(r)}_{\sigma'}=\frac{1}{d^{2}}\tr[(\mathcal{P}_{\sigma'}\mP_{0}^{-1})^{r}],  
\eeq
for $r=1,2,\ldots, d^{2}$. Note from Eq.~(\ref{spec}) that the quantity $\mathcal{F}_{\sigma'}^{(r)}$ is just the process fidelity \cite{Horodecki1999,Nielsen2002} of the map $S_{\sigma'}^{\circ r}$, with respect to the identity.~Clearly, both permutational tests, described by  Eqs.~(\ref{pd-test}) and (\ref{cycle-test-trace}), require sequences involving at least two different gates (instructions). In addition, the sequences of instructions employed should be long enough to amplify the context-dependence effects, if present, to ensure statistical significance.~A drawback of these tests is that they require guessing the right sequences and permutations in order to detect context-dependence. Nonetheless, these tests could be used to verify (or rule out)  a specific family of models, without having to worry about SPAM errors. Individual gates can be studied, and partially characterized, by means of iterative tests~\cite{veitia2017macroscopic}, which make use of sequences of the form $\mathcal{S}_{m}={\mathcal{G}\circ \mathcal{G}\circ\ldots\mathcal\circ\;\mathcal{G}}$ (i.e., instruction $\mathcal{G}$ is repeated $m$ times).
\subsection{SPAM-independent approach to CP-indivisibility}
\label{subsec:spam-cp}
{\indent}Before proceeding to discuss iterative tests for context-independence, we briefly make contact with the problem of CP-divisibility (non-Markovianity) \cite{Wolf2008,rivas2010entanglement,chruscinski2014degree,bernardes2015experimental}. It turns out that Eq.~(\ref{main}) may be used to construct a CP-indivisibility witness which does not depend on the SPAM details, provided we make an additional assumption. Recall that the relation Eq.~(\ref{main}) was derived under the assumptions that {\bf{(i)}} the process $S$ does not depend on the gates $\{G_{i}^{\text{in}}\}$ used to prepare the input states, and {\bf{(ii)}} the output gates $\{G_{k}^{\text{out}}\}$ do not depend on either the sequence $S$ or the input gates $\{G_{i}^{\text{in}}\}$ (plus the reasonable assumption that $\rho_{0}$ and $M_{0}$ are fixed). This set of assumptions is weaker than the context-independence of \emph{all} the operations on a given quantum system. Indeed, it could be the case that context-dependence is only present in the sequence of operations  $\Seq$, in which case the relation Eq.~(\ref{main}) holds. Thus, let us assume that the probabilities in scheme Fig.~\ref{fig:spam-circuit} are given by $\mP_{k|i}(\Seq)=\tr[M^{\text{exp}}_{k}S(\rho^{\text{exp}}_{i})]$ (for any $S$), where $\{\rho_{i}^{\text{exp}}\}$ and $\{M^{\text{exp}}_{k}\}$ 
are unknown, yet \emph{fixed}, input states and observables; we will refer to such state preparation and measurement as ``fixed-SPAM''. Now, following the approach of Ref.~\cite{rivas2010entanglement}, we consider two sequences of instructions $\Seq_{m_{0}}$ and $\Seq_{m}=\Seq_{m,m_{0}}\circ \Seq_{m_{0}}$ and assume that the corresponding maps $S_{m_{0}}$ and $S_{m}$ are CPTP. The problem of CP-divisibility addresses the question of whether the map corresponding to the sequence $\Seq_{m,m_{0}}$ is CPTP as well. The idea of Ref.~\cite{rivas2010entanglement} is to examine the positivity of the  Choi-Jamio{\l}kowski matrix $\rho_{S_{m,m_{0}}}$ \cite{choi1975completely,jamiolkowski1972linear} (see also Ref.~\cite{wolk2018consistency}) associated with the map $S_{m,m_{0}}:=S_{m}S_{m_{0}}^{-1}$. Because we have assumed that the operators  $\{\rho_{i}^{\text{exp}}, M^{\text{exp}}_{k}\}$ are unknown, it is clear that we cannot reconstruct the Choi-Jamio{\l}kowski matrix $\rho_{S_{m,m_{0}}}$.~However, if $\{\rho_{i}^{\text{exp}}, M^{\text{exp}}_{k}\}$ are fixed, we can make use of Eq.~(\ref{spec}), which yields the relation  $\text{Spec}(S_{m,m_{0}})=\text{Spec}[\mP(S_{m})\mP^{-1}(S_{m_{0}})].$ On the other hand, it is known that the eigenvalues of a CPTP map lie on the unit disc $|\lambda_{i}|\leq 1$ (see e.g., Ref.~\cite{rudnicki2018gauge}). In fact, this result does not require complete positivity; it suffices to assume just positivity and trace preservation~\cite{Wolf2008}. Therefore, if $S_{m,m_{0}}$ is CPTP, the pair of probabilities matrices $\mP(S_{m})$, $\mP(S_{m_{0}})$ must obey the inequality
\beq 
\label{spec-radius}
\mathcal{R}_{m,m_{0}}:=\mathcal{R}[\mP(S_{m})\mP(S_{m_{0}})^{-1})]\leq 1,
\eeq
where $\mathcal{R}[A]$ denotes the spectral radius of $A.$ Thus, we can make use of the spectral radius $\mathcal{R}_{m,m_{0}}$ to witness CP-indivisibility, in a way which is insensitive to fixed SPAM errors.\\
{\indent} A closely related approach to CP-indivisibility -- also based on spectral properties of CPTP maps -- was proposed in Ref.~\cite{Lorenzo2013}. Specifically, the idea presented in \cite{Lorenzo2013} consists in examining the behavior of the determinant  $\det(S_{t})$, where the map $S_{t}$ describes the evolution of a system. Since the moduli of the  eigenvalues of a CPTP satisfy $|\lambda_{i}|\leq 1$, it is clear that if $S_{t}$ is CP-divisible, then the determinant $\det(S_{t})$ cannot increase with $t$ \cite{Lorenzo2013}.~Moreover, the CP-indivisibility witness $\det(S_{t})$ can be interpreted geometrically as the volume of the set of accessible states \cite{Lorenzo2013}. It turns out that if we assume fixed-SPAM, the determinant of the probability matrix $\mP(\Seq)$ can be used to witness CP-indivisibility, as briefly mentioned in Ref.~\cite{veitia2017macroscopic}.~To see this, let us consider an arbitrary sequence of instructions $\Seq_{m}=\mathcal{G}_{i_{m}}\circ \mathcal{G}_{i_{m-1}}\circ \ldots \circ \mathcal{G}_{i_{1}},$~which may be divided as  $\Seq_{m}=\Seq_{m,m_{0}}\circ \Seq_{m_{0}}$, for some $m_{0}$ ($1<m_{0}<m$). Then, from Eq.~(\ref{spec-radius}) we find that $|\det(\mP(\Seq_{m})\mP^{-1}(\Seq_{m_{0}})|\leq 1$ (if the sequence is CP-divisible) and therefore
\beq
|\det(\mP(\Seq_{m}))|\leq |\det(\mP(\Seq_{m_{0}}))|. 
\eeq
Hence, if $\Seq_{m}$ can be represented as a sequence of CPTP maps, the quantity $|\det(\mP(\Seq_{m}))|$ cannot increase with $m$ (in the case of fixed SPAM).
\subsection{Iterative determinant test and unitarity measures}
\label{subsec:iter-uni}
{\indent}Finally, let us discuss the iterative determinant test (ID-test), introduced Ref.~\cite{veitia2017macroscopic}, which in the case of context-independence allows us to estimate the unitarity of a particular gate $G.$ The ID-test consist in applying sequences of instructions of the form $\Seq_{m}=\mathcal{G}\circ\ldots \circ\mathcal{G}$ ( i.e.,  $\mathcal{G}$ is applied $m$ times) and then examining how 
the quantity $\log(|\det(\mP_{m})|)$ behaves with the length of the sequence $m.$ If all the operation involved in determining the probabilities matrices $\mP_{m}$ are context-independent, then using Eq.~(\ref{main}) we readily find that the quantity $\log(|\det(\mP_{m})|)$ must decay linearly with the length $m$, regardless of the presence of SPAM errors.~More precisely, writing $S_{m}=G^{m}$ and using Eq.~(\ref{main}) we find that
\beq
\label{ID-test1}
\log(|\det(\mP_{m})|)=\log(|\det(\Phi_{\text{out}}\Phi_{\text{in}})|)+m \log(|\det(G)|),
\eeq
where the ``$y$-intercept'' $\log(|\det(\Phi_{\text{out}}\Phi_{\text{in}})|$ depends on SPAM and is given by $\log(|\det(\tr[M_{k}^{\text{exp}}\rho_{i}^{\text{exp}}]_{ki})|)$. This motivates the introduction of the context-dependence witness \cite{veitia2017macroscopic}
\beq 
\label{ID-test2}
L_{m}:=-\log(|\det(\mP_{0}^{\text{ideal}})|)+\log(|\det(\mP_{m})|).
\eeq
Here, the $(k,i)$ entry of the matrix  ${\mP_{0}^{\text{ideal}}}$ is $\Tr[\Pi_{k}\rho_{i}], $ where $\{\rho_{i}, \Pi_{k}\}$ are our target input states and measurement observables, so that, in the absence of SPAM  errors, $L_{0}=0.$ Finally note that, as in Ref.~\cite{veitia2017macroscopic}, the quantity $L_{m}$ may also be written as $L_{m}=\log(|\det(S_{m}^{\text{raw}})|),$ where $S_{m}^{\text{raw}}$ is related to $\mP_{m}$ via Eq.~(\ref{sraw}). \\
{\indent}An important feature of the ID-test Eq.~(\ref{ID-test2}) is that if a linear relationship between $L_{m}$ and the sequence length $m$ is observed (within error bars), then one can extract, from the slope of $L_{m}$, the quantity $|\det(G)|$ (see Eq.~(\ref{ID-test1})), which can be related to the degree of reversibility (unitarity) of the operation $G$. Indeed, it is known that for a positive and trace preserving map  $|\det(G)|=1$ implies that $G$ is either unitary or unitarily equivalent to the transposition map $T: A\rightarrow A^{T}$ (which is not CP) \cite{Wolf2008}. In addition, if $G$ is unitary then $\det(G)=1,$ which means that for high-fidelity gates, we will have $\det(G)>0$ and thus, in practice, we can  safely drop the absolute value sign in the slope $\log(|\det(G)|).$ Nonetheless, it should be kept in mind that a CPTP map may have a negative determinant~(see footnote \footnote{For example, consider the depolarizing channel  $D(A)=pA+(1-p)\tr[A]I_{2}/2$. Using the Choi-Jamio{\l}kowski state, it can be easily shown that $D$ is CPTP for $-1/3\leq p \leq 1$. Hence, $\det(D)=p^{3}$ can be negative}). Thus, in light of the above discussion, a natural definition of the unitarity $u'(G)$ of an operation $G $ is the following \cite{veitia2017macroscopic}
\beq 
\label{uni-def1}
u'(G)=|\det(G)|^{\frac{2}{d^{2}-1}}.
\eeq
This measure of unitarity enjoys the following easy-to-check properties: {\bf{(i)}} $0\leq u'(G)\leq 1$ for CPTP maps,~{\bf{(ii)}} $u(G)=1$ if the operation $G$ is unitary,~{\bf({iii)}} $u(G)$ is gauge-invariant (and therefore, representation independent), that is, if $G'=T_{\text{gauge}}GT^{-1}_{\text{gauge}}$ then $u'(G')=u'(G)$,~{\bf{(iv)}} $u'$ is non-increasing under composition of CPTP maps, i.e., 
\beq
u'(G_{2}\circ G_{1})=u'(G_{2}) u'(G_{1})\leq u'(G_{1}).
\eeq
The purpose of the power $2/(d^{2}-1)$ in the definition Eq.~(\ref{uni-def1}) is to relate $u'(G)$ to the unitarity measure $u(G)$, proposed in Ref.~\cite{wallman2015} (also see Ref.~\cite{roth2018recovering}), namely
\beq
\label{uni-def2}
u(G)=\frac{1}{(d^{2}-1)}\tr[W_{G}^{T}W_{G}], 
\eeq
where the $(d^{2}-1)\times (d^{2}-1)$ matrix $W_{G}$ is the unital part of $G$ (see Eq.~(\ref{matTP})). As we prove below, for trace preserving maps, the determinant-based measure $u'(G)$ provides a lower bound for $u(G)$. This can be easily shown by noticing that $\tr(W_{G}^{T}W^{\phantom{}}_{G})=\sum_{n=1}^{d^{2}-1}s_{n}^{2}$, where $\{s_{n}\}$ are the singular values of $W_{G}$. If $G$ is trace preserving, then we have $\det(G)=\det(W_{G})$ which implies that $|\det(G)|=\Pi_{n=1}^{d^2-1}s_{n}$. Finally, the sought result
\beq
u(G)\geq u'(G)
\eeq
readily follows from application of the  arithmetic mean-geometric mean (AM-GM) inequality.~Furthermore, as shown in Sec.~\ref{sec:uni}, for gates  that are ``close'' to a unitary operation, the measures $u(G)$ and $u'(G)$ yield nearly the same results.\\
{\indent}Let us now consider the framework of quantum-open-systems to further explore the 
connection between the measure $u'(G)$ and the effects of decoherence. First, we assume that the evolution of a $d$-dimensional system can described by the time-independent Lindblad master equation \cite{Gorini1976, Lindblad1976},
\beq
\label{Lindblad}
\frac{d\rho_{t}}{dt}=\mathcal{L}(\rho_{t})=\mathcal{H}(\rho_{t})+\mathcal{D}(\rho_{t})
\eeq
Here, $\mathcal{H}(\rho)=-i[H,\rho]$, where $H$ is the system's Hamiltonian, the term $ \mathcal{D}(\rho)=\sum_{k=1}^{d^{2}-1}\gamma_{k}\mathcal{D}_{k}(\rho)$ takes into account dissipative effects and each map $\mathcal{D}_{k}$ has the form
\beq 
\label{dissipatork}
 \mathcal{D}_{k}(\rho)=\frac{1}{2}([F_{k}^{\phantom{}}\rho, F_{k}^{\dagger}]+ [F_{k}^{\phantom{}}, \rho F_{k}^{\dagger}]),
\eeq
which, together with the condition $\gamma_{k}\geq 0$, guarantees that the evolution of the system is completely positive. Trace preservation follows from the fact that $\Tr[\mathcal{D}_{k}(\rho)]=\Tr[\mathcal{H}(\rho)]=0$.
{\indent}Since we have assumed that Eq.~(\ref{Lindblad}) is time-independent, the map $\rho_{0}\rightarrow \rho_{t}=S_{t}(\rho_{0})$, describing the evolution of the system, is simply given by $S_{t}=e^{t\mathcal{L}}.$ Now, using the Liouville representation Eq.~(\ref{matrep}), we express the generator $\mathcal{L}$ and the map $S_{t}$ as $d^{2}\times d^{2}$ real matrices, which allows us to write
\beq 
\det(S_{t})=e^{t\Tr[\mathcal{L}]}.
\eeq
As noticed in Refs.~\cite{Wolf2008,Hall2014}, the determinant $\det(S_{t})$ does not depend on the Hamiltonian part $\mathcal{H}(\cdot)$ of the Lindblad equation.~Indeed, choosing any operator basis $\{P_{n}\}_{n=1}^{d^{2}}$ one can easily show that  $\Tr[{\mathcal{H}}]=-\frac{i}{d}\sum_{n} \tr(P_{n}[H,P_{n}])=0$. Consequently, for a system evolving according to a Lindblad equation, the unitarity measure $u'(S_{t})$ will solely depend on the dissipative part $\mathcal{D}.$ Moreover, this result also holds in the case of a time-dependent Lindblad equation. The dynamical map $S_{t}$ for a time-dependent generator $\mathcal{L}_{t}=\mathcal{H}_{t}+\mathcal{D}_{t}$ can be written as  ${S}_{t}= \mathcal{T}_{\text{o}} e^{\int_{0}^{t}\mathcal{L}_{t'}d t'}$, where $\mathcal{T}_{\text{o}}$ is the time-ordering operator. Equivalently, discretizing the interval $(0,t)$, we can formally write
${S}_{t}=e^{\mathcal{L}_{t_{n}}\Delta{t}} e^{\mathcal{L}_{t_{n-1}}\Delta{t}}\ldots e^{\mathcal{L}_{t_{1}}\Delta{t}},$
where $t_{n}>t_{n-1} \ldots > t_{1}$, $\Delta{t}\rightarrow 0$ and  $n \rightarrow \infty.$~Finally, taking the determinant of $\prod_{k=1}^{n}e^{\mathcal{L}_{t_{k}}\Delta{t}}$ and invoking the fact that  $\Tr[\mathcal{H}_{t}]=0,$ results in the relation
\beq
\det(S_{t})=e^{\int_{0}^{t}\tr[\mathcal{D}_{t'}]d t'},
\eeq
which does not involve the system's Hamiltonian. This observation implies that if the implementation of a certain set of gates $\{G_{k}\}$, of equal duration $t_{g}$, can be described (correctly) by a Lindblad equation, with a deterministic control Hamiltonian $H_{k}(t)$ (e.g., describing microwave pulses applied to a superconducting qubit) and a time-independent term $\mathcal{D}$ (to ensure context-independence), then these gates will share the same determinant (and unitarity), namely 
\begin{eqnarray}
 \label{detDk}
 \det(G)&=&\exp (\sum_{k}(\gamma_{k}t_{g})\Tr[\mathcal{D}_{k}]),\\ 
 \label{unik}
u'(G)&=&\exp (\frac{2}{d^{2}-1}\sum_{k}(\gamma_{k}t_{g})\tr[\mathcal{D}_{k}] ).
\end{eqnarray}
 The relevance of this fact is that it can be verified, experimentally, employing the ID-test described earlier in this subsection.\\
{\indent}The traces $\tr[\mathcal{D}_{k}]$, appearing in the above expressions, can be easily related to the Lindblad operators $F_{k}$~\cite{Hall2014}.~Indeed, for any operator basis $\{P_{n}\}_{n=1}^{n^{2}}$ (such that $\tr[P_{n}P_{m}]=d\delta_{nm})$ and any $d\times d$ matrix $A$, we have the relation $\sum_{n}P_{n}AP_{n}=d \tr[A] I_{d}$ \footnote{Note that the expansion $A=1/d\sum_{n}\tr[AP_{n}]P_{n}$ (for any $A$) implies the more general relation $\sum_{n}(P_{n})_{ij}(P_{n})_{kl}=d\delta_{il}\delta_{jk}$}. This suffices to show that the trace of the map Eq.~(\ref{dissipatork}) is given by
\begin{eqnarray}
 \label{tracek}
 \Tr[\mathcal{D}_{k}]&=&\frac{1}{2}\sum_{n}(\tr(P_{n}[F_{k}P_{n}, F_{k}^{\dagger}])+\tr(P_{n}[F_{k},P_{n} F_{k}^{\dagger}])) \nonumber \\
                                  &=&-d(\Tr[F_{k}^{\dagger}F_{k}]-\frac{1}{d}|\tr[F_{k}]|^{2})\leq 0.
                                  \end{eqnarray}
Here, in the last step, we made use of the Cauchy-Schwarz inequality  $|\tr(A^{\dagger}B)|^{2}\leq \tr(A^{\dagger}A)\tr(B^{\dagger}B)$ (with $A=I_{d}$ and $B=F_{k}$).~Finally, it is worth recalling that the Lindblad operators $\{F_{k}\}$ are not unique as both the mixing transformation $F_{k}\rightarrow \sum_{l}U_{kl}F_{l}$, where $U$ is unitary, and the c-number shift transformation
\beq 
\label{c-shift}
F_{k}\rightarrow F_{k}+c_{k},\; H\rightarrow H+\frac{1}{2i}\sum_{k}\gamma_{k}(c_{k}^{*}F_{k}-c_{k}F^{\dagger}_{k}),
\eeq
leave the Lindblad equation invariant~\cite{wiseman2009quantum}.~As expected, these transformations do not modify the trace $\tr[\mathcal{D}_{k}]$ either, as apparent from Eq.~(\ref{tracek}).\\
{\indent}As an example, let us apply the Lindblad equation~(\ref{Lindblad}) to describe the evolution of a qubit undergoing energy relaxation and dephasing with decay rates $\gamma_{1}$ and $\gamma_{\phi}$, respectively. These decoherence channels can be described by means of the traceless Lindblad operators $F_{1}=\sigma_{-}$ and $F_{\phi}=Z/\sqrt{2}$ \cite{carmichael1999statistical}.~If the qubit's Hamiltonian is $H=-\frac{\omega}{2} Z$, then, in the operator basis $\{I,X,Y,Z\}$, the matrix representation of $S_{t}$ is
\beq
\label{uni-example}
S_{t} =\begin{bmatrix} 1&0&0&0\\
                                        0& \cos(\omega t)e^{-\frac{t}{T_{2}}}&  \sin(\omega t)e^{-\frac{t}{T_{2}}}& 0\\
                                         0& -\sin(\omega t)e^{-\frac{t}{T_{2}}}&  \cos(\omega t)e^{-\frac{t}{T_{2}}}& 0\\
                                          1-e^{-\frac{t}{T_{1}}}& 0& 0&e^{-\frac{t}{T_{1}}}\\
                                          \end{bmatrix}, 
                                          \eeq
where $1/T_{1}:=\gamma_{1}$ and $1/T_{2}:=\gamma_{1}/2+\gamma_{\phi}.$ Making use of Eqs.~(\ref{detDk}) and (\ref{tracek}), we immediately find that $\det(S_{t})=\text{exp}(-2t(\gamma_{1}+\gamma_{\phi})),$ which can be explicitly checked using the matrix Eq.~(\ref{uni-example}). Consider now two qubits, $A$ and $B,$ interacting independently with local environments, as discussed above. Then the matrix representation (in the product basis $\{I,X,Y,Z\}^{\otimes 2}$) of the map describing the evolution of the qubits is simply $S_{t}^{AB}=S_{t}^{A}\otimes S_{t}^{B},$ where the $4\times4$ matrices $S^{A(B)}_{t}$ are of the form Eq.~(\ref{uni-example}). Using the fact that $\det(C\otimes D)=\det(C)^{n} \det(D)^{n},$ for two $n\times n$ matrices $C$ and $D$, we get $\det(S^{AB}_{t})=\text{exp}[-8t(\sum_{k=1,\phi}(\gamma_{k}^{A}+\gamma_{k}^{B}))],$ where $\gamma_{k}^{A(B)}$ are the decay rates corresponding to the local environment of qubit $A(B)$. The same result can be obtained by setting $d=4, F_{k}^{A}=F_{k}\otimes I_{B}, F_{k}^{B}=I_{A}\otimes F_{k}$ in Eq.~(\ref{tracek}).\\
{\indent}Finally, this basic example can be further exploited to show that the ID-test can potentially detect context-dependence caused by classically correlated noise, such as random telegraph noise (RTN) \cite{cywinski2008enhance,benedetti2014non,d2013hidden,kaplan1964differential}. More precisely, consider fluctuations in the qubit's frequency of the form $\omega(t)=\omega_{0}+\eta(t),$ where the stochastic function $\eta(t)$ describes a RTN signal. Then, averaging the map Eq.~(\ref{uni-example}) (with $\gamma_{1},\gamma_{\phi}=0$) over many realizations (configurations) of the RTN process, we get $\expect{S_{t}}=1\oplus \expect{W_{t}},$ where the average of the unital part $W_{t}$ is
\beq 
\label{rtn}
\expect{W_{t}}=\begin{bmatrix}
\cos(\omega_{0}t)&\sin(\omega_{0}t)&\\
-\sin(\omega_{0}t)&\cos(\omega_{0}t)&\\
\end{bmatrix}\expect{e^{i\varphi_{t}}}\oplus 1
\eeq
and $\varphi_{t}=\int_{0}^{t}\eta(t')dt.$ Hence, $\det(\expect{S_{t}})=\expect{\text{exp}(i\varphi_{t})}^{2}$, which implies that a non-zero amplitude RTN signal $\eta(t)$ will lead to deviations from linearity in the, SPAM-insensitive, test Eq.~(\ref{ID-test2}). Here it should be noted that an alternative way of identifying this kind of correlated noise, in a SPAM-insensitive fashion, has been recently proposed in Ref.~\cite{o2015qubit}. Roughly speaking, the idea presented in \cite{o2015qubit} consists in studying the function $\expect{\text{exp}(i\varphi_{t})}$ by means of a combination of Ramsey and Hahn spin echo sequences and RB, through which $\Tr[\expect{W_{t}}]$ is extracted.
\subsection{A toy model of context-dependence}
\label{subsec:toy-model}
 \begin{figure*}[!t]
\includegraphics[width=15.4cm]{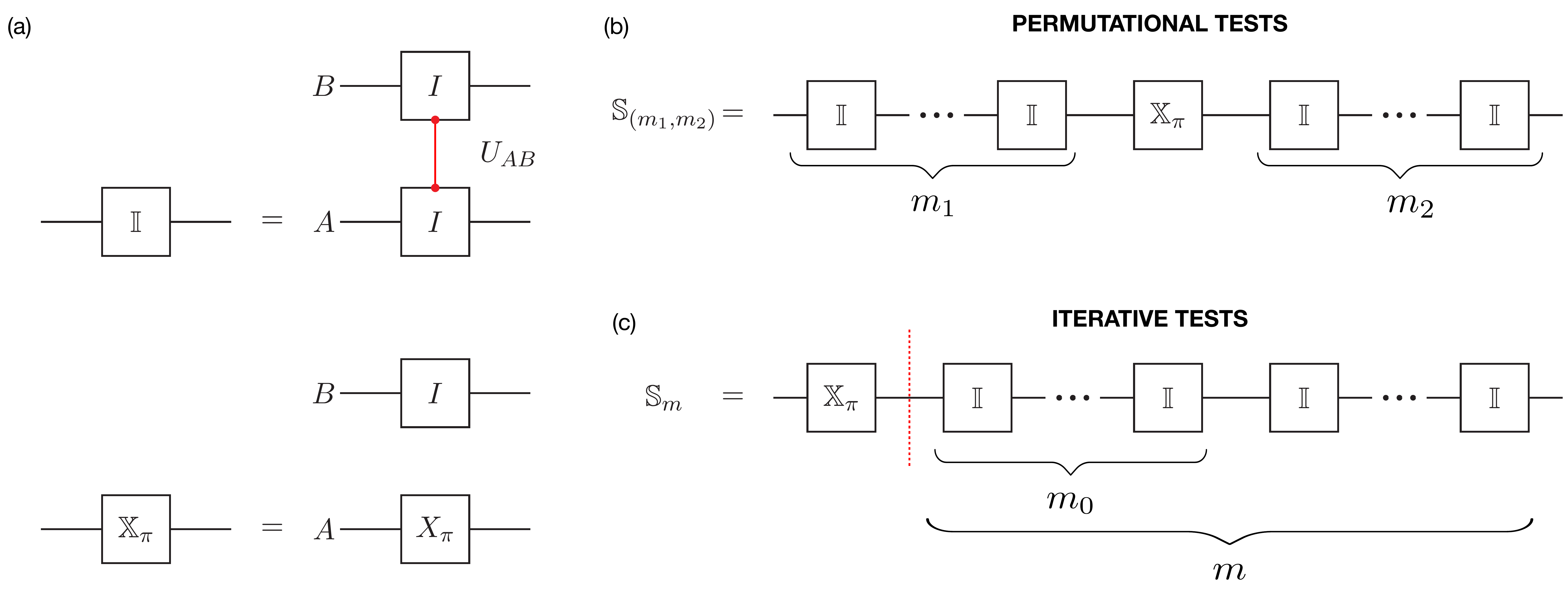}
\caption{Toy model of context-dependence based on an unwanted interaction between the system being tested ($A$) and an additional system ($B$), which acts as a memory. Panel (a) depicts two fixed operations $\mI$, $\mX_{\pi}$ acting on $AB$ and corresponding to the instructions $\mathcal{I}$ and $\mathcal{X}_{\pi}$. The sequences of instructions showed in (b) (where $m_{1}+m_{2}$ is kept fixed) and (c) 
are used to show (in a SPAM-insensitive way) that the probability matrices obtained by measuring system $A$ cannot be generated by context-independent operations.
\label{fig:toy-model}}
\end{figure*}
{\indent}To elucidate some of the points discussed in this section, we now consider a solvable toy model, wherein the context-dependence of the operations on a system $A$ is generated through an ``unwanted'' interaction with a small quantum system $B$ (which will in general introduce memory effects, see e.g., \cite{veitia2012mutual}). As mentioned earlier in subsection \ref{subsec:perm},
our permutational tests for context-dependence require testing sequences containing at least two different types of instructions. We take these instructions to be $\mathcal{I}$ and $\mathcal{X}_{\pi}$, which would ideally produce the unitary operations $I$ and $X_{\pi/2}$ on a qubit $A$. To introduce context-dependence in this model, it suffices to  assume that {\bf{(i)}} the result of the instruction $\mathcal{I}$ is an operation $\mI=U_{AB}$ acting on a larger system $AB,$ where $B$ is another qubit whose initial state is of the form  $\rho_{B}=1/2(I+n_{z}^{B}Z)$, with $|n_{z}^{B}|<1;$ {\bf{(ii)}} the instruction $\mathcal{X}_{\pi/2}$ results in the context-independent operation $\mathbb{X}_{\pi}=X_{\pi}\otimes I_{B}$ (it does not affect system $B$), as depicted in Fig.~\ref{fig:toy-model}(a); {\bf{(iii)}} finally, we take the two-qubit operation $U_{AB}$ to be given by
\beq 
U_{AB}=\exp\left(-i\frac{\varphi}{2}Z\otimes Z\right).
\eeq
{\indent}Let our target set of input states and measurement observables be $\rho_{i}=\ket{\phi_{i}}\bra{\phi_{i}}$ and $\Pi_{k}=\ket{\phi_{k}}\bra{\phi_{k}},$ where $\{\ket{\phi_{i}}\}_{i=1}^{4}$ is the ``standard'' single-qubit tomographic set $\ket{\phi_{1}}=\ket{g}$, $\ket{\phi_{2}}=\ket{e}$, $\ket{\phi_{3}}=1/\sqrt{2}(\ket{g}+\ket{e})$ and $\ket{\phi_{4}}=1/\sqrt{2}(\ket{g}+i\ket{e}).$ If the initial state of $A$ is $\rho_{0}=\ket{g}\bra{g}$ and the POVM effect is $M_{0}=\ket{e}\bra{e}$ (see the diagram Fig.~\ref{fig:spam-circuit}), then our target set can be realized using the input/output gates $\{G_{i}^{\text{in}}\}=\{I,X_{\pi},Y_{\pi/2}, X_{-\pi/2}\}$ and $\{G_{k}^{\text{out}}\}=\{X_{\pi},I, Y_{\pi/2}, X_{-\pi/2}\}.$ To keep this model simple, we assume that these gates are context-independent, which is equivalent to the fixed-SPAM assumptions discussed in subsection~\ref{subsec:spam-cp}.~We introduce SPAM errors by adding  a gate-dependent depolarizing channel $\mathcal{D}_{i}(\rho)=\alpha_{i}\rho+(1-\alpha_{i})I/2$ to each input/output gates specified above.~Specifically, we will assign $\alpha_{i}=\alpha_{\pi/2}$ to each $\pi/2$-pulse gate, $\alpha_{i}=\alpha_{\pi}$ to the $\pi$-pulse gates and $\alpha_{I}=1$ to the input/output idle gates.\\
{\indent}To show how the permutational tests work in this model, we consider the variant of the Hahn spin-echo sequence shown in Fig.~\ref{fig:toy-model}(b). More precisely, we focus 
on sequences of the form $\mS_{(m_{1},m_{2})}=\mI^{m_{2}}\mX_{\pi}\mI^{m_{1}}$, with $m_{1}+m_{2}=m$ fixed.~In other words, any $\mS_{(m_{1},m_{2})}$ is just a cyclic permutation of the sequence $\mS_{(0,m)}=\mI^{m}\mX_{\pi}.$ The corresponding probability matrices $\mP_{(m_{1},m_{2})}$ are, after tracing out the memory $B$, given by
\begin{widetext}
\beq 
\label{Pm1m2}
\mathcal{P}_{(m_{1},m_{2})}=\frac{1}{2} \begin{bmatrix} 1-\alpha_{\pi} & 1+\alpha_{\pi}^{2} &1 &1 \\
                                                       2&  1 -\alpha_{\pi}& 1 & 1\\
                                                       1& 1& 1+\alpha_{\pi/2}^{2}\cos((m_{2}-m_{1})\varphi) & 1-\alpha_{\pi/2}^{2}n^{B}_{z}\sin((m_{2}-m_{1})\varphi) \\
                                                       1& 1& 1-\alpha_{\pi/2}^{2}n_{z}^{B}\sin((m_{2}-m_{1})\varphi)& 1-\alpha_{\pi/2}^{2}\cos((m_{2}-m_{1})\varphi) \\
                              \end{bmatrix}.
\eeq
\end{widetext}
It suffices to compare two of these probability matrices, for example $\mP_{(m,0)}$ and $\mP_{(m/2,m/2)},$ to show that Eq.~(\ref{Pm1m2}) cannot be generated by context-independent operations acting on system A. Indeed, application of the PD-test Eq.~(\ref{pd-test}) to the sequences $\mS_{(m_{1},m_{2})}$ yields
\begin{eqnarray}
\label{pd-toy-model}
\det[\mP_{(m_{1},m_{2})}]&=&\frac{(1+\alpha_{\pi})^{2}\alpha_{\pi/2}^{4}}{16}\nonumber\\
 &\times&[1-(1-(n_{z}^{B})^2)\sin^{2}({\Delta}m\varphi)] \nonumber \\
 &\neq&\text{const.},
\end{eqnarray}
where $\Delta{m}:=m_{2}-m_{1}.$ Note that the initial state of the memory (qubit $B$) determines the amplitude of the variation of $\det[\mP_{(m_{1},m_{2})}].$ Since the permutations considered in Fig.~\ref{fig:toy-model}(b) are cyclic, we can also try to detect context-dependence by observing changes in the fidelities $\mathcal{F}^{(r)}_{(m_{1},m_{2})}=1/d^{2}\Tr[(\mP_{(m_{1},m_{2})}\mP^{-1}_{0})^{r}]$ (see the permutational test Eq.~(\ref{cycle-test-trace})). Here, the reference probability matrix $\mP_{0}$ (which corresponds to the instruction $\mathcal{I}_{\emptyset}$) has entries given by
\beq
(\mP_{0})_{k|i}=\Tr[M_{0}\mathcal{D}_{k}\circ G_{k}^{\text{out}}\circ \mathcal{D}_{i}\circ G_{i}^{\text{in}}(\rho_{0})],
\eeq
where $\mathcal{D}_{i}(\cdot), i=1\ldots4,$ are the depolarizing channels introduced earlier in this subsection and $G_{i}^{\text{in(out)}}$ are our ideal input (output) gates. For the sequences Fig.~\ref{fig:toy-model}(b),  we find that the fidelity $\mathcal{F}^{(1)}_{(m_{1},m_{2})}$ vanishes identically. On the other hand, from Eq.~(\ref{pd-toy-model}) we know that at least one of the higher ``moments'' $\mathcal{F}^{(r)}_{(m_{1},m_{2})}$ ($r\geq 2$) must reveal context-dependence. Indeed, for $r=2$ we have
\begin{eqnarray}
\mathcal{F}^{(2)}_{(m_{1},m_{2})}&=&\frac{1}{4}\tr[(\mP_{(m_{1},m_{2})}\mP_{0}^{-1})^{2}]\nonumber \\
                                                           &=& 1-\left(\frac{1-(n^{B}_{z})^2}{2}\right)\sin^{2}({\Delta}m\varphi) \nonumber \\
                                                           &\neq&\text{const.}
                                                           \end{eqnarray}
The form of the above context-dependence witnesses follows directly from the fact that the spectrum of the reduced map $\rho\rightarrow \Tr_{B}[\mS_{(m_{2},m_{1})}(\rho\otimes \rho_{B})]$ is $\{1,-1,|\lambda|, -|\lambda|\},$ where $|\lambda|=\sqrt{1-[1-(n_{z}^{B})^{2}]\sin^{2}(\Delta{m}\varphi)}$ (hence $\mathcal{F}^{(1)}_{(m_{1},m_{2})}=0$).\\
{\indent}The probability matrix Eq.~(\ref{Pm1m2}) may also be used to discuss the ID-test Eq.~(\ref{ID-test2}). To do this, we consider the circuit in Fig.~\ref{fig:toy-model}(c) which describes sequences of the form  $\mS_{m}=\mI^{m}\mX_{\pi}.$ Treating the operation $\mX_{\pi}$ as a SPAM error and making use of Eqs.~(\ref{Pm1m2})  and (\ref{pd-toy-model}), we readily find that \\
\begin{eqnarray}
\label{ID-test-toy-model}
L_{m}&=&-\log(|\det(\mP_{0}^{\text{ideal}})|)+\log(|\det(\mP_{(0,m)})|)\nonumber \\
          &=& 2\log\left(\frac{(1+\alpha_{\pi})\alpha_{\pi/2}^{2}}{2} \right)\nonumber \\
          &+&\log\left[1-(1-(n_{z}^{B})^2)\sin^{2}(m\varphi)\right]. 
          \end{eqnarray}
(Note that for our tomographic set $\det(|\mP_{0}^{\text{ideal}}|)=1/4.)$ The non-linear behavior of $L_{m}$ in the above equation implies that the reduced dynamics of qubit $A$ cannot be generated by  iterations of a context-independent operation  $\mI_{A}$. Furthermore, for sufficiently long sequences (with $m>m_{\text{cr}}=\lceil{\pi/(2\varphi)}\rceil$), the non-monotonicity of $L_{m}$ indicates CP-indivisibility (for fixed SPAM). In this toy model, the same conclusion is reached by examining the spectral radius Eq.~(\ref{spec-radius}). Specifically, employing the matrices $\mP_{(0,m_{0})}$ and $\mP_{ (0,m)}$, corresponding to the sequences described in Fig.~\ref{fig:toy-model}(c), we compute the spectral radius
\beq 
\mathcal{R}[\mP_{(0,m)}\mP^{-1}_{(0,m_{0})}]=\text{max}(1, |\mu|),
\eeq
where $\mu$ is given by
\beq
\mu=\frac{\cos(m\varphi)+i n_{z}^{B} \sin(m\varphi)}{\cos(m_{0}\varphi)+i n_{z}^{B}\sin(m_{0}\varphi)}.
\eeq
Hence, for $|n_{z}^{B}|\neq 1,$ the reduced dynamics of qubit $A$ between $m_{0}$ and $m$ ($m>m_{0}$) cannot be described by a CPTP map when $\sin^{2}(m\varphi)<\sin^{2}(m_{0}\varphi)$ (for which $|\mu|>1$). 
\section{The ZZ model and dissipation} 
\label{Sec:zz}
{\indent}In this section, we describe in greater detail the more realistic model of context-independence introduced in Ref.~\cite{veitia2017macroscopic}. This model will be used later in this work to explore how statistical fluctuations affect the tests described in the previous section. The main difference between the toy model presented in the previous subsection and the one we discussed here, is that the latter takes into account dissipative effects such as energy relaxation and dephasing. In addition, this model will be consistent in the sense that the form of the input and output gates, $G^{\text{in}(\text{out})}$, employed to obtain the probabilities matrices $\mP(\Seq)$, will be same as that of the gates used in the sequences we test.\\
{\indent}As the previous subsection, we consider two qubits $A$ and $B$, coupled via an Ising interaction
$V=\frac{J}{2} Z\otimes Z.$ For the sake of simplicity, we assume that single qubit gates are implemented
via the time-dependent control Hamiltonian $H_{c}(t)=\Omega_{i}\cos(\omega_{i}t+\phi_{i})X^{i}$, $i=A,B$ (here $X,Y$ and $Z$ are the standard Pauli matrices). In a frame rotating with the frequency of the qubits, this control Hamiltonian assumes the form
 \beq
\label{hrwa}
 H^{R}_{c}= \frac{\Omega_{i}}{2}(\cos(\phi_{i})X^{i}-\sin(\phi_{i})Y^{i})
 \eeq
 (after the rotating wave approximation~\cite{Geller2010}).~The time-independent Hamiltonian \(H^{R}_{c}\) can be used
 to implement a set of single qubit gates of equal duration $t_{g}$ by choosing the appropriate amplitude $\Omega_{i}$ and phase $\phi_{i}$ for each gate. On the other hand, notice that the Ising Hamiltonian $V$ and the maps $\mathcal{D}_{k}$ (see Eq.~(\ref{dissipatork})) describing energy relaxation, spontaneous excitation and dephasing (via the Lindblad operators $F_{1}=\sigma_{-}, F_{3}=\sigma_{+}$ and $F_{\phi}=Z/\sqrt{2}$), retain their form in the rotating frame (these processes commute with the free evolution of the qubit).~Based on these observations, and motivated by Eq.~(\ref{hrwa}), we will assume that the noisy implementation of a gate $G\otimes I_{B}$, in the presence of the unwanted interaction $V,$ is given by following map acting on $AB:$
 \begin{figure}[t!]
 \begin{center}
 \includegraphics[width=6cm]{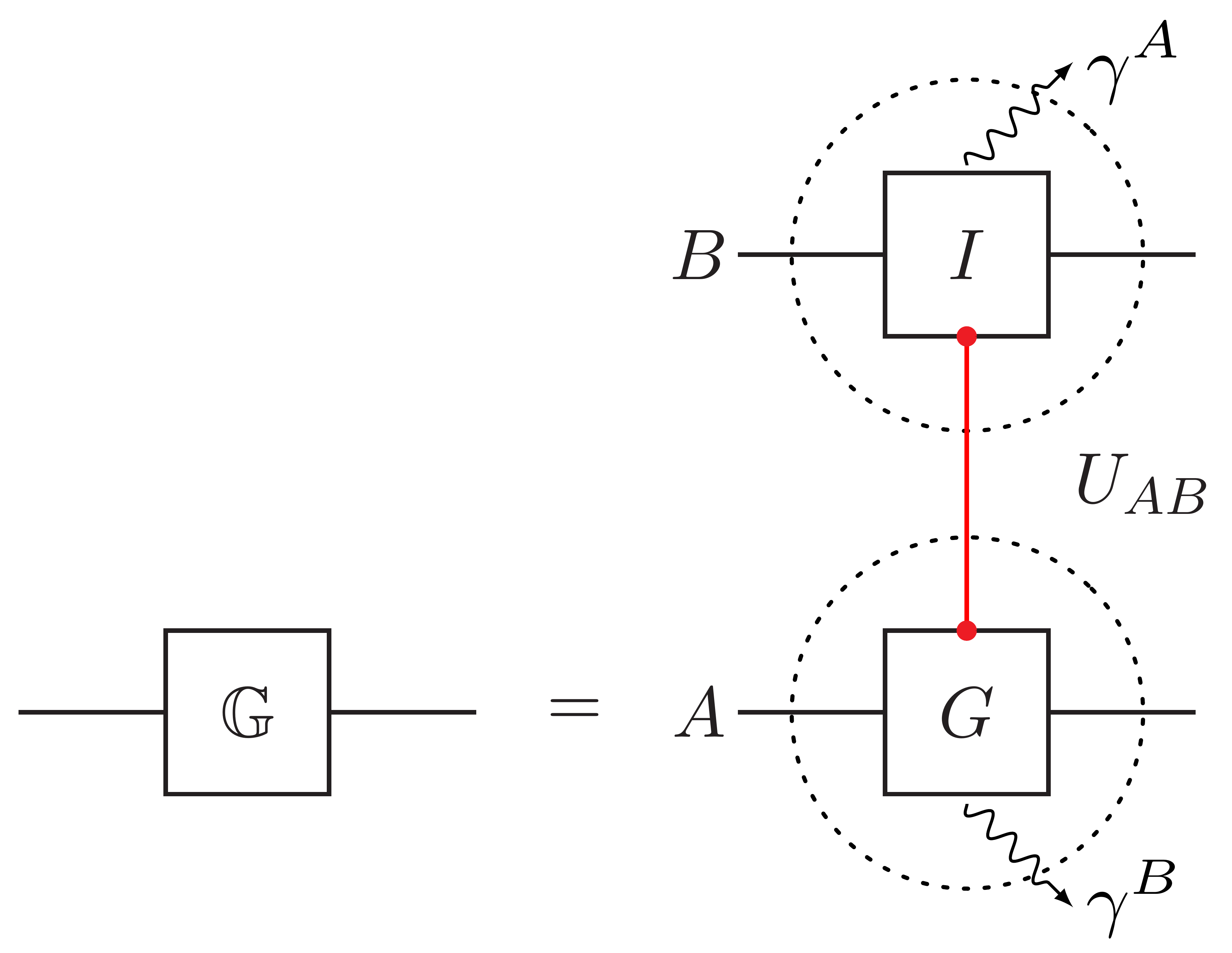}
 \end{center}
\caption{\label{fig:toy-model-diss}
Generalization of the toy-model Fig.~\ref{fig:toy-model}(a). Here, the dotted circles represent local environments contributing to the decoherence of both system $A$ and the memory $B$. The context-dependence of the operations on system $A$ is induced by the two-qubit interaction $V=\frac{J}{2} Z\otimes Z$.}
\end{figure}
 \beq
 \label{gate-model}
 \mathbb{G}=\text{exp}(\mathcal{J}_{G}+t_{g}\mathcal{V}+t_{g}{\mathcal{D}}), 
 \eeq
Here, $\mathcal{J}_{G}$ is a generator of the gate $G$, $\mathcal{V}$ represents the map $\rho\rightarrow -i[V, \rho]$, $\mathcal{D}=\sum_{k}\gamma_{k}\mathcal{D}_{k}$ describes local energy relaxation, spontaneous excitation and dephasing (with decay rates $\gamma_{1}^{i}, \gamma_{3}^{i}, \gamma_{\phi}^{i}, i=A,B$)  and
 $t_{g}$ is the gate duration. We take the generators $\mathcal{J}_{G}$ of the rotations $X_{\theta}\otimes I _{B}$ ($Y_\theta\otimes I_{B}$) to be
the Liouville representations of the maps $-i \tfrac{\theta}{2} \text{Ad}_{X(Y)}\otimes I_{B}$, where $\text{Ad}_{H}(\cdot):=[H,\cdot]$, while the noisy idle gate $\mathbb{I}$ is assumed to be generated by ${\mathcal{J}_{I}}=0$. This model of context-dependent operations is schematically summarized in Fig.~\ref{fig:toy-model-diss}.\\
{\indent}Also, for consistency, we assume that the initial state of the system is $\rho^{AB}_{0}=\rho_{0}^{A}\otimes \rho_{0}^{B}$, where $\rho^{i}=1/2(I+n_{z}^{i}Z),$ with
\beq 
\label{nz}
n_{z}^{i}=\frac{\gamma_{1}^{i}-\gamma_{3}^{i}}{\gamma_{1}^{i}+\gamma_{3}^{i}}, 
\eeq
which ensures that $\rho^{AB}_{0}$ is stationary under the action of $\mathcal{D}$ and $\mathcal{V}$. That is, $\mathcal{D}(\rho_{0}^{AB})=\sum_{k=1,\phi,3}\gamma_{k}\mathcal{D}_{k}(\rho_{0}^{AB})=\mathcal{V}(\rho_{0}^{AB})=0$. Finally, we take our POVM effect $M_{0}^{AB}$ to be $M_{0}^{A}\otimes I_{B}$ (which is equivalent to tracing out system $B$ right before measuring $M^{A}_{0}$), where $M_{0}^{A}=\eta \ket{e}\bra{e}$ ($0<\eta\leq 1$). Thus, in this model, our probability matrices $\mP(\Seq)$ will be given by
\beq
\mP_{k|i}(\Seq)=\Tr[(M^{A}_{0}\otimes I_{B})\mG^{\text{out}}_{k}\circ \mS \circ \mG_{i}^{\text{in}}(\rho^{AB}_{0})],
\eeq
where $\mG^{\text{in(out)}}$ and $\mS$ are the noisy implementations Eq.~(\ref{gate-model}) of a set of input(output) gates and a sequence of instructions $\Seq$, respectively. The sources of SPAM errors in this model are {\bf{(i)}} the errors in the gates $\mG^{\text{in(out)}}$, {\bf{(ii)}} the unknown parameter $n_{z}^{A}$ in the initial state $\rho_{0}^{A}=1/2(I+n_{z}^{A}Z)$, {\bf{(iii)}} and the ``efficiency'' $\eta$ of the POVM effect $M_{0}^{A}=\eta \ket{e}\bra{e}$. For $J=0$ (i.e., no interaction between the qubits $A$ and $B$) each operation $G$ (on $A$) is context-independent and, regardless of the SPAM errors mentioned above, we can estimate its unitarity $u'(G)$ using the ID-test Eq.~(\ref{ID-test2}). Moreover, setting $\mathcal{V}=0$ in Eq.~(\ref{gate-model}) and making use of Eqs.~(\ref{unik}) and (\ref{tracek}), we obtain the following gate-independent expression for the unitarity of $G:$
\beq
u'(G)=|\det(G)|^\frac{2}{d^{2}-1}=\text{exp}\left[-\frac{4t_{g}}{3}(\gamma_{1}^{A}+\gamma_{\phi}^{A}+\gamma_{3}^{A})\right].
\eeq
{\indent}A feature of this model is that for certain sequences of instructions the context-dependence effects will not be visible (let alone statistically significant). More precisely, if we choose $\varphi:=Jt_{g}\ll 1$, then, as shown in Ref.~\cite{veitia2017macroscopic}, application of  the ID-test to the sequence $\mS_{m}=\mX_{\pi}^{m}$ will result in a nearly linear relationship between $L_{m}$ and the length of $\mS_{m}.$ However, iterations of the form $\mS_{m}=(\mX_{-\pi/2}\mX_{\pi/2})^{m}$ or $\mS_{m}=(\mX_{2\pi}\mI)^{m}$ will lead to a marked non-linear behavior of $L_{m}$ (and even CP-indivisibility) for sufficiently long sequences. This happens because certain sequences in our model will not  amplify the context-dependence effects induced by a small parameter $\varphi=Jt_{g}$ \cite{veitia2018}.
\section{Simulation and analysis of statistical fluctuations}
\label{Sec:fluct}
\begin{figure}[t!]
\includegraphics[width=3.4in]{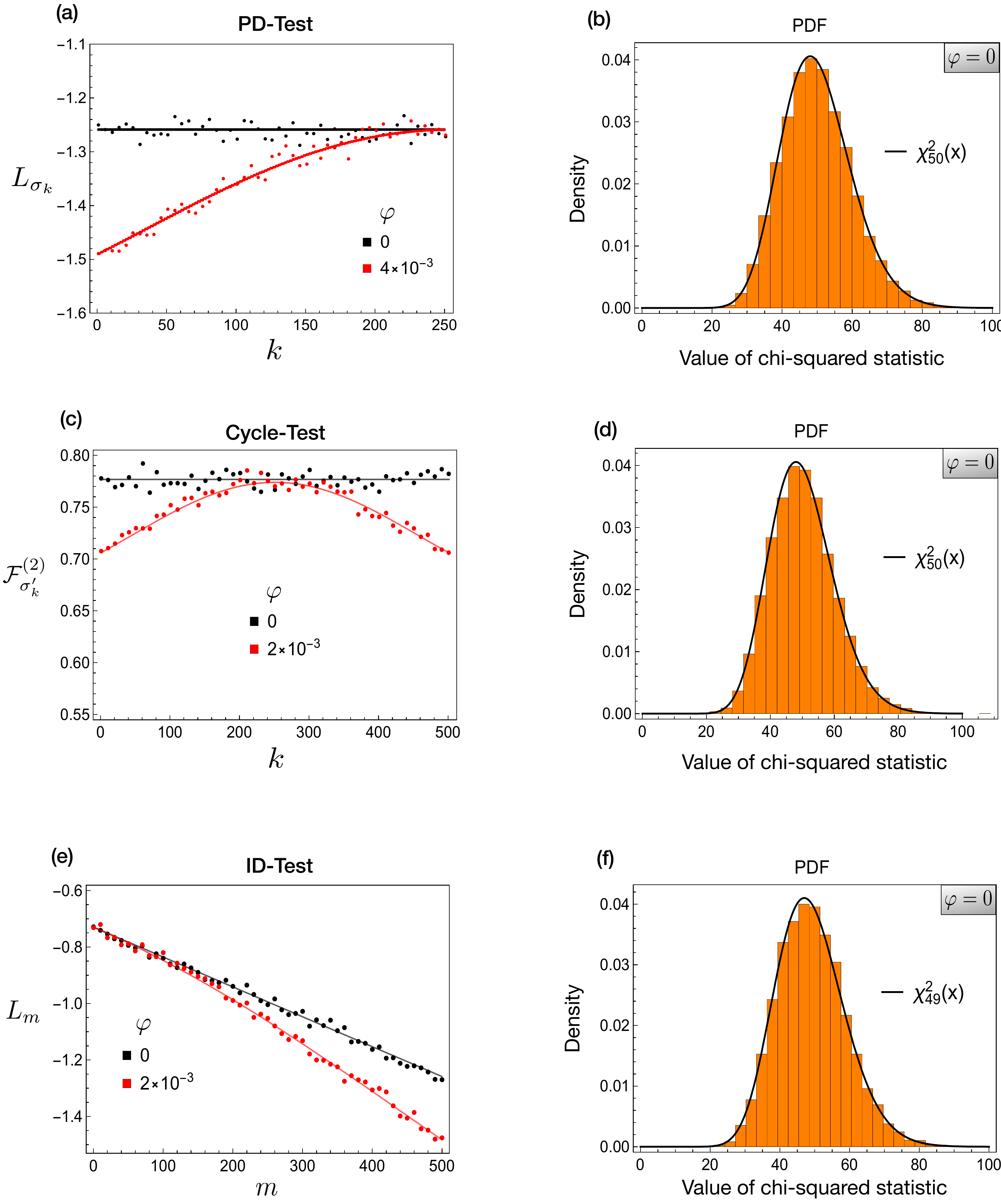}
\caption{Statistical fluctuations and context-independence test in the cases $\varphi=0$ and $\varphi\neq0$, where $\varphi=Jt_{g}$. The left panels show the statistical fluctuations (represented by dots) of the quantities $L_{\sigma_{k}}$, $\mathcal{F}_{\sigma'_{k}}^{(2)}$ and $L_{m}$, for $N_{s}=50,000$ runs per experimental configuration. The ``solid'' curves in the left panels describe our simulations in the limit $N_{s}\rightarrow \infty.$ In (a) and (c) we considered permutations of the sequence $\mI^{250}\mX_{\pi}^{250}$ and cyclic permutations of $\mX_{\pi}\mI^{500},$ respectively. Panel (e) displays $L_{m}$ for iterations of the  noisy idle gate $\mI.$ The number of points ``perturbed'' (dots) in each test is $M=51.$ The left three panels show the corresponding chi-squared distributions for each test in the case $\varphi=0.$ The histograms were obtained by considering a set of $R=10,000$  hypothetical experiments. In each test, the variances were estimated using $B=50,000$ bootstrap replicas. \label{fig:testsQ}}
\end{figure}

{\indent}We now turn our attention to examining the impact of statistical fluctuations on the test for context-dependence discussed in detail in Sec.~\ref{Sec:context}. In practice, a matrix element $\mP_{k|i}$ will be estimated by repeating the experimental configuration shown in Fig.~\ref{fig:spam-circuit} a finite number of times $N_{s}$. This will yield the estimate $\hat{\mP}_{k|i}={n_{k|i}}/{N_{s}}$, where $n_{k|i}$ is the number of times we observe the event described by the POVM effect $M_{0}$. Clearly, in the limit $N_{s}\rightarrow \infty$, the estimates $\hat{\mP}_{k|i}$ will obey  Born's rule, i.e., $\hat{\mP}_{k|i}\rightarrow \mP_{k|i}=\tr(M_{0}G^{\text{out}}_{k}\circ S \circ G^{\text{in}}_{i} (\rho_{0}))$ and thus, our tests will be exact and insensitive to SPAM errors, as shown earlier in this work. For $N_{s}\gg1$, we expect the estimates $\hat{\mP}_{k|i}=n_{k|i}/N_{s}$ to be close to the corresponding true probabilities $\mP_{k|i}.$ Thus, the goal of this, and subsequent sections, will be to explore how the fluctuations
\beq 
\delta{\hat{\mP}}_{k|i}:=\hat{\mP}_{k|i}-{\mP}_{k|i}
\eeq
affect the statistical significance of the context-dependence tests and the precision of the unitarity estimates $\hat{u}'(G)$. Providing a general answer to this question is a difficult problem because our tests are based on quantities involving all the matrix elements of $\hat{\mathcal{P}}(S)$ (e.g., $\log(|\det(\hat{\mathcal{P}}(S))|).$ For this reason, we will first restrict our discussion here to the statistical significance of context-dependence effects generated via the $ZZ$ model described in the previous section.\\
{\indent}Statistical fluctuations can straightforwardly be incorporated into our model by ``perturbing'' the true probabilities $\mP_{k|i}$ obtained from our ``ideal'' simulations (as those shown in Ref.~\cite{veitia2017macroscopic}), wherein $N_{s}=\infty$. More precisely, we will generate estimates $\hat{\mP}_{k|i}$ by sampling the number of events $n_{k|i}$ from the binomial distribution $\text{Bin}(N_{s},\mP_{k|i})$, with $N_{s}\gg 1$. For sufficiently large values of $N_{s}$,  the counts $n_{k|i}$ will be normally distributed, with mean $N_{s}\mP_{k|i}$ and variance $N_{s}\mP_{k|i}(1-\mP_{k|i})$. Hence, the fluctuations ${\delta}\hat{\mP}_{k|i}$ will be distributed as follows: 
\beq 
\label{Prob2}
{\delta}\hat{\mP}_{k|i} \sim \mathcal{N}(0, \mP_{k|i}(1-\mP_{k|i})/N_{s}),
\eeq
 where $\mathcal{N}(\mu, \sigma^{2})$ denotes a normal distribution with mean $\mu$ and variance $\sigma^{2}.$ As a result, the fluctuations of the quantities computed from $\hat{\mathcal{P}}(S)$ will be, approximately, normally distributed, for sufficiently large values of $N_{s}$ \footnote{If write an estimate of the quantity $y$ as 
$\hat{y}=y+\delta{\hat{y}}$, then for large sample sizes, the fluctuations ${\delta}\hat{y} $ will be approximately linearly related to the ${\delta}\hat{P}_{k|i}s$. Since the fluctuations ${{\delta}\hat{\mP}_{k|i}}s$ are independent and normally distributed, ${\delta}\hat{y}$ will also be a Gaussian random variable.}. \\
{\indent}Below, we describe in detail a set of simulations that will later allow us to study the power and precision of our tests.\\
 \noindent {\bf (i)} For each sequence $\mathcal{S}_{m}$, we generate an estimate $\hat{\mathcal{P}}_{m}$ by sampling the counts $n_{k|i}$ from the binomial distribution $\text{Bin}(N_{s}, \mP_{k|i})$.  We then use this estimate to calculate the quantity of interest $y_{m}$ (e.g., the log-det of $\hat{\mP}_{m}$). The purpose of this step is to simulate data corresponding to an experiment with $N_{s}$ runs per measurement configuration.\\
  \noindent {\bf (ii)} To estimate the variance of ${y}_{m}$, we resort to the bootstrap method \cite{efron1994introduction}. That is, for each $\hat{\mathcal{P}}_{m}$, we generate a set of $B$ bootstrap replicas. Each replica is generated by resampling each entry of the ``counts-matrix'' $N_{s}\hat{\mathcal{P}}_{m}$\, from the binomial distributions $\text{Bin}(N_{s}, \hat{\mathcal{P}}_{k|i}).$  For each replica $\hat{\mathcal{P}}_{m,b}$, we compute the quantity of interest ${y}_{m,b}$. Then the sample variance 
  \beq 
 \sigma_{m}^{2}:=\text{Var}[\{{y}_{m,b}\}_{b=1}^{B}], 
 \eeq
  provides a reasonable estimate of the true variance of ${y}_{m}$.  Note that, in general, the variances $\sigma_{m}^{2}$ will be heteroskedastic (i.e., $\sigma_{m}^{2}$ will depend on $m$). For the determinant-based tests, we will replace, in Sec.~\ref{subsec:log-det-dist}, the bootstrapping by a more computationally efficient, yet equivalent, method. Namely, we will work out the distribution of the quantity $\log(|\det(\hat{\mP})|)$, which will allow us estimate the variances $\sigma^{2}_{m}$ using a function $\hat{\mP}_{m}\rightarrow \tilde{\sigma}^{2}[\hat{\mP}_{m}]$.\\
 \noindent {\bf (iii)} We repeat step {\bf{(i)}} $R$ times. That is to say, we consider a set of $R$ \emph{hypothetical} experiments. We do this step to verify some of our assumptions (e.g., normality) as well as to present results (such as the power of a test) which are independent of our random number generator. The purpose of this step (and step {\bf{(ii)}}) will become more apparent in the next subsections.\\
 {\indent}The left three panels in Fig.~\ref{fig:testsQ}  show the single-qubit realization of step {\bf{(i)}}, using $N_{s}=50,000$ runs per experimental configuration. Context-dependence and SPAM errors were introduced via the ZZ model discussed in Sec.~\ref{Sec:zz}. The PD-test Eq.~(\ref{pd-test2}), displayed in Fig.~\ref{fig:testsQ}(a), was applied to permutations of the sequence $\mS_{1}=\mI^{n}\mX_{\pi}^{n},$ where $n=250$. More precisely, we considered non-cyclic permutations of the form $\mS_{\sigma_{k}}=\mI^{n-k+1}\mX_{\pi}^{n-k+1}(\mX_{\pi}\mI)^{k-1}, k=1,2,\ldots, n+1$. We then perturbed (as explained in step {\bf(i)}) the $M=51$ probability matrices $\mP_{\sigma_{1}}, \mP_{\sigma_{6}},\mP_{\sigma_{11}},\ldots ,\mP_{\sigma_{251}}$, which were used to compute the log-dets $L_{\sigma_1},\ldots, L_{\sigma_{251}}.$ Figure~\ref{fig:testsQ}(c) shows simulations of the cycle-test applied to cyclic permutations of the sequence $\mS_{1}=\mX_{\pi}\mI^{n},$ with $n=500$, that is, $\mS_{\sigma'_{k}}=\mI^{k-1}\mX_{\pi}\mI^{n-k+1}, k=1,2,\ldots, n+1.$ We computed the fidelities Eq.~(\ref{cycle-test-trace}), with $r=2,$ for the $M=51$ estimates $\hat{\mP}_{\sigma'_{1}}, \hat{\mP}_{\sigma'_{11}},\hat{\mP}_{\sigma'_{21}},\ldots,\hat{\mP}_{\sigma'_{501}}.$ To ensure that each fidelity $\mathcal{F}^{(2)}_{\sigma'_{k}}$ -- which involves a pair of probability matrices -- is unbiased, we generated a set of $M=51$ statistically independent probability matrices $\hat{\mP}^{(1)}_{0},\hat{\mP}^{(11)}_{0},\ldots, \hat{\mP}^{(501)}_{0}$ (while keeping the reference sequence $\mathcal{S}_{0}=\mathcal{I}_{\emptyset}$ fixed). The dots in Fig.~\ref{fig:testsQ}(c) represent the fidelity estimates $\mathcal{F}^{(2)}_{\sigma'_{k}}=1/4\Tr[(\hat{\mP}_{\sigma'_{k}}(\hat{\mP}^{(k)}_{0})^{-1})^2],$ whose fluctuations are now, to a good approximation, distributed as $\mathcal{N}(0,\sigma_{k}^{2}).$ Finally, the panel Fig.~\ref{fig:testsQ}(e) displays the results of the ID-test Eq.~(\ref{ID-test2}), applied to the sequences  $\mS_{m}=\mI^{m}, m=0,10,20,\ldots, 500$. The parameters we used to generate the plots showed in Fig.~\ref{fig:testsQ} were as in Ref.~\cite{veitia2017macroscopic}, namely,  $\gamma_{1}^{A}=\gamma_{1}^{B}=\gamma_{1}=1/(60~\mu \text{s})$, $\gamma_{\phi}^{A}=\gamma_{\phi}^{B}=\gamma_{1}/2$, $t_{g}=20~\text{ns}.$ (Note that these values correspond to state-of-the-art superconducting qubits~\cite{gambetta2017building}.) The initial states of $A$ and $B$ have $n_{z}^{A}=n_{z}^{B}=0.84$ (which determines the value of $\gamma_{3}$) and we took the efficiency of the POVM effect $M_{0}=\eta\ket{e}\bra{e}$ to be $\eta=0.95$. As in the toy-model discussed in Sec.~\ref{subsec:toy-model}, the probability matrices were obtained employing the noisy set of gates $\{\mG_{\text{in}}\}=\{\mI,\mX_{\pi},\mY_{\pi/2},\mX_{-\pi/2}\}$ and $\{\mG_{\text{out}}\}=\{\mX_{\pi},\mI,\mY_{\pi/2},\mX_{-\pi/2}\}.$
 
 The right three panels in Fig.~\ref{fig:testsQ}  were obtained using steps {\bf{(ii)}} and {\bf{(iii)}}. These display the distribution (histogram) of the $\mathcal{X}^{2}$-statistic (which depends on the variances $\sigma_{m}^{2}$ of the observations), obtained by considering an ensemble of $R$ of experiments, as those shown in Figs.~\ref{fig:testsQ}(a), \ref{fig:testsQ}(c) and \ref{fig:testsQ}(e). We will return to this discussion in the next subsection, where we will use the $\mathcal{X}^{2}$-statistic for hypothesis testing.\\
{\indent}The context-dependence effects, for the nonzero values of $\varphi,$ shown in Figs.~\ref{fig:testsQ}(a), \ref{fig:testsQ}(c), \ref{fig:testsQ}(e), are markedly visible and any sensible statistical tests will lead us to the conclusion that some of our gates must necessarily be context-dependent. However, for smaller values of $N_{s}$ or $\varphi$, these signals will become less discernible; thus some statistical tools will be required to assess the statistical significance of these effects. Given the simplicity of the null hypotheses associated with our tests it will be convenient to formulate the problem of detection of context-dependence in the popular framework of hypothesis testing \cite{wasserman2013}. The null hypotheses associated with the permutational tests Eqs.~(\ref{pd-test2}), (\ref{cycle-test-trace}) and the ID-test (\ref{ID-test2}) are
\begin{eqnarray} 
&L_{\sigma_{k}}&\quad \text{is constant,}\label{dperm}\\
&\mathcal{F}^{(r)}_{\sigma'_{k}}& \quad \text{is constant}, \label{cycprob}\\
&L_{m}&\quad \text{is linear in $m$.}\label{dtest}
\end{eqnarray}
Thus, our approach will be to assume that the above hypotheses are true unless the value of some statistic $T$ provides strong evidence to reject them. Finally, note that 
our SPAM-insensitive tests for CP-divisibility (see Sec.~\ref{subsec:spam-cp}) are less amenable to hypothesis testing. For example, the question of the monotonicity of the quantity $\log(|\det(\mP_{m})|)$, taking into account statistical fluctuations, could be addressed using the isotonic regression method~\cite{barlow1972statistical}, but this topic goes beyond of the scope of this work and will not be discussed further here. 
 \subsection{Weighted least squares and the chi-squared statistic}
 The variances $\sigma_{m}^2$ found via the bootstrap method (see step {\bf{(ii)}}) play a key role in assessing the goodness of fit of a particular model. Since the homoskedasticity of our observations cannot be guaranteed, we will employ the weighted least squares method (WLS) to fit $q$-parameter models of the form 
   \beq
   \label{poly}
   y_{i}(\beta)=\sum_{n=0}^{q-1} \beta_{n} i^{n},
   \eeq
  to a set of observations $\{y_{i}\}_{i=1}^{M}$ (see, e.g., the plots in Fig.~\ref{fig:testsQ}). The WLS estimate of $\beta=(\beta_{0}, \ldots, \beta_{q-1})$ minimizes the 
  weighted residual sum of squares (with weights $w_{i}:=1/\sigma_{i}^2 $). More precisely, for the model Eq.~(\ref{poly}), the WLS estimate of $\beta$ is the solution of the optimization problem
\beq
\label{WLS-MIN}
\hat{\beta}=\underset{\beta}{\text{arg min}}\sum_{i=1}^{M}\frac{(y_{i}-y_{i}(\beta))^{2}}{\sigma_{i}^{2}}. 
\eeq
\begin{figure}[!t]
\includegraphics[width=1.8 in]{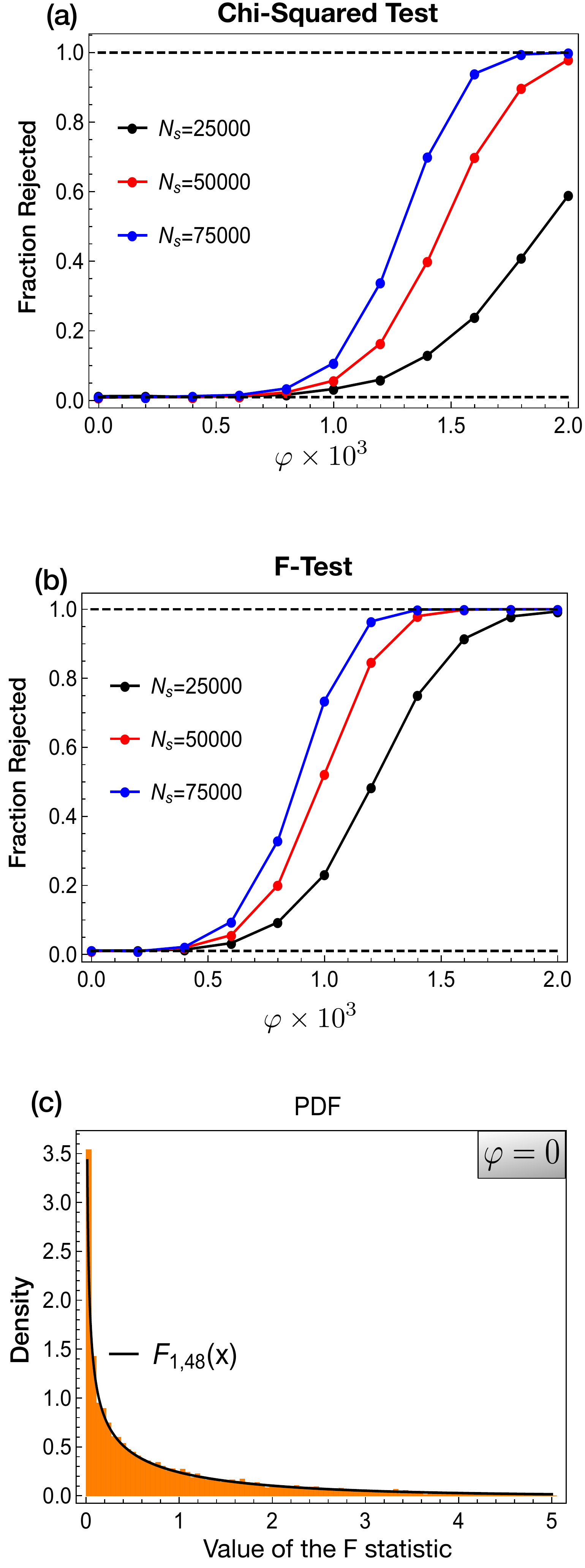}
\caption{Hypothesis testing for the ID-test applied to the sequences $\mS_{m}=\mI^{m},$ as in Fig.~\ref{fig:testsQ}(e). Panel (a) shows the fraction of times the null hypothesis is rejected (i.e., the power of test) by means of the chi-squared statistic vs. the interaction strength $\varphi$ ($\times 10^{3}$) and $N_{s}$. The number of hypothetical experiments is $R=10,000.$ Panel (b) deals with hypothesis testing based on the $F$ statistic. In both cases, we set $p_{\text{cr}}=0.01.$ Panel (c) shows the agreement between the histogram, built out the $R$ hypothetical experiments and $N_{s}=50,000$, and the target distribution $F_{1,48}$, when $\varphi=0$. 
  \label{fig:ChiFDtest}}
\end{figure}
Note now that the objective function minimized in Eq.~(\ref{WLS-MIN}), is just the chi-squared ($\mathcal{X}^{2}$) statistic (see e.g., Ref~\cite{wasserman2013}). Furthermore, if there really is an underlying relationship (i.e., a ``true'' model) of the form Eq.~(\ref{poly}), that is,
\beq 
\label{true} 
y=X_{q}\beta+\varepsilon, \quad \varepsilon_{i} \sim \mathcal{N}(0,\sigma_{i}^{2}),
\eeq
then, as discussed in appendix \ref{Sec:chi-appendix}, the $\mathcal{X}^{2}$ statistic should follow the chi-squared distribution with $M-q$ degrees of freedom, i.e., $\mathcal{X}^{2}\sim \chi^2_{M-q}.$ Consequently, if a permutational test is implemented (see Eqs.~(\ref{dperm}) and (\ref{cycprob})) using a set of context-independent gates, then the statistic $\mathcal{X}^{2}$ will be $\chi^2_{M-1}$    distributed. In the same way, for the ID-test test Eq.~(\ref{dtest}) one should observe $\mathcal{X}^{2}\sim \chi^{2}_{M-2}$, provided all the operations involved are context-independent. In this subsection we will take full advantage of this fact to test for the statistical significance of a potential deviation from context-independence. Specifically, using the PDF of the chi-squared distribution with $n$ degrees of freedom $f_{n}(x),$ we will compute the one-sided $p$-value given by the integral
\beq 
p:=\int_{\mathcal{X}^{2}}^{\infty}f_{n}(x)dx, 
\eeq
where the $\mathcal{X}^{2}$ statistic will be obtained by fitting a particular model Eq.~(\ref{poly}) to our observations $y=[y_{1},\ldots, y_{M}]^{T}.$ By setting an artificial threshold $p_{\text{cr}}$, we will reject the null hypothesis (i.e., the context-independence hypothesis) if we observe $p<p_{\text{cr}}.$

\begin{figure}[!t]
\includegraphics[width=1.8in]{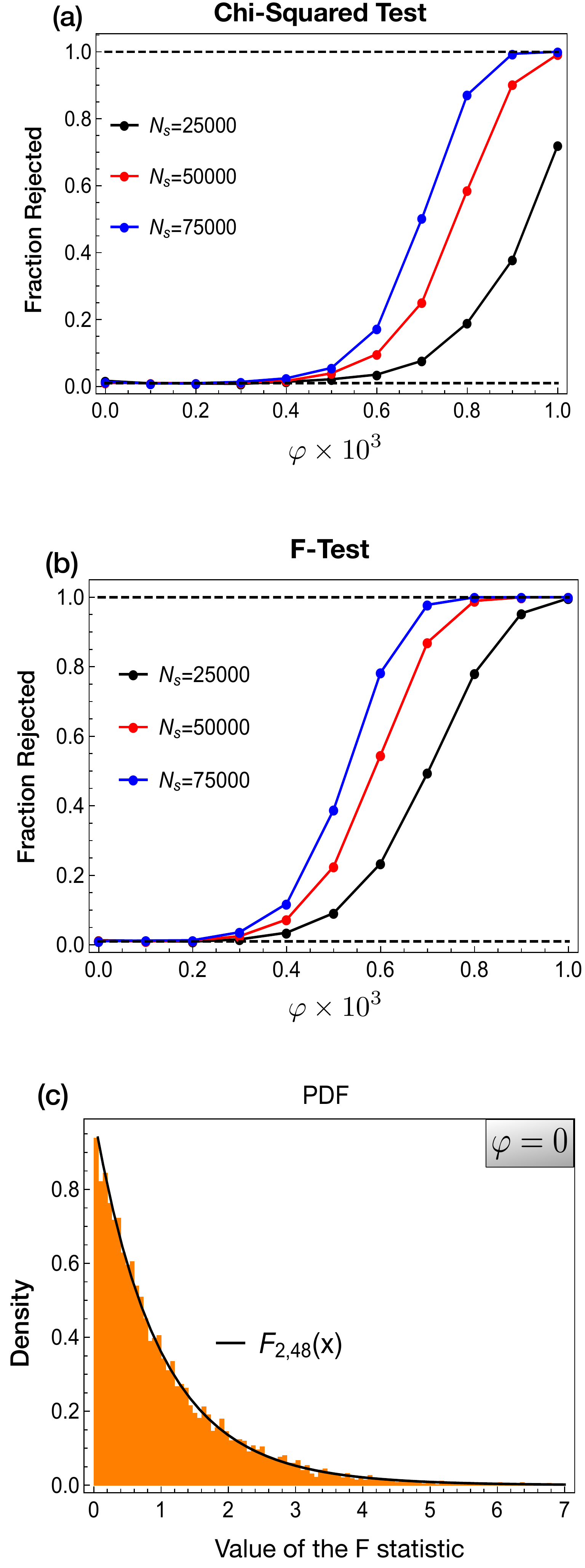}
\caption{Hypothesis testing for the cycle-test, applied to the cyclic permutations of the sequence $\mS_{1}=\mX_{\pi}\mI^{500}$ discussed earlier in Fig.~\ref{fig:testsQ}(c). The description of the panels (a), (b) and (c) is as in Fig.~\ref{fig:ChiFDtest}. 
  \label{fig:ChiFcyc}}
\end{figure}

The right panels in Fig.~\ref{fig:testsQ} show the distributions of the chi-squared statistic for the context-independent case $\varphi=0.$ Panels (b)  and (d) display the distribution of $\mathcal{X}^{2}$, resulting from fitting a constant model $y_{i}(\beta)=\beta_{0}$. Examining $M=51$ observations (i.e., $M=51$ gate sequences) and considering a set of $R=10,000$ of hypothetical experiments (used to build the histograms), we obtained the correct distribution, that is, $\chi^{2}_{51-1}.$  Panel (f) in Fig.~\ref{fig:testsQ} shows the distribution of the $\mathcal{X}^{2}$ statistic obtained from fitting the linear model $y_{m}(\beta)=\beta_{0}+\beta_{1}m$, to the observations corresponding to the ID-test. Here, using the same values of $M$ and $R$, we found, as expected, that $\mathcal{X}^{2} \sim \chi^{2}_{51-2}$. These results allowed us to verify, indirectly, that indeed the statistical fluctuations are normally distributed (with zero mean) and, more importantly, that their variances were reasonably estimated via the bootstrap method. \\
{\indent}Figures \ref{fig:ChiFDtest}(a) and \ref{fig:ChiFcyc}(a) show the fraction of times we reject the null hypothesis for the iterative determinant test, and the cycle test, as a function of $N_{s}$ and the interaction parameter $\varphi.$ More precisely, we again generated $R$ observations;  for each observation $y_{r}=(y_{1,r}, y_{2,r}, \ldots y_{M,r})$ we computed the statistic $\mathcal{X}^{2}_{r},$ which we then used to compute the $p$-value $p_{r}=\int_{\mathcal{X}^{2}_r}^{\infty}f_{M-q}(x)dx,$
with $q=2$ for the ID-test, and $q=1$ for the cycle-test. We chose the critical $p$-value $p_{\text{cr}}=0.01$ and counted the number of times $p_{r}<p_{cr}$, that is, the number of times $N_{\text{rejc}}$ we rejected the null hypothesis. In figures \ref{fig:ChiFDtest}(a) and \ref{fig:ChiFcyc}(a), we observe how the ratio $N_{\text{rejc}}/R$ increases as we either increase the interaction strength $\varphi$ or the number of number of runs $N_{s}$. Note that for $\varphi \neq 0$ and large values of $R$, the ratio $N_{\text{rejc}}/R$ approximates the \emph{power of the test} \cite{wasserman2013}, i.e., the probability of correctly rejecting the null hypothesis. In the next subsection we show how by comparing two different models we can improve the rejection ratio, for the same significance level $p_{\text{cr}}.$
 \subsection{Statistical F-Test for nested models}
 \label{Sec:F-Test}
In the previous subsection we discussed how the $\mathcal{X}^{2}$ statistic may be used to reject the context-independence hypothesis. Roughly speaking, if the WLS fit of the context-independent model was ``sufficiently bad",  we concluded that, most likely, our gates are context-dependent. Now, in the spirit of model selection \cite{claeskens2008model,akaike1998information,schwarz2013error,sheldon2016characterizing}, we show how the goodness of fit of two nested models of the form Eq.~(\ref{poly}) may be compared.\\
{\indent}The main idea is the following: Consider two nested models $\mathcal{M}_{1}$  and $\mathcal{M}_{2}$, of the form Eq.~(\ref{poly}), such that $\mathcal{M}_{1}\subset \mathcal{M}_{2}$  (i.e., $\mathcal{M}_{1}$ is a special case of the richer model $\mathcal{M}_{2}$) and $\dim{\mathcal{M}_{1(2)}}=q_{1(2)}.$ It is clear that if model $\mathcal{M}_{1}$ is correct then so is model $\mathcal{M}_{2}.$ In addition, the corresponding chi-squared statistics obey $\mathcal{X}^2_{1(2)}\sim \chi^{2}_{M-q_{1(2)}}$ and $\mathcal{X}_{1}^{2}\geq \mathcal{X}_{2}^{2}$ (which can be seen from Eq.~(\ref{WLS-MIN})). The following is a useful result (see e.g.~\cite{bingham2010regression} or appendix \ref{Sec:F-appendix}), asserting that if $\mathcal{M}_{1}$ is correct then 
\beq 
\label{diffchi}
\Delta \mathcal{X}^{2}_{12}:=\mathcal{X}_{1}^{2}-\mathcal{X}_{2}^{2}\sim \chi^{2}_{q_{2}-q_{1}}.
\eeq
Furthermore, the difference $\Delta \mathcal{X}^{2}_{12}$ and $\mathcal{X}_{2}^{2}$ are statistically independent (see appendix \ref{Sec:F-appendix}). For two nested models $\mathcal{M}_{1}, \mathcal{M}_{2},$ $\mathcal{M}_{1}\subset \mathcal{M}_{2}$, we will consider the $F$ statistic 
\beq 
\label{F}
F=\frac{M-q_{2}}{q_{2}-q_{1}}\left(\frac{\mathcal{X}_{1}^{2}}{\mathcal{X}_{2}^{2}}-1\right). 
\eeq
Then, based on the fact Eq.~(\ref{diffchi}), if model $\mathcal{M}_{1}$ is correct, the above $F$ statistic should, by definition (see e.g.~\cite{bingham2010regression}), follow the  $F_{q_{2}-q_{1}, M-q_{2}}$ distribution \footnote{The $F_{n_{1},n_{2}}$ distribution may be defined as follows: If $X_{1}$ and $X_{2}$ are independent random variables and  
$X_{1}\sim \chi^{2}_{n_{1}}$, $X_{2}\sim \chi^{2}_{n_{2}}$, then $\frac{X_{1}/n_{1}}{X_{2}/n_{2}}\sim F_{n_{1},n_{2}}$}. This result will allow us to decide whether model $\mathcal{M}_{2}$ describes our observations significantly better than $\mathcal{M}_{1}.$ If it does, then we will reject $\mathcal{M}_{1}$, i.e., the null hypothesis. As in the preceding subsection, making use of the $F_{n_{1},n_{2}}$ distribution's PDF $f_{n_{1},n_{2}}(x)$ we compute the one-sided $p$-value
\beq 
\label{pvalueF}
p=\int_{F}^{\infty} f_{q_{2}-q_{1}, M-q_{2}}(x) dx, 
\eeq
where $F$ is given by Eq.~(\ref{F}).  Here, again, in order to study the power of the F-test we considered $R=10,000$ hypothetical experiments and used Eq.~(\ref{pvalueF}) to obtain a set of $p$-values $\{p_{r}\}_{r=1}^{R}$. We then counted the number of times a $p_{r}$ is smaller than $p_{\text{cr}}=0.01$, that is to say,  the number of times we rejected the null hypothesis specified by model $\mathcal{M}_{1}.$\\
{\indent} The panels \ref{fig:ChiFDtest}(b) and \ref{fig:ChiFcyc}(b) show the power of the F-test for the iterative determinant test and the cycle-test. For the ID-test, the models compared in Fig.~\ref{fig:ChiFDtest}(b) were the null hypothesis $y_{m}=\beta_{0}+\beta_{1}m $ ($\mathcal{M}_{1}$) and $y_{m}=\beta_{0}+\beta_{1}m +\beta_{2}m^{2}$ ($\mathcal{M}_{2}$). For the cycle test, we compared the null hypothesis, i.e., the constant model $y_{k}=\beta_{0}$ ($\mathcal{M}_{1}$) and the quadratic model $y_{k}=\beta_{0}+\beta_{1}k+\beta_{2} {k^{2}}$ ($\mathcal{M}_{2}$). In both cases, we observe that the $F$ statistic performs substantially better than the $\mathcal{X}^{2}$ statistic. Finally, the panels \ref{fig:ChiFDtest}(c) and \ref{fig:ChiFcyc}(c) reflect the agreement between the histograms, built out the $R=10,000$ hypothetical experiments, and the target distributions in the case $\varphi=0.$
\section{Estimating the unitarity of a gate}
\label{sec:uni}
\begin{figure}[t!]
\includegraphics[width=2in]{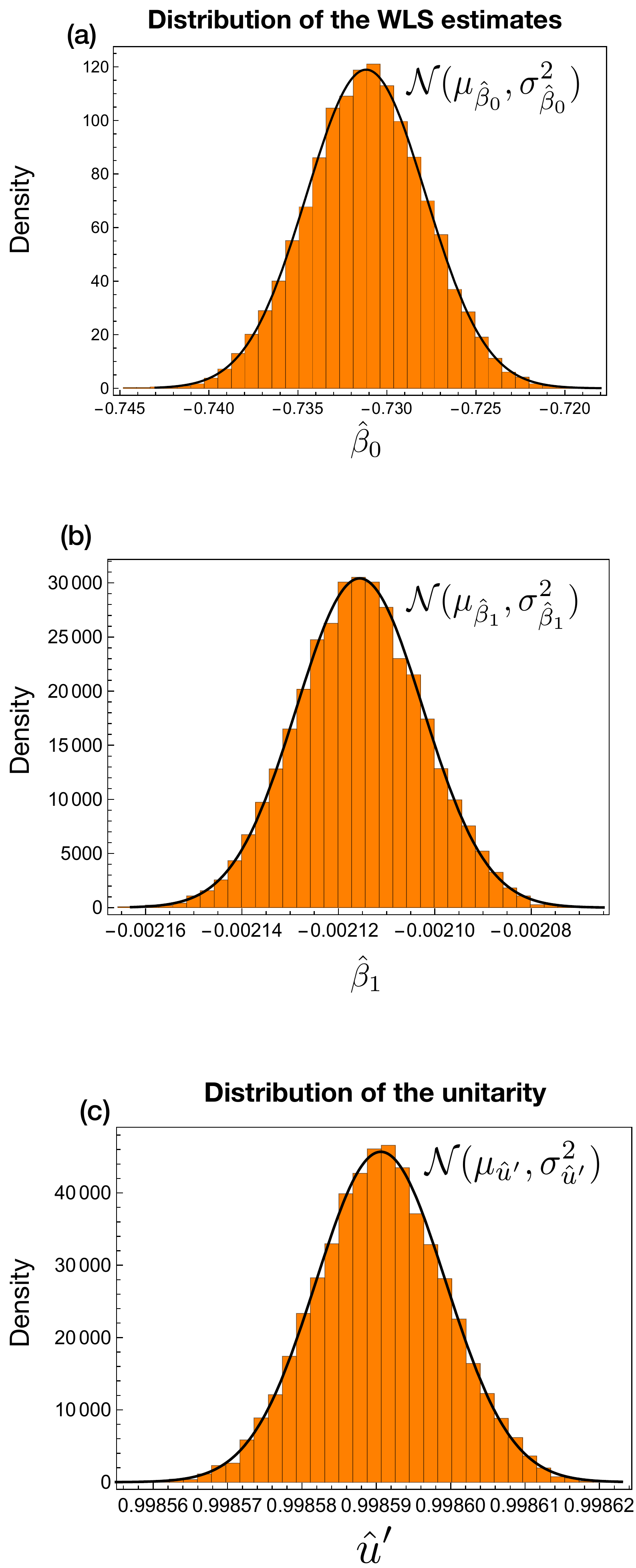}
\caption{Panels (a) and (b) display the probability distribution of the weighted least squares estimate $\hat{\beta}=[\hat{\beta}_{0},\hat{\beta_{1}}]^{T}$ for the ID-test, applied to the gate $\mI$. We considered $N_{s}=50,000$ runs per measurement configuration. The means of the estimates $\hat{\beta}_{0}$ and $\hat{\beta}_{1}$ are $\mu_{\hat{\beta}_{0}}=-0.731$ and $\mu_{\hat{\beta}_{1}}=-2.12\times 10^{-3},$ respectively. The standard deviations of the WLS estimates were found to be $\sigma_{\hat{\beta}_{0}}=3.35\times 10^{-3}$ and $\sigma_{\hat{\beta}_{1}}=1.31\times 10^{-5}.$ Panel (c) shows the distribution of the unitarity estimate $\hat{u}'(\mI)$. Here we found that $\hat{u}'\sim \mathcal{N}(\mu_{\hat{u}'}, \sigma^{2}_{\hat{u}'})$, with $\mu_{\hat{u}'}=0.998591$ and $\sigma_{\hat{u}'}=8.73\times 10^{-6}.$ The lengths of the sequences used in this simulation were $\{m_{n}\}=\{0,10,20,\ldots 500\}.$ 
 \label{fig:distbeta}}
\end{figure}

{\indent}As shown in Sec.~\ref{subsec:iter-uni}, an attractive feature of the ID-test is that even when no context-dependence is detected, it can be used to  
extract the unitarity $u'(G)$ of a gate $G$, in a robust fashion, from the slope of $\log(|\det(\mP_{m})|).$ This will be discussed in greater detail in this section, wherein we will take into account the effects of statistical fluctuations and study the precision of the estimate $\hat{u}'(G)$ obtained via the ID-test. Let us therefore assume the null hypothesis associated with the ID-test to be true (i.e., context-independence). Then our observations $y_{m}=L_{m}$ should obey the linear model 
\beq 
\label{linear-unitarity}
y_{m}:=L_{m}=\beta_{0}+\beta_{1}m+\varepsilon_{m}, \quad \varepsilon \sim \mathcal{N}(0, \sigma_{m}^{2}),
\eeq
 where $\beta_{0}$ partially characterizes SPAM errors (in the absence of SPAM errors $\beta_{0}=0$) and $\beta_{1}=\log(|\det(G)|)$. Thus, according to definition Eq.~(\ref{uni-def1}), our unitarity estimate will be
 \beq 
 \label{est-uni}
 \hat{u}'(G)=\exp\left({\frac{2\hat{\beta}_{1}}{{d^{2}-1}}}\right),
 \eeq
 where $\hat{\beta}_{1}$ is the WLS estimate discussed in the previous section. Hence, in order to find the spread of the estimates $\hat{u}'(G)$ we must first determine the probability distribution of the slope estimate $\hat{\beta}_{1}$, which is a standard problem in linear regression. Indeed, making use of equation (\ref{betaWLS}) and writing the linear model Eq.~(\ref{linear-unitarity}) as $y=X_{2}\beta+\varepsilon$, we find that $\hat{\beta}=\beta+(X_{2}^{T}WX_{2})^{-1}X_{2}^{T}\sqrt{W}(\sqrt{W}\varepsilon),$ where $W$ is the weight matrix $W :=\text{diag}(w_{1}, \ldots, w_{M}),$ with $w_{i} =1/\sigma_{i}^2$. Since $z:=\sqrt{W}\varepsilon \sim \mathcal{N}(0,I_{M})$ and $\text{Cov}(Az)=A z A^{T}$, we conclude that the  WLS estimates are distributed according to the multivariate normal distribution
 \beq 
 \label{distbeta}
 \hat{\beta}=[\hat{\beta}_{0}, \hat{\beta}_{1}]^{T} \sim \mathcal{N}(\beta, \Sigma_{2}),\quad \Sigma_{2}=(X_{2}^{T}W X_{2})^{-1}.
\eeq
  The design matrix $X_{2}$ for the linear model Eq.~(\ref{linear-unitarity}) is simply
\beq 
X^{T}_{2}=\begin{bmatrix}  1&1 \hdots & 1 \\
                                     m_{1}&m_{2} \hdots &m_{M}\\
                                             \end{bmatrix},
\eeq
where $\{m_{n}\}_{n=1}^{M}$ are the sequence lengths considered in the iterative determinant test,  for example, $\{m_{n}\}_{n=1}^{M}=\{0, 10, 20, \ldots 500\}$ (as in Fig.~\ref{fig:testsQ}(e)). Now, substituting the design matrix $X_{2}$ into Eq.~(\ref{distbeta}), we obtain the covariance matrix 
\beq 
\label{Cov}
\Sigma_{2}=\frac{\sum_{n=1}^{M}w_{n}\begin{bmatrix} m_{n}^2&-m_{n} \\
                                                      -m_{n}& 1\\ \end{bmatrix}}{(\sum_{n=1}^{M} w_{n})(\sum_{n=1}^{M}w_{n}m_{n}^{2})-(\sum_{n=1}^{M}w_{n}m_{n})^2}, 
\eeq
from which we extract the standard deviations of the WLS estimates, that is,  $\sigma_{\hat{\beta}_{0}}$ and $\sigma_{\hat{\beta}_{1}}.$ Now, since we expect $\sigma_{\hat{\beta}_{1}}\ll 1, $ we can write the unitarity estimate 
Eq.~(\ref{est-uni}) as $\hat{u}'(G)\approx \text{exp}( 2\beta_{1}/(d^{2}-1))(1+2(\delta{\hat{\beta}_{1}})/(d^{2}-1))$ from which we readily see that the unitary estimates will be, approximately, normally distributed. Specifically, when $\sigma_{\hat{\beta}_{1}}\ll 1,$ the unbiased estimator $\hat{u}'(G)$ will we distributed according to
\beq
\label{u-dist}
\hat{u}'(G)\sim \mathcal{N}\left(u'(G), \frac{4u'^2(G)}{(d^{2}-1)^2} \sigma_{\hat{\beta}_{1}}^2 \right),
\eeq
where $u'(G)$ is the true unitarity of the gate $G$. Clearly, in practice, only the estimate $\hat{u}'(G)$ will be available. Nonetheless, the latter can be used to reasonably approximate the standard deviation in Eq.~(\ref{u-dist}). That is to say, we can set  $u'(G)=\hat{u}'(G),$ which leads to the simple relation
\beq 
\label{var-uni}
\hat{\sigma}^{2}_{\hat{u}'} \approx \frac{4 \hat{u}'^2(G)}{(d^{2}-1)^2} \sigma_{\hat{\beta}_{1}}^2.
\eeq
In addition, note that for high-fidelity gates we will find that $\hat{u}'(G)\approx 1$ and therefore, $\hat{\sigma}_{\hat{u}'}\approx 2/(d^{2}-1) \sigma_{\hat{\beta}_{1}}.$\\
{\indent}Figure~\ref{fig:distbeta} shows the distributions of the estimates $\hat{\beta}_{0}$, $\hat{\beta}_{1}$ and $\hat{u}',$ obtained from our simulation of the ID-test, applied to the noisy idle gate $\mI.$ We chose the parameters of our ZZ model to be $\varphi=0$ (which implies that the linear model Eq.~(\ref{linear-unitarity}) is correct), $n_{z}=0.84$, $\eta=0.95,$ $\gamma_{1}^{-1}=60~\mu \text{s}$, $\gamma_{\phi}=\gamma_{1}/2$ and $t_{g}=40~\text{ns}$ (we assumed that both qubits $A$ and $B$ in our ZZ model are identical). The histograms shown in Fig.~\ref{fig:distbeta} were built using $R=10,000$ hypothetical experiments. Furthermore, in table~\ref{tab:distbeta} we present the results of our simulation of the ID-test, applied to various single-qubit gates. Unsurprisingly, we found the means of our estimates to be equal to the true values $\beta_{0}$, $\beta_{1}$ and $u'(G)$. The true value of the $y$-intercept is 
$\beta_{0}=L_{0}=2\log(2)+\log(|\det(\mathcal{P}_{0})|)=-0.731297,$
which only depends on SPAM errors. The true values of $\beta_{1}$ and $u'$ corresponding to iterations of gates of duration $t_{g}=40~\text{ns}$ are 
\begin{eqnarray}
\beta_{1}&=&\log(\det(G))=-2t_{g}(\gamma_{1}+\gamma_{\phi}+\gamma_{3})\nonumber \\
              &=&-2t_{g}\left(\frac{2\gamma_{1}}{1+n_{z}}+\gamma_{\phi}\right)=-2.11594 \times 10^{-3},\\
              u' &=&e^{\frac{2\beta_{1}}{3}}=0.9985904.
\end{eqnarray}
{\indent}Table~\ref{tab:distbeta} also contains a comparison between the mean of the estimates $\hat{u}'(G)$ and the unitarity Eq.~(\ref{uni-def2}) $u(G)=1/(d^{2}-1)\tr(W_{G}^{T}W_{G}).$ Note that the results presented in Table~\ref{tab:distbeta} show no discernible differences between the measures $u(G)$ and $u'(G).$ For example, for the model parameters chosen in our simulations, we find the following difference between the true values of the unitarities $u'(\mI)$ and $u(\mI)$: $u(\mI)-u'(\mI)=3.7\times10^{-10}$ (recall the inequality $u(G)\geq u'(G),$ proved earlier in Sec.~\ref{subsec:iter-uni}).
\begin{table}[t]
\begin{ruledtabular}\begin{tabular}{ccccccccc}
$\text{Gate}$ & $\mu_{\hat{\beta}_{0}}$ & $\sigma_{\hat{\beta}_{0}}$ & 
$\mu_{\hat{u}'}$& $\sigma_{\hat{u}'}$ &$u$ & \\ [3pt]
\hline 
$\mI$                                                 &     $-0.731$       &     $3.35\times 10^{-3}$      &       $0.998590$        &  $8.73\times 10^{-6}$   &   $0.998590$ \\
$\mX_{\frac{\pi}{2}}$                          &     $-0.731$       &      $3.14\times 10^{-3}$     &       $0.998590$        &  $8.83\times 10^{-6}$   &   $0.998590$ \\
$\mX_{\pi}$                                        &     $-0.731$       &      $3.14\times 10^{-3}$     &       $0.998590$        &  $8.75 \times 10^{-6}$  &   $0.998590$ \\
$ \mZ_{\pi}$                                       &     $-0.731 $      &      $3.73\times 10^{-3} $     &      $0.99718   $       & $1.08 \times 10^{-5} $  &   $0.99718   $ \\   
$\mZ_{\frac{\pi}{2}}$                          &     $-0.731 $      &      $ 3.72\times 10^{-3} $    &      $0.99578   $        &$ 1.26  \times  10^{-5} $  &   $0.99578  $ \\ [3pt]
 
    \end{tabular}\end{ruledtabular}
        \caption{Distribution of the estimates $\hat{\beta}_{0}$ and $\hat{u}'$ obtained via the ID-test, applied to various gates. The assumptions and parameters are as in Fig.~\ref{fig:distbeta}. The duration of the gates $\mI,$ $\mX_{\frac{\pi}{2}}$ and $\mX_{\pi}$ is $t_{g}$. Here, $\mZ_{\pi}=\mX_{\pi}\mY_{\pi}$ and $\mZ_{\frac{\pi}{2}}=\mX_{\frac{\pi}{2}}\mY_{\frac{\pi}{2}}\mX_{-\frac{\pi}{2}}$ and therefore, the durations of these gates are $2t_{g}$ and $3t_{g}$, respectively. Finally, the last column of the table displays the values the unitarity $u(G)=\frac{1}{d^{2}-1}\tr[W_{G}^{T}W_{G}]$.}
        \label{tab:distbeta}
            \end{table}
The fact that this difference is so small is not a mere coincidence resulting from our particular model. This, in general, will be the case for high-fidelity gates. This observation may be explained as follows: A high-fidelity gate $G$ will be close to a unitary operation, which has all singular values equal to $1.$ Therefore, we can write the singular values of $W_{G}$ as $s_{n}=1-\delta s_{n}$, where $|\delta s_{n}| \ll 1.$ Next, using the series expansions $u(G)=1-2/(d^{2}-1)\sum_{n}\delta{s_{n}}+O(\{\delta s_{n}\}^2)$ and $\det(G)=\det(W_{G})=1-\sum_{n}{\delta}s_{n}+O(\{\delta s_{n}\}^2)$ yields
\beq
u(G)-u'(G)=O(\{\delta s_{n}\}^2).
\eeq
 We thus expect, in practice, both measures to yield similar results. Recall, however, that our determinant-based protocol was devised to detect deviations from context-independence. If no context-dependence is detected, the ID-test can be used to determine the unitarity of an individual gate, unlike the RB-based protocol presented in \cite{wallman2015}, which yields the unitarity $\bar{u}$ of the average error map of a 2-design. Note that with our definition of unitarity, the knowledge of the unitarity of the noisy (and context-independent) generators of a group determines the unitarity of all elements of an implementation of that group (e.g., the Clifford group). Finally, it is worth mentioning a class of maps $G$ for which the measures $u(G)$ and $u'(G)$ coincide. Consider a trace preserving map of the form $G=U\Lambda$, such that $U$ is unitary (i.e., $U^{T}U=I $) and the unital part of $\Lambda$, which we denote by $W_{\Lambda}$, is of the form $\text{diag}(\alpha, \alpha, \ldots, \alpha)$. Then, it is clear that $|\det(G)|=|\det(\Lambda)|=|\det(W_{\Lambda})|=|\alpha|^{d^{2}-1}$ and therefore, $u'(G)=|\alpha|^2=1/(d^2-1)\tr(W^{T}_{\Lambda} W_{\Lambda})=1/(d^2-1)\tr(W_{G}^{T}W_{G})=u(G).$ This class includes the depolarizing channel and the map describing the free evolution of a qubit undergoing energy relaxation and dephasing (see the matrix Eq.~(\ref{uni-example})), in the particular case $T_{1}=T_{2}.$\\
 {\indent}The following steps summarize our unitarity estimation protocol: \\
{\bf (i)} Iterate a gate $G$ to obtain a set of $M$ probability matrices estimates $\{\hat{\mathcal{P}}_{m}\}_{m=1}^{M}$ \\
{\bf(ii)} Use these probability matrices to compute the observations $y_{m}=L_{m}.$\\ 
 {\bf(iii)} Estimate the weights $w_{m}=1/\sigma_{m}^2$, which can be done either via bootstrapping or using the method we present in the next section.\\
 {\bf(iv)} Fit a linear model $\hat{y}_{m}=\hat{\beta}_{0}+ \hat{\beta}_{1}m$ and then assess the goodness of fit.\\
 {\bf(v)}  Using Eq.~(\ref{est-uni}), compute the unitarity estimate $\hat{u}'(G)$ and estimate the standard deviation $\sigma_{\hat{u}'}$ via Eqs.~(\ref{Cov}) and (\ref{var-uni}).
 \subsection{Bounds on the precision of the WLS estimates}
{\indent} In this subsection we further examine the precision of the WLS estimates $\hat{\beta}_{0}$ and $\hat{\beta}_{1}$ ($\hat{u}$ is directly
related to $\hat{\beta}_{1}),$ which are determined by the set 
of weights $\{w_{m}=1/\sigma_{m}^{2}\}$ and the lengths $\{m_{n}\}$ of the sequences used in the ID-test. The weights $\{w_{m}\}$ are, as discussed earlier, specified by the spread of the log-det of the estimates $\hat{\mathcal{P}}_{m}$. If our observations were homoskedastic i.e., $\sigma_{m}=\sigma_{0}=\text{const.}$, then we could easily express 
the standard deviations $\sigma_{\hat{\beta}_{0}}$ and $\sigma_{\hat{\beta}_{1}}$ in terms of $\sigma_{0}$ and the lengths $\{m_{n}\}$. Indeed, if $\sigma_{m}=\sigma_{0},$ then the sums $\sum_{n} w_{n}m_{n}$ and $\sum_{n} w_{n}m_{n}^2$ in Eq.~(\ref{Cov}) can be performed explicitly using elementary methods. In particular, for $M$ evenly spaced sequence lengths $\{m_{n}=b(n-1) \}_{n=1}^{M}$, Eq.~(\ref{Cov}) yields
 \beq 
 \label{sigma0}
 \sigma_{\hat{\beta}_{0}}=\sqrt{\frac{2}{M}\left(\frac{2M-1}{M+1}\right)}\sigma_{0} \approx \frac{2}{\sqrt{M}}\sigma_{0},
 \eeq
 where the approximation is for $M\gg1$. Analogously,  for $\sigma_{\hat{\beta}_{1}}$ we find the relation  
  \beq 
  \label{sigma1}
 \sigma_{\hat{\beta}_{1}}=\frac{2\sqrt{3}}{\sqrt{M(M^{2}-1)b^2}}\sigma_{0} \approx \frac{2\sqrt{3}}{\sqrt{M}m_{\text{max}}}\sigma_{0},
   \eeq
 where $m_{\text{max}}:=b(M-1)$ is the length of the longest sequence used in the ID-test (in our simulations $b=10$ and $m_{\text{max}}=500$). In general, however, the variability of $\log(|\det(\hat{\mathcal{P}}_{m})|)$ will depend on the length of the sequence $m$, as illustrated in Fig.~\ref{fig:disp}. The usefulness of formulae (\ref{sigma0}) and (\ref{sigma1}) is that these can be used to, crudely, estimate the standard deviations  $\sigma_{\hat{\beta}_{0}}$ and  $\sigma_{\hat{\beta}_{1}}$ (i.e., their orders of magnitude) using only one standard deviation $\sigma_{m}$. For example, let us consider only the point $m=0,$ for which  $\sigma_{0}$ is approximately $10^{-2}.$ Substituting this value in  Eqs.~(\ref{sigma0}) and (\ref{sigma1}) we obtain $\sigma_{\hat{\beta}_{0}}\approx2.8\times 10^{-3}$ and $\sigma_{\hat{\beta}_{1}}\approx 9.7\times 10^{-6}$. These guesstimates are relatively close to the actual values (see the caption of Fig.~\ref{fig:distbeta}). \\
{\indent}Suppose now that we know the maximum and minimum values of $\sigma_{m},$ which we denote by $\sigma_{\text{max}}$ and $\sigma_{\text{min}},$ respectively. Then, instead of equations (\ref{sigma0}) and (\ref{sigma1}) -- which require homoskedasticity -- we can, for evenly spaced sequences, write down the following upper and lower bounds for $\sigma_{\hat{\beta}_{0}}$ and  $\sigma_{\hat{\beta}_{1}}$:
 \beq 
 \label{boundsigma0}
\sqrt{\frac{2}{M}\left(\frac{2M-1}{M+1}\right)} \frac{\sigma_{\text{min}}^2}{\sigma_{\text{max}}} \leq \sigma_{\hat{\beta}_{0}} \leq \sqrt{\frac{2}{M}\left(\frac{2M-1}{M+1}\right)} \frac{\sigma_{\text{max}}^2}{\sigma_{\text{min}}},
 \eeq
   \beq 
   \label{boundsigma1}
\frac{2\sqrt{3}}{\sqrt{M(M^{2}-1)b^2}} \frac{\sigma_{\text{min}}^2}{\sigma_{\text{max}}} \leq \sigma_{\hat{\beta}_{1}} \leq \frac{2\sqrt{3}}{\sqrt{M(M^{2}-1)b^2}}\frac{\sigma_{\text{max}}^2}{\sigma_{\text{min}}},
 \eeq
 where we assumed that $M\gg1.$ We prove these inequalities by bounding the numerators and the denominator of Eq.~(\ref{Cov}). For example, to bound the numerator of $\sigma_{\hat{\beta}_{0}}$ we use the trivial inequality  $w_{\text{min}}\sum_{n}m_{n}^2\leq \sum_{m}w_{m}m_{n}^{2} \leq w_{\text{max}}\sum_{n}m_{n}^2.$ To derive upper and lower bounds on the denominator $D_{\text{en}}:=\sum_{n}w_{n}\sum_{n}w_{n}m_{n}^2 -(\sum_{n}w_{n}m_{n})^2$  of Eq.~(\ref{Cov}), we set $\mathcal{W}:=\Tr[W]=\sum_{n}w_{n},$ $q_{n}:=w_{n}/\mathcal{W}$ (so that $\sum_{n}q_{n}=1$),  and then notice that the denominator $D_{\text{en}}$ can be rewritten as
 \beq
D_{\text{en}} =\mathcal{W}^{2}\sum_{n>n'}q_{n}q_{n'}(m_{n}-m_{n'})^2.  
 \eeq
 Hence, we have the inequality $\mathcal{W}^{2}q_{\text{min}}^2\sum_{n>n'}(m_{n}-m_{n'})^2\leq D_{\text{en}}\leq \mathcal{W}^{2}q^{2}_{\text{max}}\sum_{n>n'}(m_{n}-m_{n'})^2.$ Finally, the bounds (\ref{boundsigma0}) and (\ref{boundsigma1}) are obtained by making use of the fact that for evenly spaced lengths $m_{n}=(n-1)b, n=1,2,\ldots, M$ we have  $\sum_{n>n'}(m_{n}-m_{n'})^2=\frac{b^{2}}{12}M^{2}(M^{2}-1).$\\ 
 {\indent}From Fig.~\ref{fig:disp}(a) we find  that for the ID-test, applied to the sequences $\mS_{m}=\mI^{m},$ $m=0,10,\ldots, 500$, we have 
  $\sigma_{\text{min}}=\sigma_{0}\approx 0.011$ and $\sigma_{\text{max}}=\sigma_{500}\approx 0.018.$ For these values, the bounds derived above imply that $\sigma_{\hat{\beta}_{0}}\in[ 1.9\times 10^{-3}, 8.1\times 10^{-3}]$ and  $\sigma_{\hat{\beta}_{1}}\in[6.4\times 10^{-6}, 2.8\times 10^{-5}],$ in agreement with the values presented in the caption of Fig.~\ref{fig:distbeta}.  Finally, using the relation $\sigma_{\hat{u}'}\approx 2/(d^{2}-1)\sigma_{\hat{\beta}_{1}}$ (see discussion around Eq.~(\ref{var-uni})), we find that the standard deviation of the unitarity must  be in the range $\sigma_{\hat{u}'}\in[4.3\times 10^{-6}, 1.9 \times 10^{-5}].$ In conclusion, these bounds can be used to estimate the possible ranges of $\sigma_{\hat{\beta}_{0}}$ and $\sigma_{\hat{\beta}_{1}}$  by only studying a few points. For instance, if we hypothesize that the standard deviation of the log-det of $\mP_{m}$ increases monotonically -- due to decoherence -- with the length of the sequence (which in the absence of context-dependence seems to be a reasonable assumption), then the bounds Eqs.~(\ref{boundsigma0}) and (\ref{boundsigma1}) can be computed from the probability matrix estimates $\hat{\mathcal{P}}_{0}$ and $\hat{\mathcal{P}}_{m_{\text{max}}}.$ 
 \section{Distribution of the log-det and heteroskedasticity}
 \label{subsec:log-det-dist}
   \begin{figure}[!t]
\includegraphics[width=2.2in]{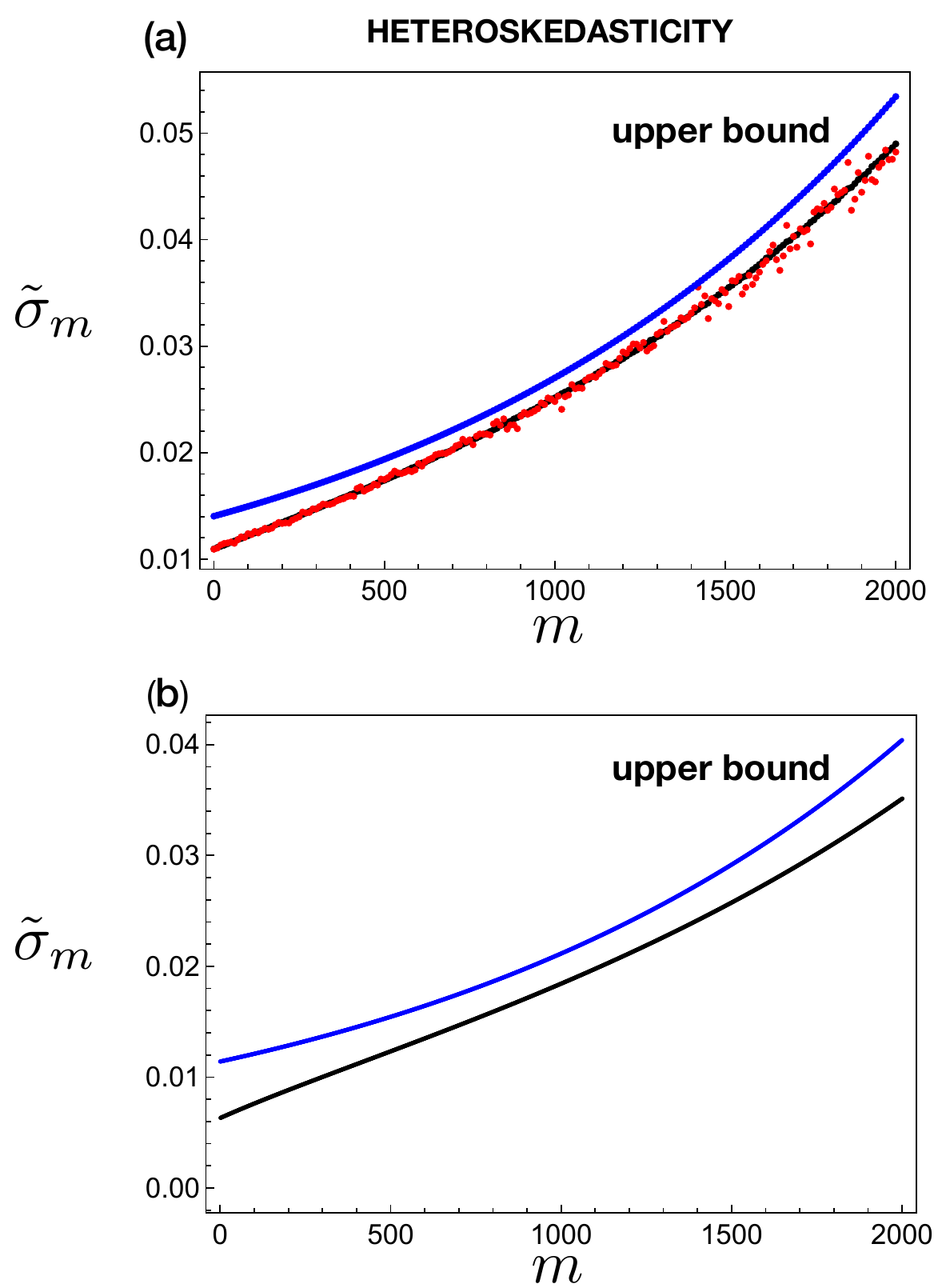}
\caption{Standard deviation of $\log(|\det(\hat{\mathcal{P}}_{m})|)$ in the iterative determinant test, applied to $\mS_{m}=\mI^{m}$. Panel (a) shows a comparison between the ``true'' standard deviations (black dots), obtained via the bootstrapping method (with $B=4\times 10^{5}$) and the estimates (red dots) $\tilde{\sigma}_{m}=\tilde{\sigma}[\hat{\mP}_{m}]$, obtained using Eq.~(\ref{sigma}). The model parameters in panel (a) are as in Fig.~\ref{fig:distbeta} and $N_{s}=50,000$. Panel (b) displays $\tilde{\sigma}[\mP_{m}]$ (black curve) for the model Eq.~(\ref {idlegamma}). Both panels display the corresponding upper bounds (blue curves) computed from Eq.~(\ref{upper-bound1}). 
  \label{fig:disp}}
\end{figure} 
{\indent}In this section we study in detail the distribution of the quantity $\log(|\det(\hat{\mP})|)$ -- which plays a major role in this work -- as a function of the true probability matrix ${\mP}$. To do so, we start by assuming that $\mathcal{P}$ is invertible, which allows us to write $\log(|\det(\hat{\mathcal{P}})|)=\log(|\det(\mathcal{P}+\delta\hat{\mathcal{P}})|)=\log(|\det(\mathcal{P})|)+\log(|\det(I+\mathcal{P}^{-1}\delta\hat{{\mathcal{P}}})|).$ Next, for sufficiently small fluctuations $\delta\hat{\mathcal{P}}_{k|i}$, we can use the well-known approximation $\det(I+{\mathcal{P}^{-1}\delta \hat{\mathcal{P}}})\approx 1+\tr[{\mathcal{P}^{-1}\delta\hat{ \mathcal{P}}}].$ Hence, for $N_{s}\gg1,$ we have
\beq 
\label{log-det-approx}
\log(|\det(\hat{\mathcal{P}})|)\approx \log(|\det(\mathcal{P})|)+\tr{[\mathcal{P}^{-1}\delta \hat{\mathcal{P}}}].
\eeq
Since the fluctuations $\delta\hat{{\mathcal{P}}}_{ki}$ are independent and gaussian random variables (for $N_{s}\gg1$), we find, making use of Eqs.~(\ref{Prob2}) and (\ref{log-det-approx}), the sought distribution
\beq
\log(|\det(\hat{\mathcal{P}})|)\sim \mathcal{N}(\log(|\det(\mathcal{P})|), \tilde{\sigma}^2[\mathcal{P}]),
\eeq
where the variance $\tilde{\sigma}^2[\mathcal{P}]$ is given by 
\begin{eqnarray}
\label{sigma}
\tilde{\sigma}^{2}[\mP]&=&\frac{1}{N_{s}}\sum_{i,k}({\mP^{-1}}_{ik})^2\mP_{ki}(1-\mP_{ki})\\
                                   &=& \frac{1}{N_{s}}\tr[(\mathcal{P}^{-1}\circ \mathcal{P}^{-1})(\mathcal{P}-\mathcal{P}\circ \mathcal{P})] \label{sigma-hhad}\\
                                   &=& \frac{1}{N_{s}}\tr[(\mathcal{P}^{-1}\circ \mathcal{P}^{-1})(\mathcal{P}\circ{Q}[\mP])].
                                                                                                         \label{sigma-had}
\end{eqnarray}
Here, $A \circ B$ denotes the entry-wise product of the matrices $A$ and $B$ (this operation is also known as the Hadamard product) and $Q[\mathcal{P}]$ is the complementary probability matrix, whose entries are ${Q[\mathcal{P}]}_{ki}=1-\mathcal{P}_{ki}$. The last two equations, expressed in terms of the Hadamard product, will prove useful in the next section. Note that in deriving the above relation, we made use of the fact that for two statistically independent Gaussian variables, $X_{1(2)}\sim \mathcal{N}(\mu_{1(2)},\sigma_{1(2)}^2),$ the probability distribution of the sum $c_{1}X_{1}+c_{2}X_{2}$ is $\mathcal{N}(c_{1}\mu_{1}+c_{2}\mu_{2}, c_{1}^{2}\sigma_{1}^2+c_{2}^2\sigma_{2}^{2}).$\\
{\indent} An  obvious application of Eq.~(\ref{sigma}) is to estimate the standard deviation of $\log(|\det(\hat{\mP})|).$ That is, given a probability matrix estimate $\hat{\mP},$ we can approximate the true standard deviation $\tilde{\sigma}[\mP]$ with $\tilde{\sigma}[\hat{\mP}].$ Note that this approach is, to a large extent, equivalent to  the parametric bootstrap method \cite{efron1994introduction}, employed earlier in Sec.~\ref{Sec:fluct}. A clear advantage of using Eq.~(\ref{sigma}), instead of the bootstrap method, is 
that it allows us to considerably save computational time because we do not need to generate $B\gg1$ replicas of $\hat{\mP},$ for each sequence.~Another useful application of Eq.~(\ref{sigma}) is that it can be used to establish bounds on  the standard deviation $\tilde{\sigma}.$ For example, it is straightforward to show that for any probability matrix $\mP,$ with singular values $\{s_{k}(\mP)\}_{k=1}^{d^{2}},$ the following inequality holds: 
\begin{eqnarray} 
\tilde{\sigma}^2[\mP]&=& \frac{1}{N_{s}}\sum_{i,k}({\mathcal{P}^{-1}}_{ik})^{2}\mP_{ki}(1-\mP_{ki})\\
                        &<& \frac{1}{4N_{s}}\sum_{i,k}({\mathcal{P}^{-1}}_{ik})^{2}=\frac{1}{4N_{s}}\sum_{k} s_{k}^{2}(\mathcal{P}^{-1})\\
                        &=&\frac{1}{4N_{s}}\sum_{k} \frac{1}{s_{k}^{2}(\mathcal{P})}=\frac{1}{4N_{s}}||\mathcal{P}^{-1}||^{2}_{\text{F}},\label{upper-bound1}
                        \end{eqnarray}
where $||\cdot ||_{F}$ denotes the Frobenius norm \footnote{For a real matrix $A$ the Frobenius norm is $||A||_{F}=\sqrt{\tr(A^{T}A)}$}. In deriving this upper bound we made use of the inequality $\mP_{ik}(1-\mP_{ik})\leq 1/4$, for all $i$ and $k$. The above bound will be further examined in Sec.~\ref{subsec:Monte-Carlo}.\\
{\indent}We used the results presented thus far in this section to study the standard deviations (SDs) associated with the ID-test applied to the gate $\mI$ (see Fig.~\ref{fig:disp}(a)). More precisely, we compared the true SDs obtained via the bootstrap method, using the true probability tables $\mP_{m}$ (obtained from our ideal simulations), with the standard deviations 
$\tilde{\sigma}(\hat{\mP}_{m})$ obtained via Eq.~(\ref{sigma}). Figure \ref{fig:disp}(a) also shows the upper bounds ${1}/(4N_{s})||\mathcal{P}_{m}^{-1}||^{2}_{\text{F}},$ computed from the true probabilities matrices $\{\mP_{m_{n}}\}_{n=1}^{M}.$ The results illustrated in Fig.~\ref{fig:disp}(a) show that, indeed, the quantities $\tilde{\sigma}[\hat{\mP}_{m}]$ can be used  as reasonably accurate estimates of the true standard deviations, when $N_{s}\gg 1$.\\
{\indent}The relation Eq.~(\ref{sigma}) can also be exploited to get a crude estimate of the magnitude of the SD of the quantity $\log(|\det(\hat{\mP})|)$ by simply considering a simplified model of the gate $\mG$ we are iterating. This approach will prove useful in reducing the SDs and, consequently, increasing the power of our context-independence tests and the precision of the unitarity estimates, as we explicitly show in the next section. But first, let us examine some simple, yet important, cases. Consider the ideal single-qubit idle gate $\mI=I$ (i.e., the identity matrix)  and suppose we can perfectly prepare the input states used thus far (i.e.,  $\{\ket{\phi_{i}}\}_{i=1}^{4}=\{\ket{g}, \ket{e},1/\sqrt{2}(\ket{g}+\ket{e}), 1/\sqrt{2}(\ket{g}+i\ket{e})\}$) and also suppose we can perfectly measure the projectors $\{\Pi_{k}\}_{k=1}^{4}=\{\ket{\phi_{k}}\bra{\phi_{k}}\}_{k=1}^{4}.$ Then the probability matrix (with entries $\mP^{\text{ideal}}_{ki}=|\expect{\phi_{k}|\phi_{i}}|^2$) reads
\beq 
\mP^{\text{ideal}}=\begin{bmatrix} 1 & 0 &1/2 &1/2 \\
                                                       0&  1 & 1/2 & 1/2\\
                                                       1/2& 1/2& 1 & 1/2\\
                                                       1/2& 1/2& 1/2& 1 \\
                              \end{bmatrix}.
\eeq
Now, making use of Eq.~(\ref{sigma}) we easily find that for this probability matrix the statistical fluctuations of the log-det (when $N_{s}\gg 1$) are characterized by the standard deviation
\beq 
\label{SD-st}
 \tilde{\sigma}[\mP^{\text{ideal}}]=\sqrt{\frac{2}{N_{s}}}.
 \eeq
 To analyze a slightly more complicated case, let us now assume that the matrix representation (in the Pauli basis) of the idle gate is 
\beq 
\label{idlegamma}
\mI_{\gamma}=\begin{bmatrix} 1& 0& 0&0 \\
                                0 & e^{-(\gamma_{1}/2+\gamma_{\phi})t_{g}} & 0&0 \\
                                0 & 0 & e^{-(\gamma_{1}/2+\gamma_{\phi})t_{g}}&0 \\
                                1-e^{-\gamma_{1}t_{g}}&0&0&e^{-\gamma_{1}t_{g}} \\                                              
                                 \end{bmatrix},                       
 \eeq
which corresponds to setting $\gamma_{3}=0$ and $\varphi=0$ in our ZZ model. Ignoring SPAM errors and using the same input states and measurements as in the previous example, we find that the variance Eq.~(\ref{sigma}), associated with the sequence $\mS_{m}=\mI_{\gamma}^{m}$, is  
\begin{eqnarray}
\label{sigmagamma}
\tilde{\sigma}_{m}^2&=&\frac{1}{N_{s}}[6 e^{2\gamma_{\phi}t_{g}m}\sinh(\gamma_{1}t_{g}m) +4e^{\gamma_{\phi t_{g} m}}\nonumber \\
&\times&(\sin(\frac{\gamma_{1}t_{g}m}{2})+e^{\gamma_{\phi}t_{g}m})+(e^{\gamma_{1}t_{g}m}-3) ].  
\end{eqnarray}
  This expression for $\tilde{\sigma}_{m}$ vs. the length sequence is plotted in Fig.~\ref{fig:disp}(b), where the relevant model parameters are $t_{g}=40~\text{ns}$,  $\gamma_{1}=40~\mu s$ and $\gamma_{\phi}=\gamma_{1}/2.$ 
From Eq.~(\ref{sigmagamma}), we find that for short sequences, i.e.,  $m \gamma_{1}t_{g}\ll 1$ and $ m\gamma_{\phi}t_{g}\ll 1,$ the standard deviation of the quantity $\log(|\det(\hat{\mP}_{m})|)$ grows linearly with $m$. Specifically, we have
\beq 
\tilde{\sigma}_{m}\approx \sqrt{\frac{2}{N_{s}}}\left(1+\frac{(9\gamma_{1}+8\gamma_{\phi})t_{g}}{4}m\right).
\eeq
For very long sequences, we learn from Eq.~(\ref{sigmagamma}) that the standard deviation $\tilde{\sigma}_{m}$ grows exponentially with $m.$ Note, however, that for sufficiently long sequences the approximation Eq.~(\ref{log-det-approx}), used to derive Eq.~(\ref{sigma}), may break down (because of large matrix elements $\mP^{-1}_{ki}$). Nonetheless, in Fig.~\ref{fig:disp}(a) we see that we can go up to $m_{\text{max}}=2000$, while still obtaining estimates $\tilde{\sigma}_{m}=\tilde{\sigma}[\hat{\mP}_{m}]$ close to the ``true'' standard deviations (found via the bootstrap method).\\
{\indent}Finally, let us consider the free evolution of a qubit in the absence of decoherence. Suppose that the qubit's Hamiltonian is $H_{0}=\omega \ket{e}\bra{e}.$ Then for the ideal input states and projectors discussed in the previous examples, we find the variance
\beq 
\tilde{\sigma}^2_{t}=\tilde{\sigma}^2[\mP_{t}]=\frac{2+\sin^{2}(2\omega t)}{N_{s}}, 
\eeq
where the entries of the probability matrix $\mP_{t}$ are ${[\mP_{t}}]_{ki}=|\expect{\phi_{k}|e^{-iH_{0}t}|\phi_{i}}|^{2}$. This basic example shows two things: (i) The fluctuations of $\log(|\det(\hat{\mP}_{m})|)$ do not necessary have to increase monotonically with the length of the sequence $m$, even when the evolution of the system is Markovian. (ii) More importantly, the standard deviation $\sigma[\hat{\mP}]$ will, in general, depend on how we prepare and measure our system, and therefore it will be affected by SPAM errors. Indeed, the last example can be reinterpreted as (i) preparing a set of input states $\text{exp}(-i\tau H_{0})\ket{\phi_{i}},$ which implies SPAM errors; (ii) iterating $m$ times the ideal idle gate $\mI=I$ (iii) measuring the ideal projectors $\{\Pi_{k}\}_{k=1}^{4}.$ Clearly, the true model for such fictitious ID-test is $y_{m}=L_{m}=0\times m +\varepsilon_{m},$ where $\varepsilon _{m} \sim \mathcal{N}(0, \tilde{\sigma}^2_{\tau})$.   
\section{Determinant-based tests and SIC-sets}
\label{Sec:SIC}
\begin{figure}[t!]
\includegraphics[width=3.2 in]{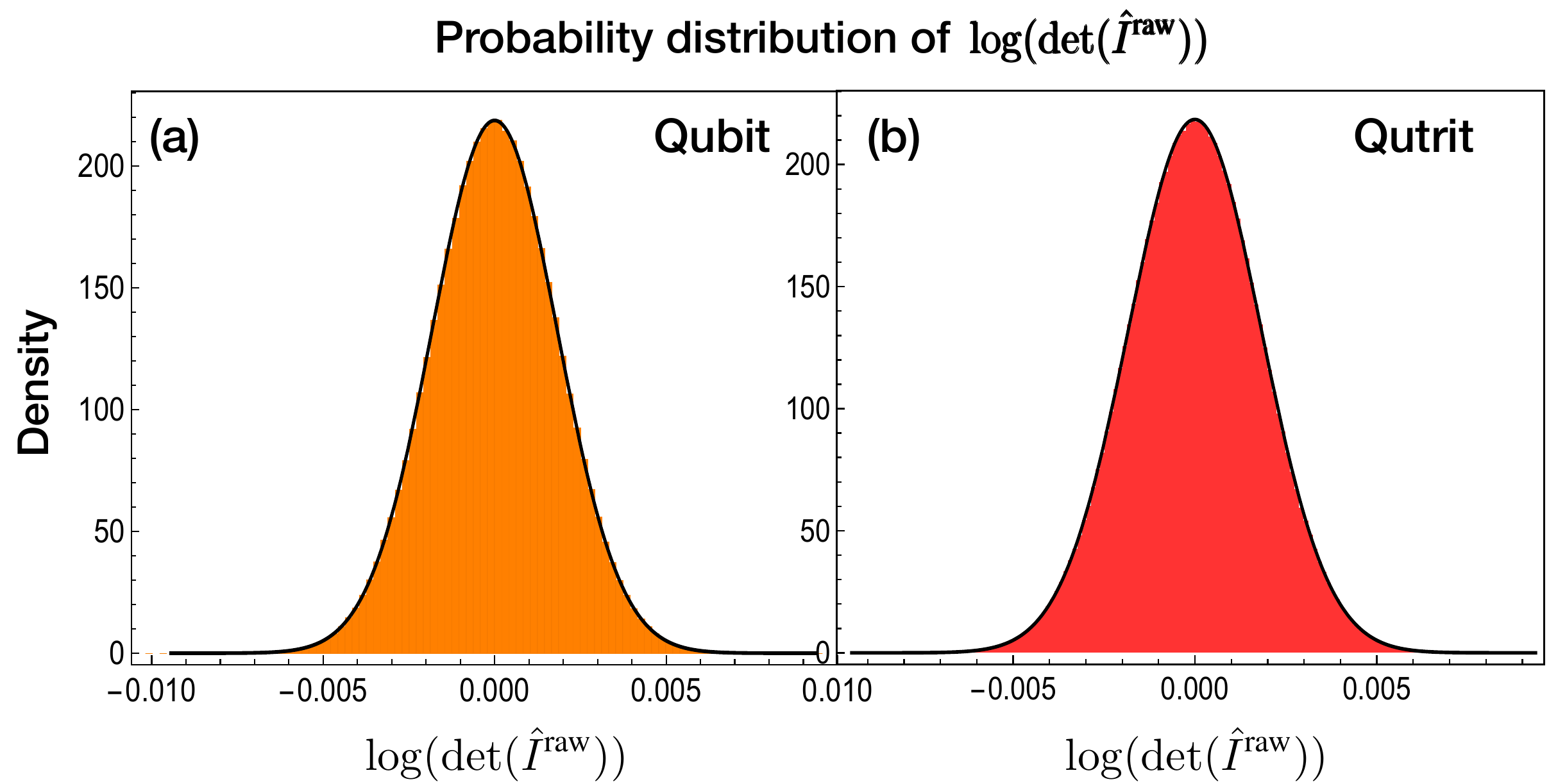}
\caption{Simulations of the distribution of $\log(\det(\hat{I}^{\text{raw}}))$ for a qubit SIC-set (panel (a)) and a qutrit SIC-set (panel (b)). To obtain the distributions, for each measurement configuration, we sampled $R=4\times 10^{6}$ counts $n_{k|i}$  from the binomial distribution $\text{Bin}(N_{s}, \mP_{k|i}^{\text{ideal}}),$ with $N_{s}=50,000.$ The simulations yielded the values $\sigma_{\text{qubit}}=1.825\times10^{-3}$ and $\sigma_{\text{qutrit}}=1.826\times10^{-3}$, in very good agreement with the predicted value $\tilde{\sigma}=1/\sqrt{6N_{s}}=1.82574\times 10^{-3}.$ 
\label{fig:SIC-dist}}
\end{figure}

{\indent}As already shown in the previous section, the standard deviation of  $\log(|\det(\hat{\mP})|)$ depends on the measurement configurations, i.e., the set of input states and measurements used to obtain the probability matrix $\hat{\mP}.$ The single-qubit tomographic set used thus far in this work was based on the states 
 \begin{eqnarray}  
\label{ini-james4}
\ket{\phi_{1}}&=&\ket{g},\\ 
\ket{\phi_{2}}&=&\ket{e},\\
\ket{\phi_{3}}&=&\frac{1}{\sqrt{2}}(\ket{g}+\ket{e}),\\ 
\label{end-james4}
\ket{\phi_{4}}&=&\frac{1}{\sqrt{2}}(\ket{g}+i\ket{e}).
\end{eqnarray} 
Sets of this form are commonly used when performing quantum process tomography (see e.g., \cite{Shabani2011,Rodionov2014}) as they can be easily prepared and measured, using basic Clifford operations. In this section, we will explore the possibility of employing more refined sets of states with the purpose of reducing the standard deviation of our estimates. Namely, we will focus on symmetric sets of states, such as  
\begin{eqnarray} 
\label{ini-sic-states}
\ket{\psi_{1}}&=&\ket{g},\\
\ket{\psi_{2}}&=& \frac{1}{\sqrt{3}}(\ket{g}+\sqrt{2}\ket{e}),\\
\ket{\psi_{3}}&=& \frac{1}{\sqrt{3}}(\ket{g}+e^{\frac{2\pi i}{3}}\sqrt{2}\ket{e}),\\ 
\label{end-sic-states}
\ket{\psi_{4}}&=& \frac{1}{\sqrt{3}}(\ket{g}+e^{-\frac{2\pi i}{3}}\sqrt{2}\ket{e}), 
\end{eqnarray}
which represents a regular tetrahedron inscribed in the Bloch sphere and constitutes a single-qubit symmetric informationally complete (SIC) set. More generally, a SIC-set is defined as a set of $d^{2}$ (where $d$ is the dimension of our Hilbert space) vectors $\ket{\psi_{i}}$ satisfying 
\beq 
\label{sic1}
|\expect{\psi_{i}|\psi_{j}}|^2=\frac{d\delta_{ij}+1}{d+1},
\eeq
(see e.g.,~\cite{appleby2014symmetric,renes2004symmetric,wootters2006quantum}). If we write $M_{k}=(1/d)\Pi_{k}$, where $\Pi_{k}=\ket{\psi_{k}}\bra{\psi_{k}},$ then the set of positive semidefinite operators $\{M_{k}\}_{k=1}^{d^{2}}$ is a SIC-POVM \cite{renes2004symmetric}. \\
{\indent}Now, as in the previous section, we consider the ideal idle gate $\mI=I$ and we assume the absence of SPAM errors. This approach will allow us to examine (via the function $\mP\rightarrow \tilde{\sigma}[\mP]$) the advantage of using SIC-sets to increase the power and precision of our tests.~Moreover, most of the interesting quantum gates satisfy (ideally) the condition $G^{n}=I_{d^{2}}$, for some $n\geq 0$, which means that in the case of high-fidelity gates and moderate SPAM errors, the results obtained by considering the identity matrix $I$ will be applicable (to some extent)  to a wide class of gate sequences e.g.,  $\mS_{k}=(\mX_{\pi/2})^{4k}, k=0,1,2\ldots$ \footnote{At least for sufficiently short sequences.}. Thus, we begin by considering the probability matrix 
\beq 
\label{ProbSic}
\mP^{\text{ideal}}_{\text{sic}}=\begin{bmatrix} 1 & 1/3&1/3 &1/3 \\
                                                       1/3&  1 & 1/3 & 1/3\\
                                                       1/3& 1/3& 1 & 1/3\\
                                                       1/3& 1/3& 1/3& 1\\
                              \end{bmatrix}, 
\eeq
whose entries are given by $(\mP_{\text{sic}}^{\text{ideal}})_{ki}=|\expect{\psi_{k}|\psi_{i}}|^{2},$ where $\{\ket{\psi_{i}}\}_{i=1}^{4}$ is a qubit SIC-set 
(such as the one presented earlier in this section). Now, using the expression Eq.~(\ref{sigma}) for the SD of the estimates $\log(|\det(\hat{\mP}^{\text{ideal}}_{\text{sic}})|),$ we readily find that
\beq 
\label{sigma3}
\tilde{\sigma}[\mP^{\text{ideal}}_{\text{sic}}]=\frac{1}{\sqrt{6N_{s}}}.
\eeq
This result, in comparison with $\tilde{\sigma}[\mP^{\text{ideal}}]=\sqrt{{2}/{N_{s}}}$ (see Eq.~(\ref{SD-st})), obtained for the states (\ref{ini-james4}--\ref{end-james4}), represents a reduction by a factor of $12$ in the number of experimental runs $N_{s}$ needed to achieve a specified level of precision $\sigma_{*}$. Furthermore, the simple structure of the probability matrices for SIC-sets, allows us to derive the following formula for $\tilde{\sigma}[\mP^{\text{ideal}}_{\text{sic}}],$ valid for any dimension $d$ (provided a SIC-set exists in that dimension \cite{fuchs2017sic}):
\newtheorem*{factSIC}{Fact 1}
\begin{factSIC} 
Let $\{\ket{\psi_{i}}\}_{i=1}^{d^{2}}$ be a SIC-set in d dimensions and let $[\mP_{\text{sic}}^{\text{ideal}}]_{ij}=|\expect{\psi_{i}|\psi_{j}}|^2.$ Then the variance of $\log(|\det(\hat{\mP}_{\text{sic}}^{\text{ideal}})|)$ is given by the relation
\beq 
 \label{sigmad}
 \tilde{\sigma}^2[\mP^{\text{ideal}}_{\text{sic}}]= \frac{d-1}{d(d+1)}\frac{1}{N_{s}}.
 \eeq
\end{factSIC}
{\noindent}{\emph{Proof}}. To prove this fact, we first notice that the inverse of a matrix $\mP$ having all diagonal entries equal to $1$ and off-diagonal entries equal to $1/(d+1)$ may be written as 
 \beq 
 \label{sicinv}
(\mP^{\text{ideal}}_{\text{sic}})^{-1}=\frac{d+2}{d}I_{d^2}-\frac{d+1}{d^{2}}\mP^{\text{ideal}}_{\text{sic}}.
\eeq
This expression may also be easily obtained by first noticing that $\mP^{\text{ideal}}_{\text{sic}}=d/(d+1)(I_{d^{2}}+dvv^{T}),$ where $v=1/d[1,1,\ldots,1]^{T},$ and then applying the Sherman-Morrison formula \cite{sherman1950adjustment} to obtain the inverse $(\mP^{\text{ideal}}_{\text{sic}})^{-1}.$ Since the diagonal elements of $\mP^{\text{ideal}}_{\text{sic}}$ are all $1$, only the off-diagonal elements will contribute to the sum $\sum_{ki}(\mP^{-1}_{ki})^2\mP_{ik}(1-\mP_{ik}),$ in Eq.~(\ref{sigma}). Hence, we find that 
\begin{eqnarray}
\tilde{\sigma}^2[\mP^{\text{ideal}}_{\text{sic}}]&=&\frac{1}{N_{s}}d^{2}(d^{2}-1)\left(\frac{1}{d^{2}}\right)^2\times \frac{1}{d+1}\times\left(1-\frac{1}{d+1}\right)\nonumber \\
                                                      &=& \frac{d-1}{d(d+1)}\frac{1}{N_{s}}.\nonumber 
\end{eqnarray}
{\indent} Setting $d=2$ in the above formula, we recover Eq.~(\ref{sigma3}), that is to say $\tilde{\sigma}[\mP^{\text{ideal}}_{\text{sic}}]={1}/{\sqrt{6N_{s}}}.$ Surprisingly, we find the same result for a qutrit ($d=3$), as confirmed by simulations (see Fig.~\ref{fig:SIC-dist}). Figure~\ref{fig:SIC-dist} shows the simulated distribution of $\log(|\det(\hat{\mP})|)$ when SIC-sets in $d=2,3$ are used to estimate the log-det of the ideal idle gate $\mI=I_{d}.$ Note that we shifted both distributions by the amounts $3\log(3)-4\log(2)$ (qubit) and $16\log(2)-9\log(3)$ (qutrit) to obtain the distributions of $\log(\det(\hat{I}^{\text{raw}}))=-\log(|\det(\mP_{\text{sic}}^{\text{ideal}})|)+\log(|\det(\hat{\mP}_{\text{sic}})|),$ for which $\mathbb{E}[\log(\det(\hat{I}^{\text{raw}}))]=0$. These shifts are easily shown to be given by $-\log(|\det(\mP^{\text{ideal}}_{\text{sic}})|)=(d^2-1)\log(d+1)-d^{2}\log(d)$ \footnote{Here we made use of the following fact: The determinant of an $n \times n$ matrix $M$ having diagonal entries $M_{ii}= b$ and off-diagonal matrix elements $M_{ij}=a$ is given by $\det(M)=( (n-1)a+b )(b-a)^{n-1}$}.  
As predicted by Eq.~(\ref{sigmad}), these two distribution are virtually identical.\\
 {\indent}Although in the simulations shown in Fig.~\ref{fig:SIC-dist} we did not make use of a concrete qutrit SIC-set, for completeness' sake, we present the following realization (see e.g., \cite{pimenta2013minimum}): Let $\{\ket{0}, \ket{1}, \ket{2} \}$ be an orthonormal basis in $d=3$ and consider the states
 \begin{eqnarray}
\ket{\Psi_{1}}&=&\frac{1}{\sqrt{2}}(\ket{0}+\ket{1}), \\
\ket{\Psi_{2}}&=&\frac{1}{\sqrt{2}}(\ket{0}+e^{\frac{2\pi i}{3}}\ket{1}),\\
\ket{\Psi_{3}}&=&\frac{1}{\sqrt{2}}(\ket{0}+e^{-\frac{2\pi i}{3}}\ket{1}). 
\end{eqnarray}
These states satisfy $|\expect{\Psi_{i}|\Psi_{j}}|^2=1/4$ for $i\neq j.$ The remaining 6 members of the SIC-set are obtained by replacing the pair of kets $\{\ket{0}, \ket{1}\}$ in the above equations by $\{\ket{0} \ket{2}\}$ and $\{\ket{1}, \ket{2}\}.$ In addition to the above SIC-set in $d=3$ we may also consider the ``standard'' tomographic set \cite{baldwin2014quantum,nielsen2010quantum}
\begin{eqnarray}
\{\ket{\Phi_{i}}\}_{i=1}^{d^2}&=&\{\ket{n}\}_{n=0}^{d-1}\cup \left\{\frac{1}{\sqrt{2}}(\ket{n}+\ket{m})\right\}_{0\leq n<m\leq d-1} \nonumber \\
&\cup & \left\{\frac{1}{\sqrt{2}}(\ket{n}+i\ket{m})\right\}_{0\leq n<m \leq d-1}, \label{standardset}
\end{eqnarray}
(with $d=3$) which is simply a natural extension of the set $\{\ket{\phi_{i}}\}_{i=1}^{4}$ (see Eqs.~(\ref{ini-james4}-\ref{end-james4})) to the $d$-dimensional case. Using Eq.~(\ref{sigma}), we readily find that this standard qutrit set yields the standard deviation   
$\tilde{\sigma}[\mP^{\text{ideal}}]=\sqrt{6/N_{s}}.$\\
{\indent}It is important to note from Eq.~(\ref{sigmad}) that $\tilde{\sigma}[\mP_{\text{sic}}^{\text{ideal}}]$ decreases as the dimension $d$ of our system increases. This result, although interesting, has limited applications mainly because our scheme is not scalable as it  requires $d^{4}$ experimental settings to estimate a probability matrix $\hat{\mP}$.
Moreover, we do not know how stable this reduction in the SD is against errors in the implementation of a SIC-set and/or decoherence, for large dimensions $d$. Hence, a reasonable approach to study operations acting on small composite systems, such as two-qubit gates, is the use of local SIC-sets. For example, to characterize two-qubit gates we will use tomographic sets of the form 
$\{\ket{\psi_{i}}_{A}\otimes \ket{\psi_{j}}_{B}\},$ where $\{\ket{\psi_{i}}_{A(B)}\}$ are single-qubit SIC-sets.~For such local tomographic sets and a gate $G\approx I_{AB}=I_{\text{A}}\otimes I_{\text{B}}$, the resulting probability matrices, in the absence of SPAM errors, will approximately be of the form $\mP^{A}_{\text{sic}}\otimes \mP^{B}_{\text{sic}}.$ In addition, it is clear that the SIC-sets used for the subsystems need not be equal, which means that the SD corresponding to $G=I_{AB},$ will only depend on the dimensions of the subsystems $A$ and $B$. This approach can be straightforwardly extended to the $n-$qubit (or qutrit) case. We will refer to this tomographic scheme, based on local SIC-sets,  as $SIC^{\otimes n}_{d}$ ($d$ is the dimension of the qudit). For this tomographic scheme, we can prove the following useful result:
\newtheorem*{factSICn}{Fact 2}
\begin{factSICn} 
Let $\text{SIC}^{\;\otimes n}_{d}$ be the $n$-fold tensor product of a SIC-set in $d$ dimensions and let $\mP_{(n)\text{sic}}^{\text{ideal}}:=({\mP_{\text{sic}}^{\text{ideal}}})^{\otimes n}.$ Then the variance of the estimates 
$\log(|\det(\hat{\mP}_{(n)\text{sic}}^{\text{ideal}})|)$ is given by the formula
\begin{eqnarray}
\label{sicn}
 \tilde{\sigma}^2[\mP_{(n)sic}^{\text{ideal}}]&=&\left[\left(d^2+2d-1-\frac{1}{d}\right)^n\right. \nonumber \\
                                                                             &-&\left. \left(d^2+2d-1-\frac{2}{d+1}\right)^n\right]\frac{1}{N_{s}}
\end{eqnarray}
\end{factSICn}
{\emph{Proof.}} First, let us introduce the shorthand notation $\mP_{n}:=({\mP_{\text{sic}}^{\text{ideal}}})^{\otimes n},$ $\mP_{n-1}=( {\mP_{\text{sic}}^{\text{ideal}}})^{\otimes n-1},$\ldots $,\mP_{1}= \mP_{\text{sic}}^{\text{ideal}}.$ We now examine the terms appearing in Eq.~(\ref{sigma-hhad}) and notice that  the matrix $\mP_{n}^{-1}\circ \mP_{n}^{-1}=(\mathcal{P}_{1}^{-1}\otimes \mathcal{P}_{n-1}^{-1})\circ (\mathcal{P}_{1}^{-1}\otimes \mathcal{P}_{n-1}^{-1})$ may be conveniently written in the block form
\beq 
\begin{bmatrix}\beta^{2}\mP^{-1}_{n-1}\circ \mP^{-1}_{n-1}&\gamma^2 \mP^{-1}_{n-1}\circ \mP^{-1}_{n-1} &\ldots & \gamma^2 \mP^{-1}_{n-1}\circ \mP^{-1}_{n-1} \\   \gamma^2\mP^{-1}_{n-1}\circ \mP^{-1}_{n-1}&\beta^{2} \mP^{-1}_{n-1}\circ \mP^{-1}_{n-1} & \vdots &\gamma^2 \mP^{-1}_{n-1}\circ \mP^{-1}_{n-1}\\
                        \vdots &\ldots &\ddots& \vdots \\
                        \gamma^2 \mP^{-1}_{n-1} \circ \mP^{-1}_{n-1}& \gamma^2 \mP^{-1}_{n-1}\circ \mP^{-1}_{n-1} &\ldots & \beta^2 \mP^{-1}_{n-1}\circ \mP^{-1}_{n-1},
\end{bmatrix} 
\eeq
where $\beta=(d^2+d-1)/d^2$ and $\gamma=-1/d^{2}$ (these coefficients follow from Eq.~(\ref{sicinv})). Likewise, we write the matrix  $\mP_{n}=(\mP_{1}\otimes \mP_{n-1})$ in block form, which leads us to the relation
\begin{eqnarray}
R_{n}:&=&\Tr[(\mP_{n}^{-1}\circ \mP_{n}^{-1})\mP_{n}]\nonumber \\
           &=&d^{2}(\beta^{2}+(d^{2}-1)\alpha \gamma^{2})R_{n-1}, 
\end{eqnarray}
where $\alpha=1/(d+1)$. From the above recursion relation, we find that $R_{n}= (d^{2}(\beta^{2}+(d^{2}-1)\alpha \gamma^{2}))^n.$ Analogously, the remaining term in Eq.~(\ref{sigma-hhad}) is found to be given by 
\begin{eqnarray}
Q_{n}:&=&\Tr[(\mP_{n}^{-1}\circ \mP_{n}^{-1})(\mP_{n}\circ \mP_{n})]\nonumber \\
          &=&(d^{2}(\beta^{2}+(d^{2}-1)\alpha^{2} \gamma^{2}))^n.
\end{eqnarray}
The difference $R_{n}-Q_{n}$ yields are the result Eq.~(\ref{sicn}) which, as the reader can easily verify, reduces to Eq.~(\ref{sigmad}) for $n=1$. Since
 $SIC^{\otimes n}_{d}$ is not a SIC-set, it comes as no surprise that the variance Eq.~(\ref{sicn}) does not scale well with the dimension of the system. For instance, for two $d$-dimensional systems, Eq.~(\ref{sicn}) assumes the form
 \beq 
\tilde{\sigma}^{2}[\mP_{(2)\text{sic}}^{\text{ideal}}]=\frac{(d-1)(2d^{4}+6d^{3}+2d^2-5d-1)}{d^{2}(d+1)^2}\frac{1}{N_{s}},
\eeq
which for $d\gg1$, grows linearly with the subsystem's dimension $d.$ Finally, we consider the tensor product of two SIC-sets, corresponding to two systems with dimensions $d_{1}$ and $d_{2}$ (e.g., a qubit-qutrit system). For the  probability matrix  $\mP^{\text{ideal}}_{d_{1}{d_{2}\text{sic}}}:=\mP^{\text{ideal}}_{d_{1}\text{sic}}\otimes \mP^{\text{ideal}}_{d_{2}{\text{sic}}}$ we have the following result:
\begin{eqnarray}
\tilde{\sigma}^2[\mP^{\text{ideal}}_{d_{1}{d_{2}\text{sic}}}]=[&\prod \limits_{i=1,2}&d^2_{i}(\beta_{i}+(d_{i}^{2}-1)\alpha _{i}\gamma_{i}^{2})- \nonumber \\
                                                         &\prod \limits_{i=1,2}&d^2_{i}(\beta_{i}+(d_{i}^{2}-1)\alpha^2 _{i}\gamma_{i}^{2})]\frac{1}{N_{s}},\end{eqnarray}
where $\beta_{i}= (d_{i}^2+d_{i}-1)/d_{i}^2$, $\gamma_{i}=-1/d_{i}^{2}$ and $\alpha_{i}=1/(d_{i}+1)$. This expression can be derived along the same lines as those used to obtain Eq.~(\ref{sicn}).\\
{\indent} The results obtained so far allow us to compare the standard deviation obtained using local SIC-set schemes and local standard schemes. Table \ref{tab:sic} shows such comparison for some low-dimensional systems. In particular, for the important case $2\times2$, we find that the $SIC_{2}\otimes SIC_{2}$  scheme reduces approximately the SD by a factor of 6, that is, $\tilde{\sigma}^{(I)}/\tilde{\sigma}^{(II)}\approx 6$ (see table \ref{tab:sic}). For the $2\times 2$ case it is also worth discussing the possibility of employing the global standard set Eq.~(\ref{standardset}). This set involves 4 product states and 12 entangled states, which can be prepared employing a single entangling gate plus local gates. Interestingly, for this tomographic set  -- which involves entanglement -- we find by means of Eq.~(\ref{sigma}) the standard deviation $\tilde{\sigma}[\mP^{\text{ideal}}]=2\sqrt{3/N_{s}}\approx 3.5/\sqrt{N_{s}}$, which is still larger than that obtained for to the local scheme $SIC_{2}\otimes SIC_{2}$\\
\begin{table}[!t]
\begin{ruledtabular}\begin{tabular}{ccc}
$\text{Dimension}$ & $\qquad \sqrt{N_{s}}\tilde{\sigma}^{(I)} (\text{standard}) $ &  $\sqrt{N_{s}}\tilde{\sigma}^{(II)} (\text{sic}) $  \\ [3pt]
\hline 
2                                    &     $\sqrt{2}\approx 1.4 $                          &     $\frac{1}{\sqrt{6}}\approx 0.4 $       \\ [7pt]
$3$                                &     $ \sqrt{6}\approx 2.4 $                              &      $\frac{1}{\sqrt{6}} \approx 0.4 $    \\ [7pt]
$2\times2$                                                                      &  $2\sqrt{19}\approx8.7$     &      $\frac{\sqrt{77}}{6}\approx 1.5 $   \\[7pt]
$2\times 3 $                  &  $3\sqrt{26}\approx 15.3 $                            &  $\sqrt{\frac{10}{3}} \approx 1.8$                                       \\[7pt]
$2\times 2\times 2 $     & $2\sqrt{542}\approx 46.6$     &      $ \frac{\sqrt{4447}}{6\sqrt{6}}\approx 4.5 $  \\ [7pt]
$3\times 3$                   &    $12\sqrt{5}\approx 26.8$                            &      $\frac{\sqrt{163}}{6} \approx 2.1  $    \\  [3pt] 
\end{tabular}\end{ruledtabular}
\caption{Comparison between the standard deviations $\tilde{\sigma}^{(I)}$ and $\tilde{\sigma}^{(II)}$ of the estimates $\log(|\det(\hat{\mP})|)$ corresponding to the standard and SIC tomographic sets, respectively. For the composite systems considered in the table (e.g., $2\times 3$ and $2\times2\times2$), we compared the SDs associated with tensor products of the simpler 1-qubit and 1-qutrit sets discussed in this section (the standard set for  $d=3$ is specified by Eq.~(\ref{standardset})). A comparison between the second and third columns shows the advantage of using local SIC-sets. For example, for the three qubit system ($2\times2\times2$) the use of tensor product of single-qubit SIC-sets reduces the standard deviation by a factor of $10$ (approximately). 
\label{tab:sic}}
\end{table}
{\indent}Given that most of the results presented in this section rest upon the implementation of a SIC-set in $d=2$, we will now briefly discuss how to construct such set, starting from the computational state $\ket{0}$ (for example, the ground state of our system). First, we notice that if a qubit is initialized in the magic state $\ket{T_{+}}=\cos(\theta/2)\ket{0}+e^{\frac{i\pi}{4}}\sin(\theta/2)\ket{1},$ where $\theta=\cos^{-1}(1/\sqrt{3})$ \cite{bravyi2005universal}, then a SIC-set can be generated by applying the Clifford gates $I,X_{\pi},Y_{\pi}$ and $Z_{\pi}$ \cite{planat2017magic}.  It is now clear that in order to produce a qubit SIC-set it suffices to implement a single non-Clifford gate $U_{3}$ such that $U_{3}\ket{0}=\ket{T_{+}}.$ A natural choice for this unitary is $U_{3}=\ket{T_{+}}\bra{0}+\ket{T_{-}}\bra{1},$ where $\ket{T_{-}}=\sin(\theta/2)\ket{0}-e^\frac{i\pi}{4}\cos(\theta/2)\ket{1},$ so that $\expect{T_{-}|T_{+}}=0$. Thus, in terms of the general 3-parameter single-qubit unitary $U(\theta,\phi,\lambda)$ \cite{williams2010explorations}, we can write 
 \beq 
U_{3}=\begin{bmatrix} \cos(\theta/2) & -e^{-i\lambda} \sin(\theta/2) \\
e^{i\phi}\sin(\theta/2) & e^{i(\lambda+\phi)}\cos(\theta/2) 
                              \end{bmatrix}, \eeq
where $\cos(\theta/2)=\sqrt{(1+1/\sqrt{3})/2}$, $\phi=\pi/4$ and $\lambda=\pi$.\\
\subsection{Examining the optimality of SIC-sets in $d=2$}
\label{subsec:Monte-Carlo}
\begin{figure}[t!]
\includegraphics[width=2.4 in]{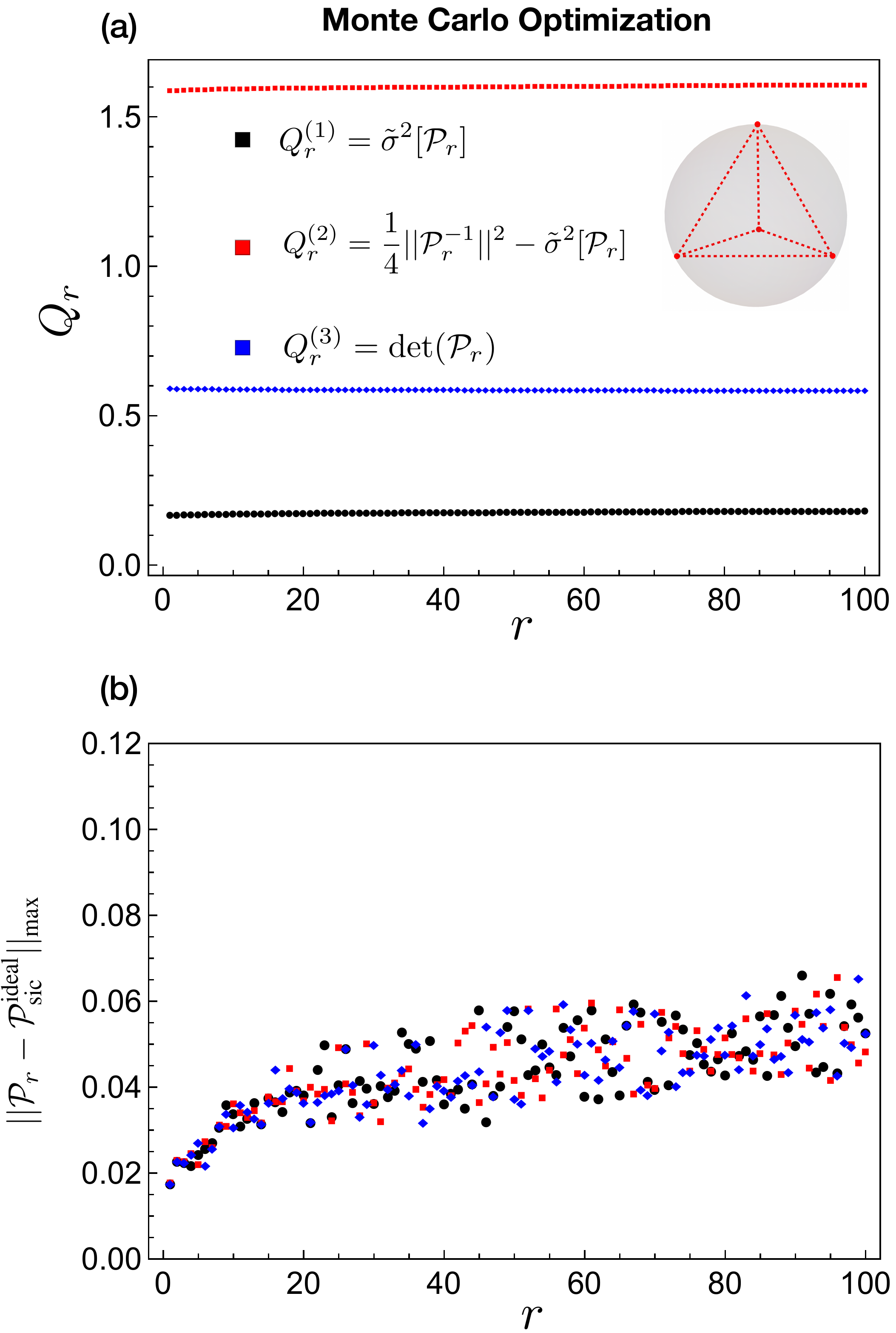}
\caption{Monte Carlo optimization method applied to the quantities $Q^{(1)} =\tilde{\sigma}^{2}[\mP]$, $Q^{(2)} = \Delta_{F}[\mP]:=1/(4N_{s})||\mP^{-1}||^2-\tilde{\sigma}^{2}[\mP]$ and $Q^{(3)}=\det(\mP)$. In order to search for the global optima of these quantities, we generated a set of $R=5\times 10^{7}$ probability matrices $\{\mP_{r}\}_{r=1}^{R}$ corresponding to states uniformly distributed on the Bloch sphere. We discarded the frames $\{\ket{\phi}_{i}\}_{i=1}^{4}$ producing ill-conditioned probability matrices by choosing only those having $\det(\mP)>10^{-5}.$ Panel (a) shows the $K=100$ smallest values of the quantities $\tilde{\sigma}^{2}[\mP_{r}]$, $\Delta_{F}[\mP_{r}]$ (with $N_{s}=1$) and the $K=100$ largest values of $\det(\mP_{r}).$ The minima found for $\tilde{\sigma}[\mP]$, $\Delta_{F}[\mP]$ and the maximum for $\det(\mP)$, are close to those corresponding to the matrix $\mP_{\text{sic}}^{\text{ideal}}$, namely $1/6$, $19/12$ and $16/27$, respectively. The inset in panel (a) shows the frame (after a rotation about the $z$-axis) corresponding to the minimum variance found in our simulation. Panel (b) shows the distance between the probability matrices $\{\mP_{r}\},$ yielding the smallest/largest values of the quantities discussed in (a), and $\mP_{\text{sic}}^{\text{ideal}}$. To quantify the distance between these probabilities, we used the matrix norm $||\cdot||_{\text{max}}$, defined as $||A||_{\text{max}}=\underset{i,k}{\max}|A_{ki}|$. 
 \label{fig:monte-carlo}}
\end{figure}

Finally, we address the question of the optimality of the single-qubit SIC-set. That is, we would like to verify that this set leads to the smallest possible variance $\tilde{\sigma}[\mP]$ in $d=2$. Unfortunately, we were not able to solve this problem analytically. However, this, and other related questions, can be easily explored through Monte Carlo simulations because for $d=2$, generating a random set $\{\ket{\phi_{i}}\}_{i=1}^{4}$ (a frame) reduces to picking random points on the two-dimensional sphere (the Bloch sphere). To generate a random point on the sphere, we first generate three random, and independent, standard normal variables $x=(x_{1}, x_{2}, x_{3})$ from which we obtain a state $\ket{\phi_{i}^{r}}$, pointing in the direction $x/||x||_{E}$. This procedure is known to produce a set of uniformly distributed points (or states) on the two-dimensional sphere \cite{muller1959note}. 
Using this method, we generated a set of  $R=5\times10^{7}$ probability matrices $\{\mP_{r}\}_{r=1}^{R}$ (with entries given by $|\expect{\phi^{r}_{k}|\phi^{r}_{i}}|^{2}$), which we then used to determine the $K=100$ smallest values of the variances $\{\tilde{\sigma}^2[\mP_{r}]\}_{r=1}^{R},$ as shown in Fig.~\ref{fig:monte-carlo}. In addition, we made use of  the set of probability matrices  $\{\mP_{r}\}_{r=1}^{R}$ to search for the maximum value of the determinant $\det(\mP_{r})$ and to study the inequality Eq.~(\ref{upper-bound1}), that is, the difference $\Delta_{F}:=1/(4N_{s})||\mP^{-1}_{r}||_{F}^{2}-\tilde{\sigma}^{2}[\mP_{r}].$ The results obtained from this simulations suggest that, indeed, in $d=2$ the SIC-set minimizes both the variance $\tilde{\sigma}[\mP]$ and $\Delta_{F},$ and maximizes the determinant $\det(\mP)$ (see Fig.~\ref{fig:monte-carlo}(b)). Moreover, these results indicate that in $d=2$ the inequality Eq.~(\ref{upper-bound1}) is not tight for probability matrices of the form $\mP_{k|i}=|\expect{\phi^{r}_{k}|\phi^{r}_{i}}|^{2}.$ Extending this analysis to $d>2$ is beyond the scope of this work. 
\subsection{Simulations of single-qubit determinant-based tests and SIC-sets}
\label{Q}
\begin{figure}[!t]
\includegraphics[width=2.4in]{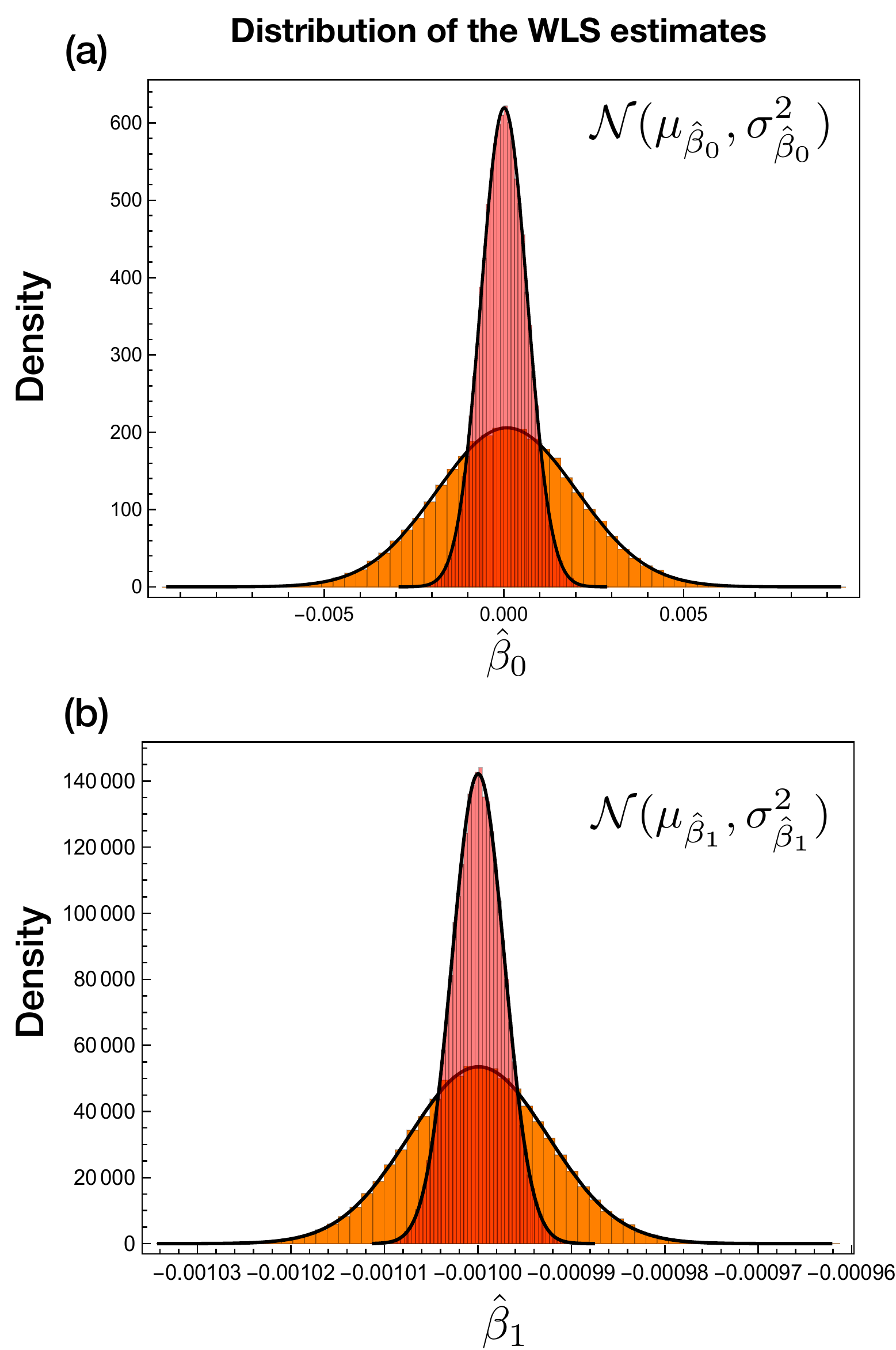}
\caption{Probability distribution of the WLS estimates $\hat{\beta}_{0}$ and $\hat{\beta}_{1}$ corresponding to the standard set $\{\ket{\phi_{i}}\}_{i=1}^{4}$ (orange histograms) and the SIC-set $\{\ket{\psi_{i}}\}_{i=1}^{4}$ (red histograms). The sequence lengths considered were $m=\{0, 10, 20,\ldots, 500\}$, the number of hypothetical experiments was $R=10^5$ and $N_{s}=50,000$. The SDs of the estimates $\log(|\det(\hat{\mP}_{m})|)$ were obtained from Eq.~(\ref{sigma}).
 \label{fig:tetrabeta}}
\end{figure}

{\indent}We now apply the ideas discussed in the previous subsection to a more realistic setting, which will include decoherence. To do so, we make use of our ZZ model and focus on our determinant-based tests  Eq.~(\ref{dperm}) (permutational) and Eq.~(\ref{dtest}) (iterative). But first, we will slightly modify our model.~Namely, we will assume perfect state preparation and measurements (i.e., no SPAM errors). The reason why we do not model SPAM errors here is because we will be comparing different tomographic sets (specifically, the set Eqs.~(\ref{ini-james4}-\ref{end-james4}) and Eqs.~(\ref{ini-sic-states}-\ref{end-sic-states})) and including SPAM errors might favor one set over the other, even in the case of the ideal idle gate $\mI=I$. We will model our gates as in Eq.~(\ref{gate-model}), that is, we will assume that the noisy implementation of an ideal gate $G$ is $\mG=\text{exp}(\mathcal{J}_{G}+t_{g}\mathcal{V}+t_{g}\mathcal{D}),$ where $t_{g}$ is the duration of the gate, $\mathcal{J}_{G}$ generates the ideal gate $G$ and $\mathcal{V}$ describes the Ising interaction $V=(J/2)Z\otimes Z$, respectively. The superoperator $\mathcal{D}$ accounts for the local decoherence of qubit $A$ and the memory (qubit $B$); the only difference between the superoperator $\mathcal{D}$ we consider here and that used in previous simulations, is that we will now set the thermal excitation rate $\gamma_{3}^{A}$ of the qubit $A$ equal to zero. We will thus assume that the initial state of the system is $\rho^{AB}_{0}=\ket{g}\bra{g}\otimes 1/2(I+n_{z}^{B}Z)$ where $n^{B}=(\gamma_{1}^{B}-\gamma_{3}^{B})/(\gamma_{1}^{B}+\gamma_{3}^{B})$, which ensures the stationarity of the initial state of the system $\rho^{AB}_{0}$.\\
{\indent} We first consider the context-independent case $J=0$ and, as in Sec.~\ref{sec:uni}, we study the distribution of the WLS estimates obtained by means of the single-qubit ID-test, applied to the noisy idle gate $\mI$. The results are displayed in Fig.~\ref{fig:tetrabeta}, for the model parameters $t_{g}=20~\text{ns}$, $n_{z}^{B}=0.84$, $\gamma^{A(B)}_{1}=\gamma_{1}=(60~\mu\text{s})^{-1}$ and $\gamma_{\phi}^{A(B)}=\gamma_{1}/2.$ Figure~\ref{fig:tetrabeta} compares the  distributions of the WLS estimates obtained using the standard set $\{\ket{\phi_{i}}\}_{i=1}^{4}$ and the SIC-set $\{\ket{\psi_{i}}\}_{i=1}^{4}.$ In this simulation the use of the SIC-set $\{\ket{\psi_{i}}\}_{i=1}^{4}$ reduced the standard deviations $\sigma_{\hat{\beta}_{1}}$ (which sets the precision of the unitarity estimate $\hat{u}'$) and $\sigma_{\hat{\beta_{0}}}$ by the factors 2.7 and 3, respectively.
 \begin{figure}[t!]
\includegraphics[width=3.32in]{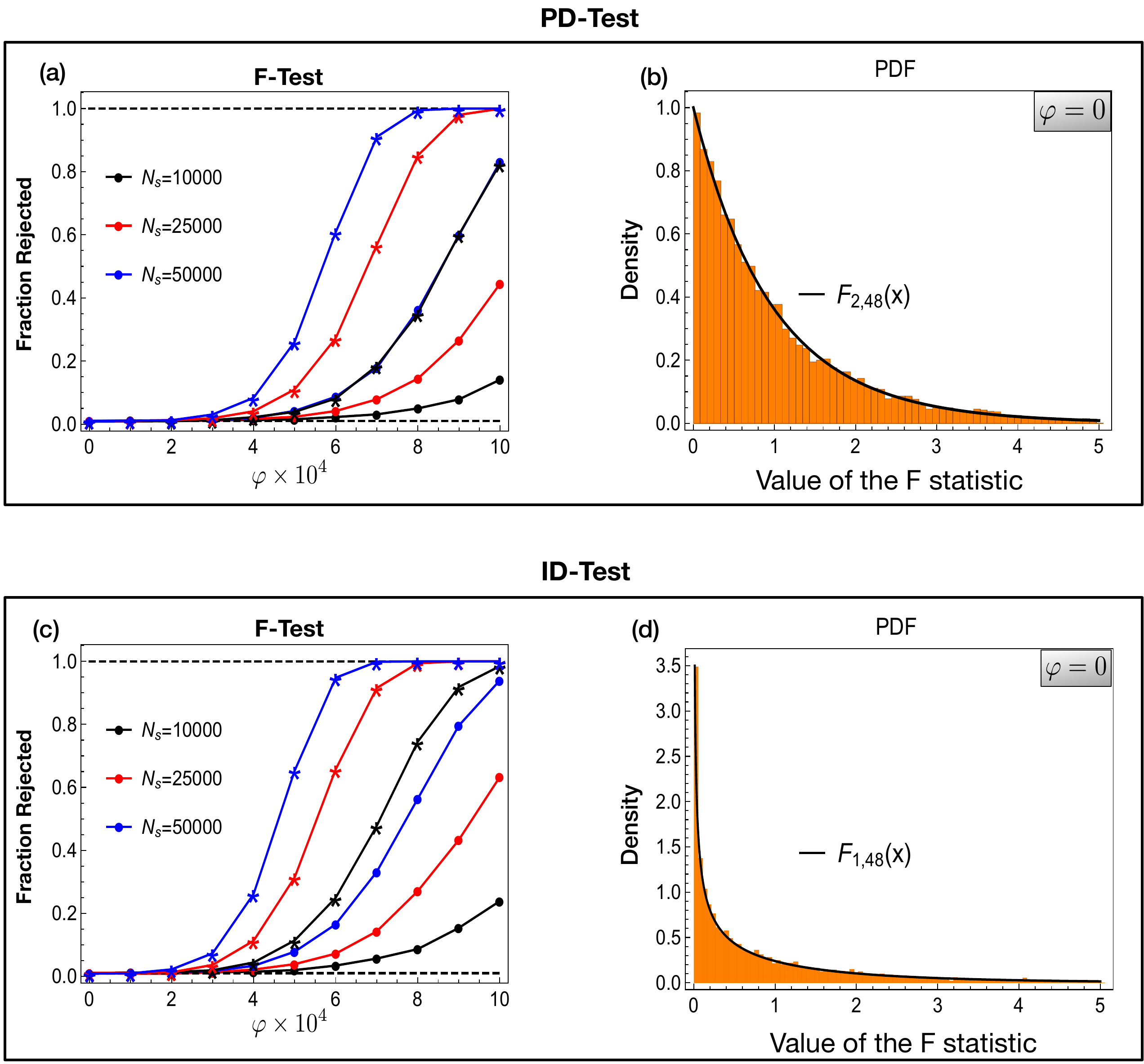}
\caption{Hypothesis testing for the determinant-based tests. The upper panels (a) and (b) display the results of the statistical F-test applied to the PD-test, discussed earlier in Fig.~\ref{fig:testsQ}(a). More precisely, we considered the following set of $M=51$ non-cyclic permutations of the sequence  $\mS_{1}=\mI^{n}\mX_{\pi}^{n}$, with $n=250$: $\mS_{1}$, $\mS_{6}$,  $\mS_{11}$, \ldots, $\mS_{251}$, where $\mS_{k}=\mI^{n-k+1}\mX_{\pi}^{n-k+1}(\mX_{\pi}\mI)^{k-1}.$ Panel (a), shows the power of the F-test for our SIC-set (asterisks) and the standard SPAM scheme (dots), for various values of the interaction parameter $\varphi$ (times $10^{4}$) and sample sizes $N_{s}$. Similarly, panel (c) displays the power of the F-test corresponding to the iterative determinant test, applied to the $M=51$ sequences $\mS_{0}=\mI^{0}, \mS_{10}=\mI^{10}, \mS_{20}=\mI^{20},\ldots, \mS_{500}=\mI^{500}.$ Panels (b) and (d) show the target $F$-distributions for both determinant-based tests when $\varphi=0$ and $N_{s}=50,000.$ To determine the power of the tests and to construct the histograms, we considered an ensemble of $R=10,000$ hypothetical experiments.
 \label{fig:TetrafishD}}
\end{figure} 
Since our tests are, in the limit $N_{s}\rightarrow \infty,$ insensitive to SPAM, the form of the true (underlying) model is the same for both sets of states, namely $y_{m}:=L_{m}=-2(\gamma_{1}+\gamma_{\phi})t_{g}m +\varepsilon_{m},$ $\varepsilon_{m} \sim \mathcal{N}(0, {\sigma}_{m}^2).$ That is to say, only the standard deviations ${\sigma}_{m}$ depend on the SPAM specifics, as dictated by Eq.~(\ref{sigma}).\\
{\indent} Let us now set $\varphi=Jt_{g}\neq 0$ and focus on detecting deviations from the null hypotheses associated with the PD-tests Eq.~(\ref{dperm}) and the ID-test Eq.~(\ref{dtest}). Although we have already addressed this problem earlier in this work, we will now concentrate on showing that the use of the SIC-set Eqs.~(\ref{ini-sic-states}-\ref{end-sic-states}), leads to an increase in the power of the statistical F-test (discussed in Sec.~\ref{Sec:F-Test}). Figures \ref{fig:TetrafishD}(a) and \ref{fig:TetrafishD}(b) show our results for the PD-test, applied to a class of permutations of the sequence $\mS_{1}=\mI^{250 }\mX^{250}$ (see caption of Fig.~\ref{fig:TetrafishD}). The nested models compared via the $F$ statistic were the null hypothesis $y_{k}=\beta_{0}=\text{const}$ (model $\mathcal{M}_{1}$) and the quadratic model $y_{k}=\beta_{0}+\beta_{1} k +\beta_{2}k^{2}$ (model $\mathcal{M}_{2}$). As explained in Sec.~\ref{Sec:F-Test}, if the $F$ statistic obtained by comparing the goodness of fit of these two models (see Eq.~(\ref{F}))  is greater than certain value $F_{\text{cr}}$,  specified by our artificial significance level $p_{\text{cr}}=0.01$ (i.e., $1\%$), we reject the null hypothesis associated with model $\mathcal{M}_{1}$ and conclude that the gates involved in the test are \emph{highly likely} to be context-dependent. As expected, we found that the use of a SIC-set leads to a noticeable increase in the power of the F-test, as observed in Fig.~\ref{fig:TetrafishD}(a).\\
{\indent} In much the same way, we analyzed the ID-test, applied to the gate $\mI$. The results are displayed in the panels \ref{fig:TetrafishD}(c) and \ref{fig:TetrafishD}(d). The nested models we compared, by means of the $F$ statistic, in this test are the null hypothesis $y_{m}=\beta_{0}+\beta_{1}m$ (model $\mathcal{M}_{1}$) and the quadratic model $y_{m}=\beta_{0}+\beta_{1}m+\beta_{2}m^{2}$ (model $\mathcal{M}_{2}$). For this iterative test, we also observed that the use the SIC-set  $\{\ket{\psi_{i}}\}_{i=1}^{4}$ boosts the power of the F-test (see Fig.~\ref{fig:TetrafishD}(c)). Finally, the panels $\ref{fig:TetrafishD}(b)$ and $\ref{fig:TetrafishD}(d)$ show that we, indeed, reproduce the correct $F$-distributions (using the SIC-set), when the gates are context-independent, i.e., $\varphi=0$. Then the corresponding PDFs are of the form $F_{q_{2}-q_{1}, M-q_{2}}(x)$ (see Eq.~(\ref{FPDF})), where $q_{1(2)}=\text{dim}(\mathcal{M}_{1(2)})$ and $M$ is the number of observations (points) used to fit the models.
\subsection{Estimating the unitarity of a two-qubit gate}
\label{QQ}
\begin{figure}[t!]
\includegraphics[width=2.2in]{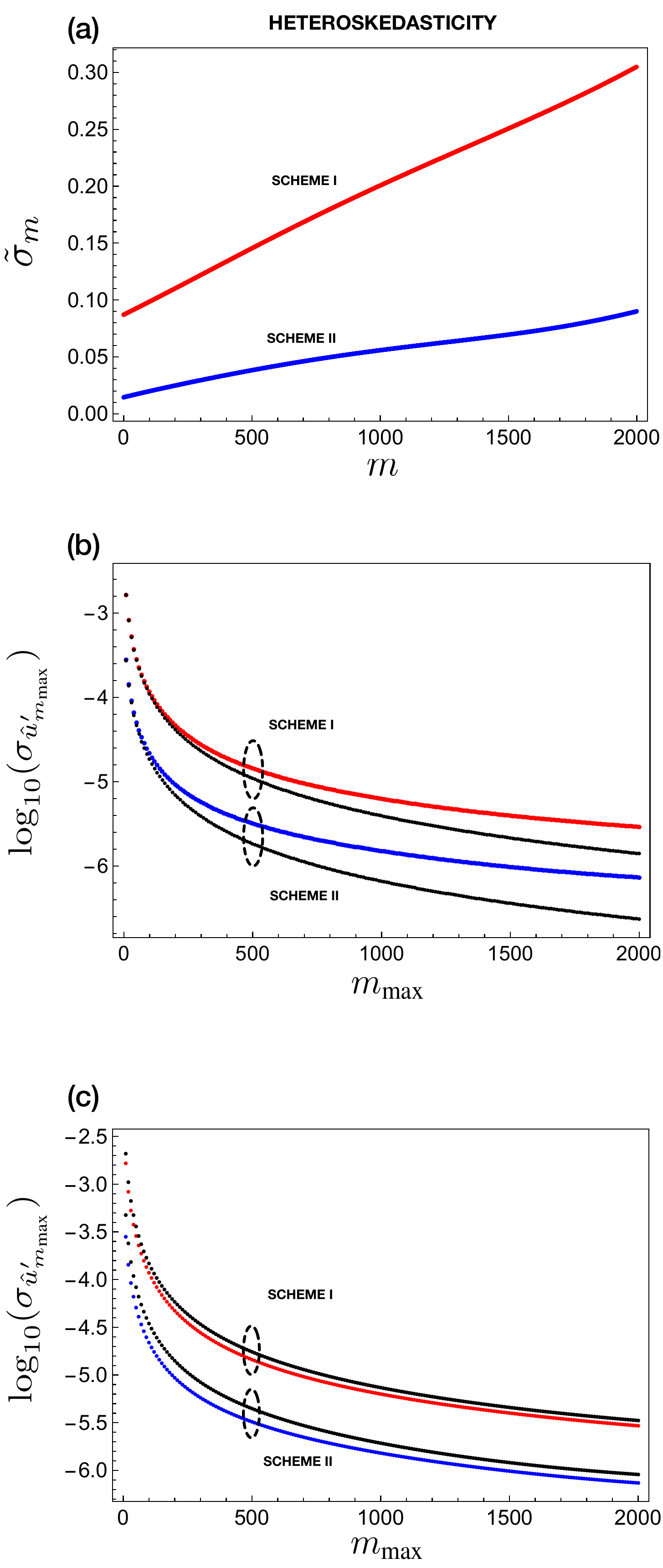}
\caption{Heteroskedasticity of $\log(|\det(\hat{P}_{m})|)$ and precision of the unitarity estimates for the ID-test, applied to the two-qubit gate 
$\mI^{\otimes 2}$. Panel (a) shows the behavior of the true standard deviations  $\tilde{\sigma}_{m}=\tilde{\sigma}[\mP_{m}]$ for SPAM schemes I and II. Panel (b) displays the standard deviation $\sigma_{\hat{u}'_{\text{max}}}$, corresponding to the set of evenly spaced sequence lengths $\{0,10,20,\ldots m_{\text{max}}\}$, versus the maximum sequence length $m_{\text{max}}$ for schemes I (red) and II (blue). The black curves in (b), show the SDs of the unitarity estimates in the hypothetical situation in which $\tilde{\sigma}_{m}$ is constant and equal to $\tilde{\sigma}_{0}$. In panel (c) we introduced SPAM errors by treating the first $n=200$ iterations of $\mI^{2\otimes}$ as part of state preparation. In other words, we considered noisy input states of the form $|\rho'_{i})=(\mI^{2\otimes})|\rho_{i})$, where $\rho_{i}$ are the ideal two-qubit input states discussed earlier in this subsection. The inset shows the structure of the SPAM matrix $\mI^{200}=(\mI^{\otimes 2})^{200}$, written in the two-qubit operator basis $\{I,X,Y,Z\}^{\otimes 2}.$
  \label{fig:hetero}}
\end{figure}

{\indent}Finally, in this last subsection, we simulate the ID-test, applied to a context-independent two-qubit gate with the purpose of estimating its unitarity. More precisely, we will consider iterations of the noisy two-qubit idle gate $\mI^{\otimes 2},$ given by $\mI^{\otimes 2}:=e^{t_{g}\mathcal{V}+t_{g}\mathcal{D}}.$ Note that, by construction, the two-qubit gate $\mI^{\otimes 2}$ is context-independent on $AB$, even when $\varphi \neq 0.$ If we denote by $\mS_{m}$ the sequence corresponding to $m$ iterations of $\mI^{\otimes 2}$, then $L_{m}=\log(\det(\mS_{m}))=mt_{g}\tr(\mathcal{D})=4mt_{g}[\tr(\mathcal{D}_{A})+\tr(\mathcal{D}_{B})],$ where we have assumed that the qubits decohere locally, i.e., $\mathcal{D}=\mathcal{D}_{A}\otimes I_{B}+I_{A}\otimes \mathcal{D}_{B}.$ Thus, assuming that the qubits, and their respective environments, are identical, we find that the slope of $L_{m}$ is given by $\beta_{1}=-16(\sum_{k}\gamma_{k})t_{g}.$ Therefore the true value of the unitarity of the two-qubit gate $ \mI^{\otimes 2}$ is given by 
 \beq
 u'(\mI^{\otimes 2})=\text{exp}\left(\frac{2\beta_{1}}{d^2-1}\right)=\text{exp}\left(-\frac{32 t_{g}}{15}\sum_{k}\gamma_{k}\right).
 \label{uniQQ}
\eeq
{\indent}It is tempting to use a SIC-set in $d=4$ to increase the precision of the unitarity estimate $\hat{u}'$ (a construction of a SIC-set in $d=4$ can be found in Ref.~\cite{renes2004symmetric}). However, such approach is somewhat impractical since it would require implementing highly nontrivial operations to prepare this two-qubit SIC-set. On the other hand high-fidelity single-qubit non-Clifford operations have been already successfully implemented (see e.g.,~\cite{barends2014rolling}). Therefore, as discussed earlier in this section, we will consider employing local tomographic sets of the form $SIC_{2}\otimes SIC_{2}$ (scheme-II) and compare the results thus obtained with those corresponding to the use of  tensor products of the single-qubit standard set  (scheme-I). As in the previous subsection, in order to meaningfully compare these tomographic schemes (as in table \ref{tab:sic}) we will assume that there are no SPAM errors. From table
 \ref{tab:sic} we know that for $\mI^{\otimes 2}=I$ (the identity matrix) a comparison between the SDs of the unitarity estimates found via the ID-test, would yield the ratio ${\sigma^{(I)}_{\hat{u}'}}/{\sigma^{(II)}_{\hat{u}'}}=12\sqrt{19/77}\approx 6$ (assuming the same $N_{s}$ for both schemes). Clearly, due to decoherence, we expect our simulations to yield a smaller ratio ${\sigma^{(I)}_{\hat{u}'}}/{\sigma^{(II)}_{\hat{u}'}}$.\\
\begin{figure}[t!]
\includegraphics[width=3.3in]{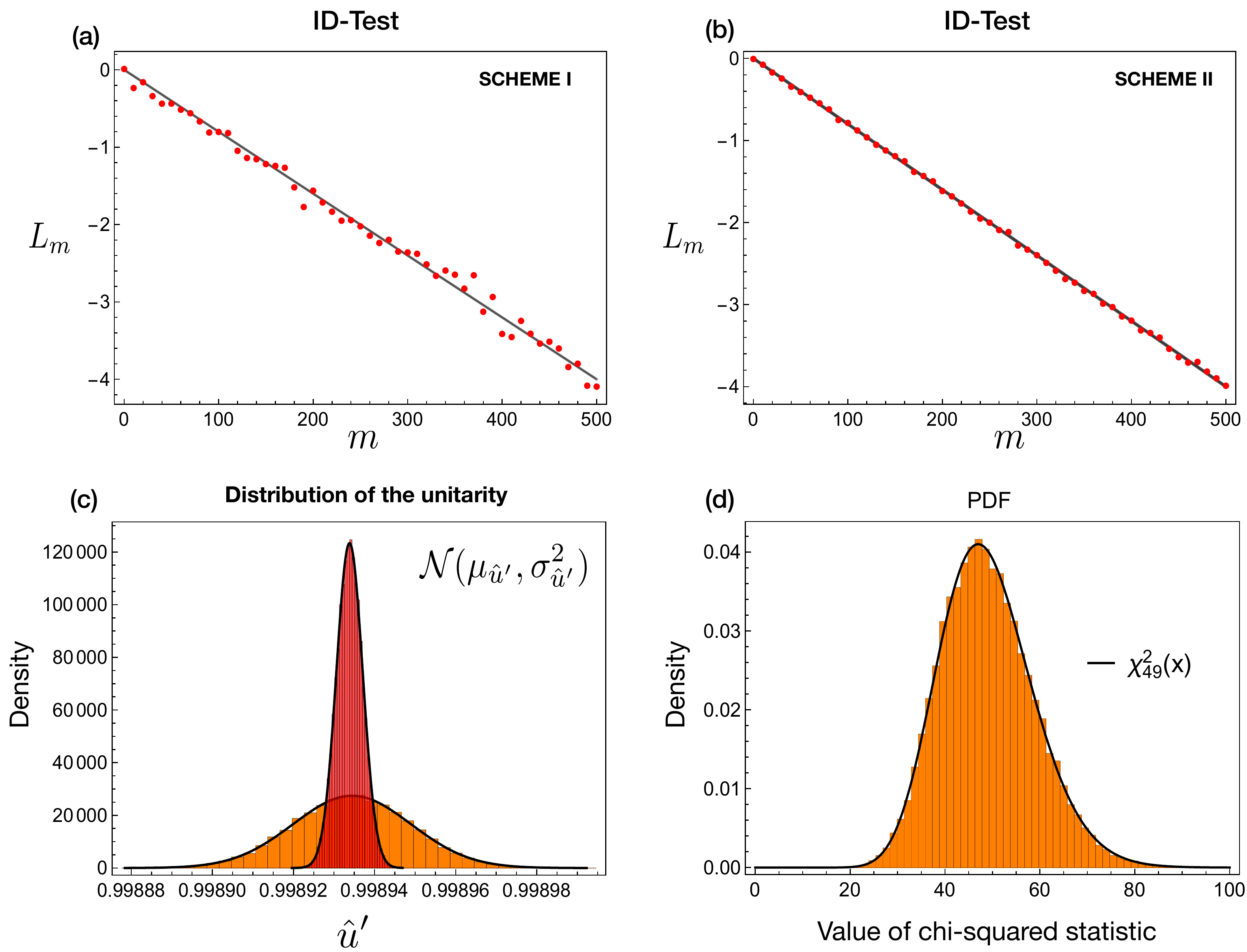}
\caption{ID-tests applied to the two-qubit gate $\mI^{\otimes 2}.$ Panels (a) and (b) show the results of a simulation (dots) of 
a hypothetical experiment employing (a) the SPAM scheme based on tensor products of standard states $\{\ket{\phi}_{i}\}$ (scheme-I) and (b) the ${SIC}_{2}\otimes {SIC}_{2}$ set (scheme-II). The ``solid'' lines in (a) and (b) represent the true values $L_{m}=\beta_{1}m$ (where $\beta_{1}=-16t_{g} (\gamma_{1}+\gamma_{\phi})$). 
Panel (c) displays the distribution of the unitarity estimates $\hat{u}'$ for the scheme-I (orange) and scheme-II (red). Panel (d) shows the distribution of the target $\mathcal{X}^{2}$ statistic obtained by considering $R=40,000$ hypothetical experiments, based on the SPAM scheme-II. The sequence lengths used in these simulations were $\{m_{n}\}_{n=1}^{51}=\{0, 10, 20,\ldots, 500\}$. The number of runs per experimental setting used in the simulation was $N_{s}=10,000.$
  \label{fig:DQQ1}}
\end{figure} 
{\indent}Figure~\ref{fig:DQQ1} shows the results of our simulations of the ID-test applied to the two-qubit gate $\mI^{\otimes 2},$ with parameters $t_{g}=20~\text{ns}$, $\varphi=Jt_{g}=1\times 10^{-3}$, $\gamma_{1}=(60~\mu{s})^{-1}$, $\gamma_{\phi}=\gamma_{1}/2$ and $\gamma_{3}=0$. Panels (a) and (b) display the decay of $L_{m}$ for the standard set (scheme-I) and the $SIC_{2}\otimes SIC_{2}$ set (scheme-II), respectively. The statistical fluctuations of the probability matrices estimates $\hat{\mP}^{(I(II))}_{m},$ corresponding to scheme-I(II), were simulated by sampling from the binomial distribution $\text{Bin}[N_{s},(\mathcal{P}^{(I(II))}_{m})_{k|i}],$ with $N_{s}=10,000.$ The resulting estimates $\{\hat{\mP}^{(I(II))}_{m_{n}}\}_{n=1}^{M}$ were used to compute the log-dets $\hat{L}^{(I)}_{m}=16\log(2)+\log(|\det(\hat{\mP}^{(I)}_{m})|)$  and $\hat{L}^{(II)}_{m}=8(3\log(3)-4\log(2))+\log(|\det(\hat{\mP}^{(II)}_{m})|),$ shown in panels (a) and (b), respectively. Comparing Figs.~\ref{fig:DQQ1}(a) and \ref{fig:DQQ1}(b) we observe the expected reduction in the magnitude of the statistical fluctuations, achieved  thanks to the use of the $SIC_{2}\otimes SIC_{2}$ set. Naturally, this reduction translates into more precise unitarity estimates, as illustrated in Fig.~\ref{fig:DQQ1}(c). Using $R=40,000$ hypothetical experiments to determine the distributions of the unitarity estimates $\hat{u}^{(I)}$ and $\hat{u}^{(II)}$ shown in Fig.~\ref{fig:DQQ1}(c), we found 
\begin{eqnarray}
\label{u1}
\hat{u}'^{(I)}&\sim& \mathcal{N}(0.99893, (1.5\times 10^{-5})^2),\\ 
\hat{u}'^{(II)}&\sim& \mathcal{N}(0.998934, (3.2 \times 10^{-6})^2).\label{u2}
\end{eqnarray}
Thus, by using scheme-II we improved the precision of the unitarity estimates by, approximately, a factor of $\sigma^{(I)}_{\hat{u}'}/\sigma^{(II)}_{\hat{u}'}\approx 4.5.$ As discussed in Sec.~\ref{sec:uni}, we expect our unitarity estimates to be close to the unitarity $u(G),$ introduced in Ref.~\cite{wallman2015}. Indeed, for our two-qubit gate we have 
 \beq 
u(\mI^{\otimes 2})=\frac{1}{d^2-1}\tr[W^{T}_{\mI^{\otimes {2}}}W^{\phantom{}}_{\mI^{\otimes 2}}]=0.998934.
\eeq 
The difference between the unitarity $u (\mI^{\otimes 2})$ and the true value of our unitarity measure $u'(\mI^{\otimes 2})=\text{exp}(-32/15 t_{g}(\gamma_{1}+\gamma_{\phi}))$ (see Eq.~(\ref{uniQQ})) is $9.8\times 10^{-7}.$\\
{\indent}Figure \ref{fig:hetero}(a) shows the variability (i.e., heteroskedasticity) of the SDs of the estimates $\log(|\det(\hat{\mP}_{m})|)$ for SPAM schemes I and II. These SDs were computed from Eq.~(\ref{sigma}) using the true probability matrices $\mP_{m}$ for each sequence length $m.$ We see in Fig.~\ref{fig:hetero}(a) that the use SPAM scheme II  leads to smaller standard deviations even for longer sequences than those considered in Fig.~\ref{fig:DQQ1} (wherein $m_{\text{max}}=500$). As discussed in Sec.~\ref{sec:uni}, it suffices to know only the SDs corresponding to minimum and maximum sequence lengths to bound $\sigma_{\hat{\beta}_{1}}$, provided $\sigma_{m}$ increases monotonically with $m$ (as it is in our case). Thus, setting $m_{\text{min}}=0,$ $m_{\text{max}}=500$, $M=51$ (as in the simulation Fig.~\ref{fig:DQQ1}), $\hat{u}'=1$ (because $\hat{u}'\approx 1$) we find, using Eqs.~(\ref{var-uni}), (\ref{boundsigma1}) and Fig.~\ref{fig:hetero}(a), the following bounds:
\begin{eqnarray}
\sigma^{(I)}_{\hat{u'}}&\in &[6.8\times 10^{-6},\hspace{0.15 cm}3.1 \times 10^{-5}],\\
\sigma^{(II)}_{\hat{u'}}&\in&[7.0\times 10^{-7}, \hspace{0.15 cm}1.3\times 10^{-5}], 
\end{eqnarray} 
which are compatible with the true uncertainties in Eqs.~(\ref{u1}) and (\ref{u2}). Fig.~\ref{fig:hetero}(b) displays the uncertainty in the unitarity estimate $u'(\mI^{\otimes 2})$ after applying the ID-test to sequences of lengths $\{0,10,20, \ldots m_{\text{max}}\}.$ The SDs in this figure were computed directly from the covariance matrix Eq.~(\ref{Cov}) (see Sec.~\ref{sec:uni}), with weights $w_{m}={1}/{\tilde{\sigma}_{m}^2},$ where the standard deviations $\tilde{\sigma}_{m}$ are  as in Fig.~\ref{fig:hetero}(a). 
Furthermore, in order to find out how much the heteroskedasticity of the observations limits the precision of our unitarity estimates, we plotted (on log scale) the following expression, which assumes homoskedasticity i.e., $\tilde{\sigma}_{m}=\tilde{\sigma}_{0}=c_{0}/\sqrt{N_{s}}$ for all $m$:  
\beq 
\label{uhom}
\sigma^{(\text{homosk})}_{\hat{u}'_{m_{\text{max}}}}=\frac{4\sqrt{3}}{15}\frac{\sqrt{b}}{\sqrt{m_{\text{max}}(m_{\text{max}}+b)(m_{\text{max}}+2b)}}\frac{c_{0}}{\sqrt{N_{s}}},
\eeq
where $c_{0}$ equals $2\sqrt{19}$ (for scheme I)  or $\sqrt{77}/{6}$ (for scheme II). The above equation was obtained via Eq.~(\ref{sigma1}) and it gives the SD of the unitarity estimate resulting from testing all the sequence lengths $\{m_{n}\}_{n=1}^{M}=\{(n-1)b\}_{n=1}^{M}$ ($m_{\text{max}}$ is related to the number of observations $M$ via $m_{\text{max}}=(M-1)b$). The black lines in Fig.~\ref{fig:hetero}(b) represent the relations Eq.~(\ref{uhom}) for schemes I and II. As expected, the fact that $\tilde{\sigma}_{m}$ increases with $m$ leads to less precise unitarity estimates. Fortunately, the effect of heteroskedasticity in our model is rather moderate, i.e., it does not lower the precision of the estimates by orders of magnitude, provided the sequences considered are not excessively long. For example, for $m_{\text{max}}=500$ (as in Fig.~\ref{fig:DQQ1}), we find that the ratio $\sigma_{\hat{u}'_{\text{max}}}/\sigma^{\text{(homosk)}}_{\hat{u}'_{\text{max}}}$ equals 1.3 for scheme I, and  1.7 for scheme II. It is also worth observing that the difference between $\log_{10}(\sigma^{(I)}_{\hat{u}'_{\text{max}}})$ (red curve) and  $\log_{10}(\sigma^{(II)}_{\hat{u}'_{\text{max}}})$ (blue curve) does not vary appreciably with $m_{\text{max}}.$ This, of course, implies that the ratio $\sigma^{(I)}_{\hat{u}'_{\text{max}}}/\sigma^{(II)}_{\hat{u}'_{\text{max}}}$ is, roughly, independent of $m_{\text{max}}$. \\
{\indent} Finally, we can also exploit the data obtained from our simulations to explore a situation involving SPAM errors. This can be done by noticing that we can treat the first $n$ gates in the sequence $\mI^{\otimes 2}$ as errors in state preparation; conversely we could also treat the last $n$ gates as measurement errors. Note that we can express the probability matrix corresponding to a sequence $\mS_{m}=\mG^{m}$ as 
\begin{eqnarray}
\mP_{k|i}&=&(M_{k}|\mG^{m}|\rho_{i})=(M_{k}|\mG^{m-n}\mG^{n}|\rho_{i})\nonumber \\
&=&(M_{k}|\mG^{m'}|\rho'_{i})=(M'_{k}|\mG^{m'}|\rho_{i}), 
\end{eqnarray}
where $m'=m-n$, $|\rho'_{i})=\mG^{n}|\rho_{i})$, $(M'_{k}|=(M_{k}|\mG^{n}$ and $(A|B):=\tr[A^{\dagger}B]$. Therefore, shifting our list of probability matrices, i.e., treating the {$nth$} iteration of the gate $\mI^{\otimes 2}$ as our reference sequence, is equivalent to introducing some amount of SPAM errors, which can quantified through the process fidelity 
\begin{eqnarray}
\mathcal{F}_{n}&=&\frac{1}{16}\Tr[(\mI^{\otimes 2})^n]=\frac{1}{16} (1+2e^{-\frac{nt_{g}}{T_{2}}}+e^{-\frac{nt_{g}}{T_{1}}})^{2}\nonumber \\
&-&\frac{1}{2}e^{-\frac{nt_{g}}{T_{2}}}(1+e^{-\frac{nt_{g}}{T_{1}}})\sin^{2}(\frac{n\varphi}{2}) \label{fid}
\end{eqnarray}
where $T_{1}^{-1}=\gamma_{1}$ and $T_{2}^{-1}=\gamma_{1}/2+\gamma_{\phi}.$ Figure \ref{fig:hetero}(c) shows the impact of these SPAM errors on the precision of the unitarity estimates for schemes I and II. We assumed the value $n=200$, for which the fidelity of the SPAM matrix is $\mathcal{F}_{200}\approx 0.9.$ A comparison of  panels \ref{fig:hetero}(b) and \ref{fig:hetero}(c) shows that the inclusion of this particular form of SPAM errors does not affect dramatically the precision of the unitarity estimates; it is also evident from these figures that the use of scheme-II, based on local SIC-sets, still leads to more precise estimates. 
\section{Conclusions}
In this work, we examined the effect of statistical fluctuations on the power and precision of the tests for context-dependence proposed in Ref.~\cite{veitia2017macroscopic}. We began this paper by highlighting the clear connection between the permutational test, the iterative determinant (ID) test, and the experimental data. More precisely, we showed how these tests can be formulated, in a natural way, in terms of $d^{2}\times d^{2}$ probability matrices corresponding to $d^{4}$ measurement configurations, whose details need not be known precisely. In addition, we showed that in the limit $N_{s}\rightarrow \infty$ these tests for context-dependence are exact, the only assumption being that the initial state of the system $\rho_{0}$ and the POVM effect $M_{0}$ (i.e., our detector) do not depend on the context. Also, making use of the fact that our tests are based on spectral properties of quantum maps, we constructed a CP-indivisibility witness which does not depend on the SPAM specifics. Finally, we discussed the multipurpose ID-test, which can be used to both detect non-Markovian errors and to characterize the unitarity of a specific gate.\\
{\indent}The rest of the paper takes into account the fluctuations in the probability (frequency) matrices, due to finite measurement repetitions $N_{s}.$ First, we
formulated our tests for context-independence in the framework of hypothesis testing. We then discussed the weighted least squares (WLS) method and two related statistics (the $\mathcal{X}^{2}$ and the $F$ statistic), which we used to test the null hypothesis (i.e., the context-independence hypothesis) in the ZZ model introduced in Ref.~\cite{veitia2017macroscopic}. We presented the power of the tests, i.e., the probability of correctly rejecting the null hypothesis, for various ``small'' interaction parameters $\varphi$ and sample sizes $N_{s}.$ Here we found that using the $F$ statistic (which compares two nested models) for hypothesis testing leads to higher powers than those obtained via the $\mathcal{X}^{2}$ statistic. Clearly, the analysis of the statistical significance of  context-dependence effects can also be carried out employing other powerful techniques such as cross validation (CV) \cite{arlot2010survey}, model selection criteria (e.g., AIC)  \cite{claeskens2008model} or Bayesian inference methods (see e.g. \cite{hincks2018bayesian}).\\
{\indent}In order to obtain results  that do not depend on our model of context-dependence (i.e., the ZZ model), we focused on studying the distribution of the log-det estimates. Specifically, we derived a useful formula for the standard deviation of $\log(|\det(\hat{\mP})|)$ in terms of the true probability matrix ${\mP}$ and $N_{s}.$ We used this result to study the heteroskedasticity of the log-det and to compare the performance of various tomographic sets in $d=2,3$ and $4$. This analysis allowed us to improve the power of our tests for context-dependence and the precision of the unitarity estimates using SIC-sets. Finally, we simulated the ID-test for a two-qubit context-independent gate with the purpose of examining the precision of the unitarity estimates vs. the maximum sequence length considered in the test. The results of this simulation, together with other bounds presented in this work, suggest that in the absence of context-dependence, the ID-test has the potential to yield very precise unitarity estimates for single- and two-qubit gates. This can be easily understood by noticing that a ``quick'' estimate of the unitary of a gate $G$ can be obtained by simply considering two probability matrices $\hat{\mP}_{0}$ and $\hat{\mP}_{m_{\text{max}}},$ corresponding to the gate sequences $S_{0}=G^{0}$ and $S_{m_{\text{max}}}=G^{m_{\text{max}}}.$ Then the standard deviation of the unitarity estimate, obtained via WLS, will be of the order of $\alpha/(m_{\text{max}}\sqrt{N_{s}})$, where $\alpha$ depends on SPAM and heteroskedasticity, and it will typically be of  the order of 1 (or less, provided that SPAM errors and decoherence are moderate). Hence, long gate sequences will considerably reduce the uncertainty of the unitarity estimates. Note that this boost in precision for a given $N_{s}$ is analogous to the \emph{hyper-accuracy} observed in GST~\cite{blume2017demonstration}.
\section*{acknowledgments}
This work was partially supported by ARO under Contract No. W911NF-14-C-0048. We thank discussions with Robin Blume-Kohout and Marcus P. da Silva.
\appendix
\section{The chi-squared statistic}
\label{Sec:chi-appendix}
In this appendix we include some useful facts dealing with the relationship between the weighted least squares (WLS) method and the chi-squared distribution.

We start by noting that the WLS estimate $\hat{\beta},$ defined via the minimization problem (\ref{WLS-MIN}), can be written in closed-form as 
 \beq 
  \label{betaWLS}
  \hat{\beta}=(X_{q}^{T}WX_{q})^{-1} X_{q}^{T}W y. 
  \eeq
Here $y=[y_{1},\ldots, y_{M}]^{T}$ is the observation vector, $X_{q}$ is the $M\times q$ design matrix (which depends on the model we are fitting) and $W:=\text{diag}(w_{1},\ldots, w_{M}),$ where $w_{i}=1/\sigma_{i}^{2}$, is the weight matrix. For observations $y$  generated by a linear model (linear in $\beta$)
\beq
\label{true} 
y=X_{q}\beta+\varepsilon, \quad \varepsilon_{i} \sim \mathcal{N}(0,\sigma_{i}^{2}).
\eeq
 the WLS estimate Eq.~(\ref{betaWLS}) coincides with maximum likelihood estimate (MLE). In addition, the $\mathcal{X}^{2}$ statistic should follow the chi-squared distribution with $M-q$ degrees of freedom, i.e., $\mathcal{X}^{2}\sim \chi^2_{M-q}.$ For completeness of exposition, we provide here a proof of this well-known fact. First, using Eqs.~(\ref{betaWLS}) and (\ref{true}), we express the $\mathcal{X}^{2}$ statistic as follows: 
\begin{eqnarray} 
\label{proof1}
\mathcal{X}^{2}&=&||\sqrt{W}(y-X_{q}\hat{\beta})||^{2}_{E}\\
        &=&||(I_{M}-\sqrt{W}X_{q}(X_{q}^{T}WX_{q})^{-1}X_{q}^T\sqrt{W})\sqrt{W}\varepsilon||_{E}^{2}, \nonumber
        \end{eqnarray}
where $||\cdot||_{E}$ is the standard Euclidean norm. We now note that the operator, $P_{w}:=I_{M}-\sqrt{W}X_{q}(X_{q}^{T}WX_{q})^{-1}X_{q}^T\sqrt{W}$ enjoys the following properties: \\
\noindent{(a)} $P_{w}^{T}=P_{w}$\\
\noindent{(b)} $P_{w}^{2}=P_{w}$ \\
\noindent{(c)} $ \tr(P_{w})=M-q. $ \\
Hence,  without loss of generality, the matrix $P_{w}$ is of the form $P_{w}= O \text{diag}(\underbrace{1,\ldots 1}_{M-q},\underbrace{ 0,\dots,0}_{q}) O^{T}$, with $OO^{T}=I_{M}$ and  
\begin{eqnarray}
\mathcal{X}^2&=&||P_{w}\sqrt{W}\varepsilon||_{E}^{2}\nonumber \\
                        &=&||O \text{diag}(\underbrace{1,\ldots 1}_{M-q},\underbrace{ 0,\dots,0)}_{q} O^{T}\sqrt{W}\varepsilon)||^{2}_{E}.
\end{eqnarray}
Finally, notice that $z:=\sqrt{W}\varepsilon \sim \mathcal{N}(0, I_{M})$, and therefore $z'=O^{T}z$ is also standard normally distributed. Since $||\cdot ||_{E}$ is invariant under 
orthogonal transformation, we have
$\mathcal{X}^{2}=\sum_{k=1}^{M-q} z'^2_{k}$. Thus, by definition \cite{wasserman2013}, $\mathcal{X}^{2}$ will be distributed as
\beq
\mathcal{X}^{2}\sim \chi^{2}_{M-q}.
\eeq
The probability density function (PDF) $f_{n}(x)$ corresponding to the chi-squared distribution with $n$ degrees of freedom is 
\beq 
f_{n}(x)=\frac{1}{2^{n/2}\Gamma(n/2)}x^{\frac{n}{2}-1}e^{-\frac{x}{2}}, x>0,
\eeq
where $\Gamma(z)$ is the gamma function.
\section{The F-statistic}
\label{Sec:F-appendix}
In this appendix we briefly discuss the $F$-statistic, which for two nested models $\mathcal{M}_{1},$ $\mathcal{M}_{2}, \mathcal{M}_{1}\subset \mathcal{M}_{2}$ is given by 
\beq 
\label{FAppendix}
F=\frac{M-q_{2}}{q_{2}-q_{1}}\left(\frac{\mathcal{X}_{1}^{2}}{\mathcal{X}_{2}^{2}}-1\right). 
\eeq
Here $q_{1(2)}=\dim(\mathcal{M}_{1(2)})$ and $\mathcal{X}^2_{1(2)}$ are obtained by fitting models $\mathcal{M}_{{1}(2)}$ to an observation vector $y=[y_{1},\ldots, y_{M}]^{T}.$

First, we present a sketch of the proof of the fact Eq.~(\ref{diffchi}). As mentioned above, we assume that the observations, i.e., the data is generated by the model $y=X_{q_{1}}\beta+\varepsilon.$ It will prove convenient to set $y=\sqrt{W^{-1}}y'$, $X_{q_{1(2)}}=\sqrt{W^{-1}}X'_{q_{1(2)}}$ and $\varepsilon=\sqrt{W^{-1}}\varepsilon'$, where the weights matrix $W$ is as in Eq.~(\ref{proof1}). We can now rewrite the difference $\Delta \mathcal{X}^{2}_{12}$ in terms of $y'$ as 
\beq
\label{proof2}
\Delta \mathcal{X}^{2}_{12}=y'^{T}(1-H_{1})y'-y'^{T}(1-H_{2})y'=y'^{T}(H_{2}-H_{1})y', 
\eeq
where $H_{1(2)}=X'_{q_{1(2)}} (X'^{T}_{q_{1(2)}} X'_{q_{1(2)}})^{-1} X'^{T}_{q_{1(2)}}$ are symmetric and idempotent $M\times M$ matrices (these are the so-called hat matrices), which satisfy $H_{1(2)}X'_{q_{1(2)}}=X'_{q_{1(2)}}.$ Notice now that since $\mathcal{M}_{1}\subset \mathcal{M}_{2},$ the $M\times q_{2}$ matrix $X'_{q_{2}}$ contains the columns of $X'_{q_{1}}.$ Hence, $H_{2}X'_{q_{1}}=X'_{q_{1}}$ and $H_{2}H_{1}=H_{1}H_{2}=H_{1}.$ These facts can be used to express Eq.~(\ref{proof2}) as 
\beq 
\Delta \mathcal{X}^{2}_{12}=\varepsilon'^{T}(H_{2}-H_{1})\varepsilon',
\eeq
and also allow us to show that $(H_{2}-H_{1})^2=H_{2}-H_{1}$, that is to say, $H_{2}-H_{1}$ is a projector. Equation~(\ref{diffchi}) follows from the fact that $\tr(H_{2}-H_{1})=q_{2}-q_{1}$ and $\varepsilon'=\mathcal{N}(0, I_{M})$ (see the derivation of Eq.~(\ref{proof1})). Finally, the statistical independence of $\mathcal{X}^{2}_{2}$ and $\Delta \mathcal{X}^{2}_{12}$ can be shown by writing $\mathcal{X}^{2}_{2}=\varepsilon'^{T}(I_{M}-H_{2})\varepsilon'$ and noticing that $(H_{2}-H_{1})(I_{M}-H_{2})=0$ \footnote{Consider two random quadratic forms $Q_{1}=z^{T}A_{1}z$ and $Q_{2}=z^{T}A_{2} z,$ such that $z\sim \mathcal{N}(0, I_{M})$ and $A_{1(2)}$ are symmetric and idempotent. Then $Q_{1}$ and $Q_{2}$ are independent iff $A_{1}A_{2}=0$.}.\\
{\indent}Hence, for two nested models $\mathcal{M}_{1}, \mathcal{M}_{2}$, $\mathcal{M}_{1}\subset \mathcal{M}_{2}$, one may test the correctness of $\mathcal{M}_{1}$ employing the $F$ statistic Eq.~(\ref{FAppendix}). The context-independence hypothesis can then be tested using the PDF of the $F_{n_{1},n_{2}}$ distribution
\beq
\label{FPDF}
f_{n_{1},n_{2}}(x)=\frac{{n_{1}}^{\frac{n_{1}}{2}} {n_{2}}^{\frac{n_{2}}{2}}}{B(\frac{n_{1}}{2}, \frac{n_{2}}{2})}\frac{x^{\frac{n_{1}}{2}-1}}{(n_{1}x+n_{2})^{\frac{1}{2}(n_{1}+n_{2})}},
\eeq
where $B(\cdot, \cdot)$ is Euler's beta function, $n_{1}=q_{2}-q_{1}$ and $n_{2}=M-q_{2}.$

\bibliography{testing-cd.bib}
\end{document}